\documentclass[submission,copyright,creativecommons]{eptcs}

\RequirePackage{etex}

\usepackage{breakurl}             % Not needed if you use pdflatex only.
\usepackage{underscore}           % Only needed if you use pdflatex.

\usepackage[margin=3cm]{geometry}
\usepackage{algcompatible}
\usepackage{mathpazo}
\usepackage{algpseudocode,algorithm,algorithmicx}
\usepackage{graphicx}
\usepackage{subcaption}

\usepackage{fullwidth}
\usepackage{float}
\usepackage{placeins}
\usepackage{tikz}
\usepackage{multicol}
\usepackage{wrapfig}
\usepackage{tkz-graph}
\usepackage{circuitikz}
\usetikzlibrary{circuits.ee.IEC}
\usepackage{bussproofs,varwidth}
\usetikzlibrary{arrows,automata}
\usetikzlibrary{arrows,shapes,positioning,scopes}
\usetikzlibrary{tikzmark,decorations.pathreplacing,fit}
\usetikzlibrary{calc}
\usetikzlibrary{arrows,automata}
\usetikzlibrary{arrows,petri}
\GraphInit[vstyle = Shade]
\usepackage{ulem}
\usepackage[latin1]{inputenc}
\usepackage{ifthen}
\usepackage{mathrsfs}
\usepackage{proof}
\usepackage{romannum}
\usepackage{multirow}
\usepackage{rotating}
\usetikzlibrary{matrix,arrows}
\usepackage{bussproofs}
\usepackage{qtree}
\usepackage{easyReview}
\usepackage{enumitem}
\usepackage{totcount}
\usepackage{tikz}
\usepackage[export]{adjustbox}

\usetikzlibrary{arrows,petri}

\usepackage{pgf}
\usepackage{times}
\usepackage[T1]{fontenc}
\usepackage{amsfonts,amsmath,amssymb,amsthm}
\usepackage[all]{xy}

\usepackage{dingbat}
\usepackage{cmll}
\usepackage{xassoccnt}

\setlist[description]{font=\itshape\textbullet\space,leftmargin=\parindent,labelindent=\parindent}

\NewTotalDocumentCounter{totalfigures}
\DeclareAssociatedCounters{figure}{totalfigures}

\floatstyle{plain}
\floatstyle{ruled}
\restylefloat{algorithm}
\newfloat{myalgo}{tbhp}{mya}
\newcommand{\INDSTATE}[1][1]{\State\hspace{1cm}}
\newlength\indensp
\newlength\myindent
\newcommand*\Let[2]{\State #1 $\gets$ #2}

\algrenewcommand\algorithmicrequire{\textbf{Precondition:}}
\algrenewcommand\algorithmicindent{1mm}

\newcommand{\size}[1]{\ensuremath{\mid#1\mid}}

\newcommand{\imply}{\supset}

\newcommand{\str}[1]{\ensuremath{''#1''}}

\newboolean{numberspec}
\setboolean{numberspec}{false}
\newcounter{specline}

\newcommand{\spec}[3][{}]{%
   \setcounter{specline}{0}%
   \ifthenelse{\equal{#1}{numbers}}%
      {\setboolean{numberspec}{true}}%
      {\setboolean{numberspec}{false}}%
   \ensuremath{
      \begin{array}{l}
         \ifthenelse{\equal{#2}{}}
            {\numberedline}
            {#2\nl}
         #3
      \end{array}
   }
}

\newcommand\Ground{%
\mathbin{\text{\begin{tikzpicture}[circuit ee IEC,yscale=0.6,xscale=0.5]
\draw (0,2ex) to (0,0) node[ground,rotate=-90,xshift=.65ex] {};
\end{tikzpicture}}}%
}

\newcommand{\thiscirc}[1]{%
\texttt{ }\kern5pt\hfill
\begin{circuitikz}%
\draw (0,0) node[ #1 ] {};%(2,0); 
\end{circuitikz}%
{\hspace{5mm}}%
}

{\begin{myalgo}[#1]
\centering
\begin{minipage}{#2}
\begin{algorithm}[H]}%
{\end{algorithm}
\end{minipage}
\end{myalgo}}

\input{tcilatex}

\title{\bf On the horizontal compression of dag-derivations in minimal purely implicational logic}          

\author{Edward Hermann Haeusler
\institute{PUC-Rio, Rio de Janeiro, Brasil}
\institute{Departamento de Inform\'{a}tica\\
Pontifical University Catholic of Rio de Janeiro\thanks{CAPES}\\
Rio de Janeito, RJ, Brasil}
\email{hermann@inf.puc-rio.br}
\and
Jos\'{e} Fl\'{a}vio Cavalcante Barros Jr.
\institute{Quixad\'{a} - UFC\thanks{CAPES}\\
Quixad\'{a}, Brasil}
\email{\quad flavio@gmail.com}
\and Robinson Callou de M.B. Filho
\institute{PUC-Rio, Rio de Janeiro, Brasil}
\institute{Departamento de Inform\'{a}tica\\
Pontifical University Catholic of Rio de Janeiro\thanks{CAPES}\\
Rio de Janeito, RJ, Brasil}
\email{rfilho@inf.puc-rio.br}
}
\def\titlerunning{On the horizontal compression}
\def\authorrunning{E.H.Haeusler, J.F.C. Barros Jr. and R.C.M. Brasil Filho}

\newsavebox{\mypt}
\begin{document}

\maketitle

\pagenumbering{arabic}

\begin{abstract}
In this report, we define (plain) Dag-like derivations in the purely implicational fragment of minimal logic $M_{\imply}$. Introduce the horizontal collapsing set of rules and the algorithm {\bf HC}.   Explain why {\bf HC} can transform any polynomial height-bounded tree-like proof of a $M_{\imply}$ tautology into a  smaller dag-like proof.   Sketch a proof that {\bf HC}  preserves the soundness of any tree-like ND in $M_{\imply}$  in its dag-like version after the horizontal collapsing application. We show some experimental results about applying the compression method to a class of (huge) propositional proofs and an example, with non-hamiltonian graphs, for qualitative analysis. The contributions include the comprehensive presentation of the set of horizontal compression (HC), the (sketch) of a proof that HC rules preserve soundness and the demonstration that the compressed dag-like proofs are polynomially upper-bounded when the submitted tree-like proof is height and foundation poly-bounded. Finally, in the appendix, we show an algorithm that verifies in polynomial time on the size of the dag-like proofs whether they are valid proofs of their conclusions. In the conclusion we discuss what part of the formal results on the {\bf HC}-compressed dag-like proofs have been proved with the use of Interactive Theorem Prover assistance.     
\end{abstract}

\section{Introduction}

In a series of articles, \cite{GH2019, GH2020, GH2022} L.Gordeev and the first author provided a proof that $NP=PSPACE$ based on the transformation of tree-like Natural Deduction derivations for purely intuitionistic minimal logic ($M_{\imply}$) into Direct Acyclic Graphs (Dags). This transformation acts by identifying equal occurrences of formulas at the same level of the tree. This identification, or collapse, changes the tree-like derivation to preserve information about the logical consequence after the collapse. It turns the tree into a Dag. Since the publication of \cite{GH2019}, many readers of it demanded a computer-assisted proof of the main result due to its underlying combinatorial structure being hard to follow. The size of the entire collapsed (Dag-like) proof is upper-bounded by a polynomial on the conclusion of the proof whenever the set of formula occurrences in the original tree-like proof and the height of the tree-like proof also is. Using \cite{Hudelmaier}, the article published in Studia Logica, \cite{GH2019}, closes the requirements on a polynomial bound on both the height and the number of formula occurrences in any ND tree-like proof of a tautology in $M_{\imply}$. Demands of some readers about the need to read another proof-theoretical paper to understand a proof of a result on computational complexity moved the research to a further improvement. In \cite{GH2022}, we drop out the need of \cite{Hudelmaier} for the special case $NP=CoNP$. We show that the horizontal compression can be applied directly to a relevant and specific class of height and formula occurrences poly-bounded Natural Deduction proofs in $M_{\imply}$, namely the class of proofs of non-hamiltonicity for (non-Hamiltonian) graphs. Thus, \cite{GH2022} shows a proof o $CoNP=NP$ that does not need Hudelmaier linear bound. Finally, in \cite{GH2022arxiv} we can find an overview of these proofs that proposes a  non-deterministic approach that takes into account the Dag-like proofs without the need for a deterministic way of reading it. It should be observed that many colleagues and readers have asked to prove most of, maybe all, the results in this report in an Interactive Theorem Prover. The first and the third author of this report are working in the formalization of part of results in this report to publish it here in Arxiv.  

In this report, we describe some experimental results on the application of an implementation of the compression method to a class of (huge) propositional proofs, as well as we show an illustrative example, with non-hamiltonian graphs, to provide a qualitative analysis of {\bf HC}, see section~\ref{sec:Experimental}. The contributions of this report are along with the respective sections. It conveys a comprehensive presentation of the set of horizontal compression (HC) in section~\ref{sec:HC}. It shows the (sketch) of a proof that HC rules preserve soundness in section~\ref{sec:HC-Soundness}. We show, in section~\ref{sec:Upper-HC}, a demonstration that the compressed dag-like proofs are polynomially upper-bounded when submitted to height and foundation poly-bounded tree-like ND proofs. In section~\ref{sec:Experimental} we discuss the experimental results we have on the implementation of the HC compression algorithm and how it performs when compared with the Huffman compression of strings benchmark. Finally, in the appendix~\ref{Algorithm} we show an algorithm that verifies in polynomial time on the size of the dag-like proofs whether they are valid proofs of their conclusions. We want to inform the reader that we are, hopefully,  developing a proof assisted version of some sections of this report. The initial section provides the basics of understanding the report. The reader should have a basic knowledge of Natural Deduction and proof-theory to appreciate better what we convey here. It is interesting mentioning that in \cite{IfCologHaeusler} there is a proof-theoretical argument that provides a good explanation of reasons for the excellent compression rate of HC when compared to traditional benchmarks in terms of string compression. 

We start this report by showing an example of the {\bf HC} compression on a very simple and, we hope, illustrative example. This is the content of the following section~\ref{sec:motiv-example}. Our logical language is restricted to the implication. The logic is the purely implicational propositional logic. We use the symbol $\imply$ for the implication to avoid confusion with the $\rightarrow$ symbol, largely used in graphs pictures and representations. As a matter of fact, many readers and colleagues ask for a {\bf Computer Assisted}  proof of this result, since it is based on the algorithmic application of a set of 

\section{Horizontally compressing Natural Deduction derivations into a Dag-like Derivations}
\label{sec:motiv-example}
\input{Feb2023subsection-motiv-example}

\FloatBarrier

By analysing the dag-like derivation top-down, and, considering that the leaf-nodes determine the initial dependency sets, we can calculated the dependecy set induced by the dag-like derivation. This must be done anyway to verify that the dag-like derivation is correct. 

To verify that the dag-like derivation is sound we perform an algorithm that is linear on the size of the dag-like derivation. The algorithm uses a set of registers, each register stores a pair formed by a dependency set and a path downwards. There is one register for each leaf-node and path that labels the ancestor edge that arrives in this leaf-node. The algorithm goes top-down following the paths in each register and updating the respective dependency set and verifying the correct application of the corresponding rules. If all rules are correct the algorithm ends with the dependency set that is associated to the derivation. This algorithmm its correctness and complexity analysis is shown in appendix~\ref{Algorithm}.

\section{Dag-like proofs basic definitions}

In this section, we define dag-like proofs and derivations in $M_{\imply}$, the minimal purely implicational logic. Dag-like proofs and derivations are data structures resulting from the compression of tree-like Natural Deduction proofs and derivations by the method that we initially defined in \cite{GH2019} based on the collapsing of nodes of the underlying tree of given derivations or proofs.  

This section contains all the basic definitions and concepts used by the horizontal compression method and most of the results, namely theorems and other related results. They are: 1- The method preserves soundness of the original derivation, and consequently proofs also; 2- The method is terminating, and. 3- The size of the compressed proof belongs to  $\mathcal{O}(H\times F^{3})$, where $H$ is the height of the original proof tree and $F$ is the size of the foundation of the proof tree, i.e., the set of all formulas that occurs in the derivation or proof. The minimal implicational logic has only implicational propositional formulas. The Natural Deduction for $M_{\imply}$ has only the $\imply$-Elimination and $\imply$-Introduction rules of N.D. Prawitz for minimal logic, see~\cite{Prawitz}. 

\begin{definition}
  Let $\Pi$ be a Natural Deduction derivation in $M_{\imply}$. Each formula $\alpha$ occurring in $\Pi$ is called a formula occurrence. Each formula occurrence is unique in $\Pi$.
\end{definition}

It is important to emphasize that two formula occurrences in a derivation $\Pi$ may be an instance of the same formula. However, in a derivation, they are unique. We can formalize the feature we mentioned by enumerating the set of all formulas in the derivation so that each formula occurrence in $\Pi$ is a pair $\langle i,\beta\rangle$, where $i$ corresponds to the enumeration of $\beta$. Thus, $\langle i,\beta\rangle$ is unique in the derivation. We omit this pair notation whenever we explicitly state that $\beta$ is a formula occurrence.

\begin{definition} A rooted Dag $\langle V,D,r\rangle$ is a direct acyclic connected graph $\langle V,E\rangle$ together with a distinguished node $r\in V$ that has out-degree 0. $r$ is the only node in $V$, such that $out(r)=0$.  
\end{definition}

In what follows, if $\langle V,D,r\rangle$ is a rooted dag, for each $v\in V$, $In(v)$ is de number of incoming edges to $v$ and $out(v)$ is the number of edges outgoing from $v$. In the case of colored edges, $In_{c}(v)$ and $Out_{c}(v)$ is the number of incoming and outgoing edge to and from the vertex $v$ with color $c$. A colored graph is a structure $\langle V,E_1,\ldots,E_n,r\rangle$, such that, for each $i=1,n$, $E_i\subseteq V\times V$ is the sets of edges colored with color $i$. Moreover, for each $i\neq j$, we have $E_i\cap E_j=\emptyset$. We say that a colored graph has no cycles, iff, each subgraph of color $i$, for each $i=1,n$, has no cycles. Finally, a path from $v$ to $w$ in a colored graph is a non-empty sequences of edges of the same color,linking $v$ to $w$. When such a path exists for a color $c$ we say that the predicate $path_{c}(v,w)$ holds on the respective colored graph.

\begin{definition}\label{def:tree-likederivation}
A tree-like derivation of a formula $\alpha$ in $M_{\imply}$ is a  rooted Dag $\langle V,E_{D},E_{d},r,l\rangle$, bi-colored, with colors $D$ and $d$, where $\langle V,E_{D},r\rangle$ is a rooted Dag with no cycles,  and $l:V\rightarrow L$,  such that:
\begin{enumerate}
\item $\forall v\in V(v\neq r\Rightarrow (in_{D}(v)=1\lor in_{D}(v)=2\lor in_{D}(v)=0))$;
\item $\forall v\in V(in_{D}(v)=1\Rightarrow \exists w(E_D(w,v)\land \forall v_d\in V(E_d(v_d,v)\Rightarrow (l(v)=\str{l(v_d)\imply l(w)}))))$;
  \item $\forall v\in V\forall v_d\in V(E_d(v_d,v)\Rightarrow path_D(v_d,v)\land in_D(v_d)=0)$
\item $\forall v\in V(in_(v)=2\Rightarrow \exists w_1\exists w_2(E_D(w_1,v)\land E_D(w_2,v)\land (l(w_2)=\str{l(w_1)\imply l(v)})))$;
  \item $L\subset Formulas(M_{\imply})$ is a superset of the set of subformulas of $\alpha$. 
% falta falar que in
\end{enumerate}
We call $L$ the foundation of the derivation. Moreover, $l(r)$ is said to be the conclusion of the derivation and the set $\{l(v)/(in(v)=0)\land \neg\exists w(E_d(v,w))\}$ is the set of assumptions of the derivation.\end{definition}

The computational verification that a structure of the form $\langle V,E_{D},E_{d},r,l\rangle$ is a tree-like derivation can be performed in time $\mathcal{O}((\size{E_{D}}+\size{E_{d}}))$. Of course,  with data structures improved with pre-compiled features, such as its list of leaves,  we have that the computational complexity can reach the linear time on the number of all edges. However, this is not essential for us because we want to show that the upper bound is polynomial, which is already the case. 

\begin{definition}
  We say that a tree-like derivation $\mathcal{T}=\langle V,E_{D},E_{d},r,l\rangle$, is a derivation of $\alpha$, a $M_{\imply}$ formula, from a set of assumptions $\Gamma$, a set of $M_{\imply}$ formulas, if and only if, $l(r)=\alpha$ and the set of assumption of $\mathcal{T}$, namely $\{l(v)/(in(v)=0)\land \neg\exists w(E_d(v,w))\}$,  is $\Gamma$.
\end{definition}

In the sequel, we define a mapping from Natural Deduction derivations of $\alpha$ from $\Gamma$ into tree-like derivations of $\alpha$ from $\Gamma$.
It is straighforward to show that all conditions stated in definition~\ref{def:tree-likederivation} holds. 
Finally, The proposition~\ref{prop:labeling} concerns the mapping from tree-like derivations to $M_{\imply}$ N.D. derivations, i.e.,  in the other directions. 

 \begin{proposition}\label{prop:ND2TreelikeDerivs}
   Let $\Pi$ be a Natural Deduction derivation of $\alpha$ from $\Gamma$ in $M_{\imply}$. There is a tree-like derivation $\mathcal{T}=\langle V,E_D,E_d,r,l\rangle$, such that:
   \begin{enumerate}
   \item $\langle V,E_D,\alpha$ is the underlying tree of $\Pi$;
   \item $r\in V$ is the root of $\langle V,E_D,\alpha$;
   \item $l$ labels each node $v\in V$ to is formula occurrence in $\Pi$, in particular $l(r)=\alpha$;
   \item The edges $E_D$ link the nodes regarded to formula occurrences that are premisses of a natural deduction rule application in $\Pi$  to its respective node that is its conclusion;
     \item The edges $E_d$ represent the discharging function associated to the application of an $\imply$-Introduction, such that $E_d(v,w)$ if and only if, $l(w)$ is and implication that is proved via an application of a $\imply$-rule that discharges the occurrence of $l(v)$ located at $v$. 
   \end{enumerate}
   \end{proposition}

\begin{proposition}\label{prop:labeling}
There is a bijective mapping $F$ between finite tree-like derivations and derivations in $M_{\imply}$, such that:
\begin{itemize}
\item If $\mathcal{T}=\langle V,E_{D},E_{d},r,l\rangle$ is a tree-like derivation of $\alpha$ from $\Gamma$, then $F(\mathcal{T})$ is a natural deduction derivation $\Pi$ of $\alpha$ from $\Gamma$ 
\item $l(r)$ is the conclusion of $\Pi$;
\item $\{l(v)/(in(v)=0)\land \neg \exists w(E_d(v,w))\}$ is the set of open assumptions of $\Pi$.
\item For each node $v\in V$, there is a unique formula occurrence $\beta$ in $\Pi$, such that $l(v)=\beta$, and:
  \begin{enumerate}
  \item If $in(v)=0$ and $\neg\exists w\in V(E_d(v,w))\}$ then $\beta\in\Gamma$;
  \item If $in(v)=0$ and $\exists w\in V(E_d(v,w))$ then the $\beta$ occurrence is discharged by a $\imply$-introduction rule that has $l(w)$ as premiss;
  \item If $in(v)=1$ then there is $w\in V$, such that $E_D(w,v)$. So, $\beta$ is conclusion of an $\imply$-introduction in $\Pi$ that has an occurence $l(w)$ as premiss;
  \item If $in(v)=2$ then there are $w_1,w_2\in V$, such that $E_D(w_1,v)$ and $E_D(w_2,v)$ holds and $l(w_2)=''l(w_1)\imply l(v)''$. So, $\beta$ is the conclusion of an $\imply$-elimination rule in $\Pi$ that has an occurrence of $l(w_1)$ as minor-premiss and an occurrence of $l(w_2)$ and major-premiss.
  \end{enumerate}
\item For each formula occurrence $\beta$ in $\Pi$ there is a unique node $v\in V$, such that $l(v)=\beta$, and:
  \begin{enumerate}
  \item If $\beta$ is the conclusion of $\Pi$ then $l(v)=\beta$ and $out(v)=0$;
  \item If $\beta$ is a discharged assumption then $in(v)=0$ and $\exists w\in V(E_d(v,w))$ and $l(w)=$''$\beta\imply\beta^{\prime}$'' for some $\beta^{\prime}$;
  \item If $\beta$ is an open assumption then $in(v)=0$ and $\neg \exists w\in V(E_d(v,w))$;
  \item If $\beta$ is $\beta_1\imply \beta_2$ that is the conclusion of an $\imply$-introduction rule in $\Pi$ then there is $w\in V$, such that, $l(w)=\beta_2$ and $E_D(w,v)$;
  \item If $\beta$ is the conclusion of an $\imply$-elimination rule in $\Pi$ then there is  
  \end{enumerate}
  \end{itemize}
\end{proposition}

\section{Dropping out the discharging ($E_d$) edges}
\label{sec:Drop-Discharging-Edges-Out}

We will drop out the discharging edges by assigning to each deduction edge the string of bits that represents the set of assumptions from which the formula that labels the target of it depends on.

Let $L$ be the foundation of a derivation $\Pi$ in $M_{\imply}$. Consider a linear ordering $\mathcal{O}(L)=\{\beta_0,\beta_1,\ldots,\beta_k\}$ on $L$.   

\begin{definition}\label{def:Bitstrings}
  Let $L$ be a set of formulas in $M_{\imply}$ and $\mathcal{O}(\alpha)$ be a linear  order $\mathcal{O}(L)=\{\beta_0,\beta_1,\ldots,\beta_k\}$ . A bitstring on $\mathcal{O}(L)$ is any string $b_0b_1\ldots b_k$, such that $b_i\in\{0,1\}$, for each $i=1\ldots k$. There is a bijective correspondence between bitstrings on $\mathcal{O}(\alpha)$ and subsets of $L$, given by $Set(b_0b_1\ldots b_k)=\{\beta_i / b_i=1\}$.
\end{definition}

The bitstring on $\mathcal{O}(L)$ will drop out the discharging edges and make explicit the information on formula dependencies in a derivation. The set of bitstrings on $\mathcal{O}(L)$ is denoted by $Bits(L,\mathcal{O}(L))$. The inverse function of $Set$ is well-defined and is denoted by $Set^{-1}$
The following result shows that the set of theorems is not changed when considering a restricted form of $\imply$-intro rules. This restricted form of a $\imply$-intro provides a sound way to remove the discharging edges from the tree-like proofs.

\begin{definition}
   Consider a derivation $\Pi$ of $\beta$ having $\Delta$ as assumptions. Let $\alpha\in\Delta$ be a (open) formula assumption in $\Pi$. A  $\imply$-intro application in $\Pi$ is greedy if and only if it produces $\alpha\imply\beta$ as a conclusion and discharges every open occurrence of $\alpha$ from which its premise $\beta$ depends on.
\end{definition}

An application of an $\imply$-intro in a tree-like derivation is greedy, if and only if, its corresponding application in a Natural Deduction derivation is also greedy.

The following function takes an arbitrary Natural Deduction proof of $\Pi$ of $\alpha$ and produces a greedy proof $G(\Pi)$ of $\alpha$. With it we can proof lemma~\ref{Greedy} by a simple induction on the size of the proof.  

\[
G(\Pi)=
\begin{cases}
  \alpha & \text{if $\Pi=\alpha$} \\
  \AxiomC{$G(\Pi^{\prime})$}
  \RightLabel{$\imply$-greedy}
  \UnaryInfC{$\beta_1\imply\beta_2$}
  \DisplayProof & 
  \text{ if $\Pi$=\AxiomC{$\Pi^{\prime}$}
            \RightLabel{$\imply$-intro}
            \UnaryInfC{$\beta_1\imply\beta_2$}
            \DisplayProof
  } \\
 %% $\begin{prooftree}
  \AxiomC{$G(\Pi_1)$}
  \AxiomC{$G(\Pi_2)$}
  \RightLabel{$\imply$-Elim}
  \BinaryInfC{$\beta$}
  \DisplayProof & \text{ if $\Pi$= %\begin{prooftree}
                            \AxiomC{$\Pi_1$}
                            \AxiomC{$\Pi_2$}
                            \RightLabel{$\imply$-Elim}
                            \BinaryInfC{$\beta$}
                            \DisplayProof
                  }
 %% \end{prooftree}$}
\end{cases}
\]

The following algorithm also produces from a proof $\Pi$of $\alpha$ a $M_{\imply}$ greedy proof of $\alpha$. Consider that $n$ is the level of the highest branch in $\Pi$ and let $P$ be  $\Pi$ in the code:
\begin{verbatim}
j=n
While 0<j do:  
  For each branch B of level j in P: 
       replace each intro-app downwards by a greedy intro-app
       possibly discharging more formula occurrences in B
  j=j-1
end
\end{verbatim}

\begin{lemma}[Greedy $\imply$-Intro are complete]\label{Greedy}
  Let $\Pi$ be a proof of $\alpha$. Let $\Pi^{\prime}=G(\Pi)$ be the result of the above function applied to $\Pi$. We conclude that
  $\Pi^{\prime}$ is also a valid proof of $\alpha$ in $M_{\imply}$.
\end{lemma}

Proof. By induction on the size of the argument $\Pi$ we show that every $\imply$-intro rule in $G(\Pi)$ is a greedy application. So we are done.  
\begin{flushright}
  $\Box$
\end{flushright}

We will remove the discharging function used in ND derivations to mark the formula occurrences discharged by a $\imply$-intro application. We will use labelling of formula occurrences in the ND derivations producing a decorated ND derivation that contains all the information regarding dependencies of any formula in the derivation. Below we define decorated greedy Natural Deduction derivations. We use the notation $\vec{b}_{\alpha}$ to denote the bitstring $b_0b_1\ldots b_k$, such that $Set(b_0b_1\ldots b_k)=\{\alpha\}$, i.e., the only $b_i=1$ is when $i$ is the order of $\alpha$.  

\begin{definition}[Decorated Greedy N.D. derivation]
Let $\Gamma$ be a set of formulas and let $\mathcal{O}_{\Gamma}$ be any linear ordering on these formulas. Let $\mathcal{B}(\mathcal{O}_{\Gamma})$ be the set of bitstrings on the ordering $\mathcal{O}_{\Gamma}$. Let $\Pi$ be a $M_{\imply}$ derivation having an element of $\mathcal{B}(\mathcal{O}_{\Gamma})$ attached as a label to each formula occurrence in $\Pi$. We say that $\Pi$ is a Decorated Greedy N.D. derivation, iff, the following conditions hold to every formula occurrence $\beta$, labelled with $\vec{b}$, i.e. $\beta^{\vec{b}}$ in symbols, in $\Pi$:
  \begin{enumerate}
  \item If $\beta^{\vec{b}}$ is a top-formula (leaf) in $\Pi$ then $\vec{b}=\vec{b}_{\beta}$;
  \item If $\beta^{\vec{b}}$ is the conclusion of a $\imply$-Elim rule having $\beta_1^{\vec{b_1}}$ and $(\beta_1\imply\beta)^{\vec{b2}}$ as premisses then $\vec{b}=\vec{b_1}\oplus\vec{b_2}$;
  \item If $\beta^{\vec{b}}=(\beta_1\imply\beta_2)^{\vec{b}}$ is the conclusion of a Greedy $\imply$-Intro rule application that has $\beta_2^{\vec{b^{\prime}}}$ as premiss then $\vec{b}=\vec{b^{\prime}}-\vec{b}_{\beta_1}$
  \end{enumerate}
\end{definition}

\begin{definition}
  Given a derivation $\Pi$ in $M_{\imply}$, let $\Gamma$ be the set of formulas occurring in $\Pi$ and $\mathcal{O}_{\Gamma}$ be any linear ordering on these formulas.  Let $L$ be the a mapping from the set of formula occurrences in $\Pi$ to $\mathcal{B}(\mathcal{O}_{\Gamma})$. This mapping is called a {\bf labeling} on its formulas.
\end{definition}

\begin{definition}[Adequate Labeling]\label{def:adequatelabeling}
  Let $\Pi$ be a derivation in $M_{\imply}$ and $L$ be a labeling on its formula occurrences. We say that $L$ is adequate to $\Pi$, if and only if, for every formula occurrence $\alpha$ of $\Pi$, $Set(L(\alpha))$ is the set of formulas from which this occurrence of $\alpha$ depends on.
  \end{definition}

\begin{lemma}\label{lemma:adequatelabel}
  For every greedy $M_{\imply}$ derivation $\Pi$, there is a labeling $L$, such that, $L$ is adequate to $\Pi$.
\end{lemma}

{\bf Proof} The fact that $\Pi$ has only greedy $\imply$-intro applications implies that for every conclusion $\alpha\imply\beta$ of an $\imply$-intro application in $\Pi$ has no $\alpha$ belonging to its dependencies. Thus, by induction on the size of the derivation $\Pi$ we prove that there is an adequate labeling for $\Pi$. 
\begin{flushright}
  $\Box$
\end{flushright}

The above lemma and lemma~\ref{Greedy} supports the following proposition.

\begin{proposition}\label{prop1}
  For every set of formulas $\Gamma$ and formula $\alpha$ in $M_{\imply}$, such that $\Gamma\vdash_{M_{\imply}}\alpha$, there is a decorated greedy derivation of $\alpha$ from $\Gamma$. In fact the label of $\alpha$ is $\vec{b}_{\Gamma}$.
\end{proposition}

\begin{definition}
  A decorated tree-like (greedy) derivation with adequate labeling is a structure $\mathcal{T}=\langle V,E_D,E_d,r,l,L\rangle$, where $F=\bigcup_{v\in V}\{l(v)\}$ is the foundation of $\mathcal{T}$, $\mathcal{O}_{F}$ an linear ordering on $F$ and $L:E_D\rightarrow \mathcal{B}(\mathcal{O}_{F})$ is an adequate labeling.
  \end{definition}

And using the above proposition we obtain:

\begin{proposition}\label{prop2}
  For every set of formulas $\Gamma$ and formula $\alpha$ in $M_{\imply}$, such that $\Gamma\vdash_{M_{\imply}}\alpha$ there is a tree-like (greedy) derivation with adequate labeling $\Pi$ of $\alpha$ from $\Gamma$ and a labeling $L$ of the deduction edges, i.e. the elements of $E_{D}$, by elements of $\mathcal{B}(\mathcal{O}_{S})$, for some ordering $\mathcal{O}_{S}$ of the set $S$ of formulas occurring in $\Pi$. $L$ is such that $Set_{\mathcal{O}_{S}}(L(\langle v_1,v_2\rangle))$ is the set of formulas from which the formula occurrence $l(v_1)$ depends on inside $\Pi$. 
\end{proposition}

{\bf Proof} We apply proposition~\ref{prop1} and lemma~\ref{lemma:adequatelabel} to obtain a decorated greedy derivation $\Pi$ with an adequate labeling $L^{\prime}$. Then, an application of 
Proposition~\ref{prop:ND2TreelikeDerivs} obtains a greedy tree-like derivation $\mathcal{T}=\langle V,E_D,E_d,r,l\rangle$ of $\alpha$ from $\Gamma$. Finally, we define $L(\langle v,w\rangle)=L^{\prime}(v)$, for every $\langle v,w\rangle\in E_D$ to obtain a tree-like (greedy) derivation with adequate labeling.     
  \begin{flushright}
    $\Box$
  \end{flushright}

  \begin{definition}
    The tuple $\langle V,E_{D},E_{d},r,l,L\rangle$ and the order $\mathcal{O}_{S}$ on the range of $l$ is called a decorated greedy tree-like derivation of $l(r)$ from $\{l(v)\mid in(v)=0\}$, iff, every application of $\imply$-introduction is greedy, $Set{\mathcal{O}_{S}}(L(\langle v_1,v_2\rangle))$ is the set of formulas from which the formula occurrence $v_1$ depends on, for every edge $\langle v_1,v_2\rangle$ form $E_{D}$.
  \end{definition}

  From the last two propositions, proposition~\ref{prop1} and~\ref{prop2}, we obtain the following theorem. 

\begin{theorem}
  For every derivation $\Pi$ of $\alpha$ from $\Gamma$ that only has Greedy $\imply$-intro rules and $\imply$-Elim rules there is a decorated greedy tree-like derivation (DGTD) $\Pi_T$ of $\alpha$ from $\Gamma$.
\end{theorem}

\begin{corollary}
  In the above statement, $size(\Pi)=size(\Pi_T)$, and if $\Pi$ is normal, then $\Pi_T$ is normal too.
  \end{corollary}

We observe that the discharging edges $E_{d}$ provide information on which formula occurrence is discharged by which rule. This information is subsumed by the labeling $L$, as shown by the corollary below.

\begin{corollary}\label{dischargecorolario}
  Let $\langle V,E_{D},E_{d},r,l,L\rangle$ be a decorated greedy tree-like derivation, $\mathcal{O}_{S}$ be the order on  the range of $l$, and $L$ the associated labeling of dependencies. We can conclude that the information provided by $E_{d}$ can be determined using the other components of the tuple.
\end{corollary}

\section{The Horizontal compressing procedure}
\label{sec:HC}
This section defines the horizontal compression procedure and proves some properties about it. The compression follows the
definition of horizontal collapse. A horizontal collapse applies to a dag-like decorated greedy derivation. It aims to identify two or more nodes in the rooted dag-like derivation at the same deduction level. The collapsing applies from the conclusion level, namely the zero level, towards the assumptions levels. When applied to tree-like rooted, and decorated, derivations, it yields dags instead of trees. The following definition formally defines this operation as a cases style definition. Forty rules form a case analysis that applies to a Dag-like derivation to yield a (new) dag-like derivation. The horizontal collapsing initially transforms tree-like derivations into dag-like derivations. Additional structure is needed to allow us to verify that a particular dag-like derivation is a (correct) derivation, indeed. We define a dag-like derivability structure as the underlying structure to encode dag-like derivations. Thus, a dag-like derivation is a dag-like derivability structure (DLDS) instance and a condition that should be true about this DLDS instance.

\begin{definition}[Dag-like derivability structures DLDS]\label{DLDS}
  Let $\Gamma$ be a set of $M_{\imply}$ formulas and  $\mathcal{O}_{\Gamma}$ an arbitrary linear ordering on $\Gamma$ and $\mathcal{O}_{\Gamma}^{0}=\mathcal{O}_{\Gamma}\cup\{0,\lambda\}$\footnote{$0<n$, for every $n\in \mathcal{O}_{\Gamma}$}. A dag-like derivability structure, {\bf DLDS for short}, is a tuple $\langle V , (E_{D}^{i})_{i\in\mathcal{O}_{\Gamma}^{i}} , E_{A}, r, l, L, P\rangle$, where:
  \begin{enumerate}
  \item $V$ is a non-empty set of nodes;
  \item For each $i\in \mathcal{O}_{\Gamma}^{0}$, $E_{D}^{i}\subseteq V\times V$ is the family of sets of edges of deduction;
  \item $E_{A}\subseteq V\times V$ is the set of edges of ancestrality;
  \item $r\in V$ is the root of the {\bf DLDS};
  \item $l:V\rightarrow \Gamma$ is a function, such that, for every $v\in V$, $l(v)$ is the (formula) label of $v$;
  \item $L:\bigcup\limits_{i\in \mathcal{O}_{\Gamma}^{0}}E_{D}^{i}\rightarrow \mathcal{B}(\mathcal{O}_{S})$ is a function, such that, for every $\langle u,v\rangle\in E_{D}^{i}$, $L(\langle u,v\rangle)$ is a bitstring. 
  \item $P:E_A\rightarrow \{1,\ldots,\mid\mid\Gamma\mid\mid\}^{\star}$, such that, for every $e\in E_A$, $P(e)$ is a string of the form $o_1;\ldots;o_n$, where each $0_i$, $i=1,n$ is an ordinal in $\mathcal{O}_{\Gamma}$;
  \end{enumerate}
\end{definition}

Note that for each $i\in\mathcal{O}_{\Gamma}^{0}$, and, $\langle u,v\rangle\in E_{D}^{i}$, $i$ can be seen as the color of the edge $\langle u,v\rangle$. Thus, each deduction edge is coloured with formulas from $\Gamma$ or the 0 colours. A colour different from $0$  is introduced every time a collapsing of nodes is performed. A tree-like greedy derivation has only 0 coloured deduction edges. The algorithm below is responsible for the horizontal compression.
Afterwards we ask that the $Set(\langle u,v\rangle)$, $\langle u,v\rangle\in E_{D}^{i}$ is a dependency set and $L$ is an adequate labeling, according to definition~\ref{def:adequatelabeling}. Edges in $E_D^{\lambda}$ are the result of the collapse of the edges. The $\lambda$ label should be read as an intentional absence of labelling. In the algorithm to verify that a {\bf DLDS} is valid or not, the dependency set associated with an edge labelled with $\lambda$ should be calculated dynamically, while all the other members of $\bigcup\limits_{i\in \mathcal{O}_{\Gamma}^{0}}E_{D}^{i}$ have it statically assigned. 

\begin{algorithm}[H]
  \caption{Horizontal Compression}
  \label{HorCom}
  \begin{algorithmic}[1]
    \Require{A tree-like greedy derivation $\mathcal{D}$}
    \Ensure{The DLDS that is $\mathcal{D}$ compressed} 
    \For{$lev$ from 1 to $h(\mathcal{D})$}
    \For{$u$ and $v$ at $lev$}
    \State $HCom(u,v)$
    \EndFor
    \EndFor
  \end{algorithmic}
\end{algorithm}

\subsection{The compression rules}
\label{subsec:Rules}

Below we show~\total{rule}~compression rules that define $HCom(u,v)$. Some of the rules already used and presented  in the example in section~\ref{sec:motiv-example} are not repeated here. Some symmetric\footnote{As an example, $\mathbf{R_e2EI}$ is symmetric to $\mathbf{R_e2IE}$}  rules may be omitted. Each of them applies to a specific case,  the DLDS $\mathcal{D}$, that matches the left-hand side. The effect of collapsing two nodes (vertexes) of a {\bf DLDS} that machtes the left-hand side of the rule produces a new {\bf DLDS} depicted by the right-hand side of the rewritting rule. It is worth noting that every decorated greedy tree-like derivation is a {\bf DLDS} having $P(v)=\epsilon$, for every $v\in V$. Figure~\ref{R0IE}, i.e., rule {\bf R0IE}  represents the rewrite rule that collapses the conclusion of an application of $\imply$-Intro with an application of $\imply$-Elim. Figure~\ref{DOIS}, i.e., rule {\bf R0HE}, represents the collapse of a hypothesis (assumption or top-formula) with the conclusion of a $\imply$-Elim. Figure~\ref{TRES}, rule{\bf R0IH}, represents the conclusion of a $\imply$-Intro with a hypothesis. Finally, figure~\ref{QUATRO}, rule {\bf R0HH}, represents the collapse of two hypotheses. Other combinations are:   the collapse of the conclusions of two $\imply$-Elim rules; the collapse of a conclusion of a $\imply$-Elim with a $\imply$-Intro; the collapse of two conclusions of $\imply$-Intro rules; the collapse of the conclusion of a $\imply$-Elim with a hypothesis. Finally, we have the collapse of a hypothesis and a $\imply$-Intro. Of course, these are all the nine possibilities. We will see that we have three four classes of rules. The naming of the rules follow a schema. They are named as $Rim_lm_r$, where $i=0\ldots 3$ and $m_l,m_r\in \{I,E,H,*\}$, such that, $i$ is the type of the rule, $m_l$ and $m_r$ means the kind of the labels in the left and right nodes that will collapse. For example, the rule {\bf R0IE} is the official name of the rule in figure~\ref{UM}. Note all names correspond to rules. For example there is no rule with the name {\bf R3II}. This will be better explained in the sequel. It is worth noticing that when one of the nodes to be collapsed is already the result of a previous collapse then there are more than one rule applied to have it as conclusion. In this case we use the letter $X$ in the name formation of the rule. For example, {\bf R1XE} is the name of the rule that collapses a left node that is already the result of a collapse with a right node that is conclusion of an $\imply$-Elim. The rule {\bf R1XE} appears in figure~\ref{R1XE}. There is a small variation in the naming of rules. There are names such as $R_{v}2m_lm_r$ and $R_e2m_lm_r$. The indexes $v$ and $e$ indicates that the respective rule collapses only vertexes ($v$) or edges ($e$) too. See figures~\ref{Rv2EE} and~\ref{Re2EE} for a better understanding of this. For example, the rule $\mathbf{R0EE}$, used in the compressing example in figure~\ref{UM-EE-motiv-section} collapes the conclusions of two elimination rules.

Figure~\ref{R0IE} shows a Type=\Romannum{1} rule, as defined below, named {\bf R0IE}. We use this figure to re-emphasize how to read the pictorial representation of each horizontal compression rule. Both the left and the right-hand sides are subgraphs of, respectively, the {\bf DLDS} $\mathcal{D}$ and $\mathcal{D}^{\prime}$. The meaning of this rule is to replace the subgraph represented by the left-hand side by the right-hand side graph in $\mathcal{D}$, resulting in $\mathcal{D}^{\prime}$,defined below, where $\bullet_a$ is the left $\bullet$ in the figure, while $\bullet_b$ is the right one.  
\[
\mathcal{D}^{\prime}=\langle V^{\prime} , (E_{D}^{i})^{\prime}_{i\in\mathcal{O}_{\Gamma}^{i}} , E^{\prime}_{A}, r, l^{\prime}, L^{\prime}, P^{\prime}\rangle
\],
considering $\mathcal{D}=\langle V , (E_{D}^{i})_{i\in\mathcal{O}_{\Gamma}^{i}} , E_{A}, r, l, L, P\rangle$, and:
\begin{itemize}
\item $V^{\prime}=V-\{v\}$,
\item $(E_D^{0})^{\prime}=E_D^{0}-\{\langle u,\bullet_a\rangle,\langle v,\bullet_b\rangle,\langle p_2,v\rangle, \langle p_3,v\rangle \}$,
\item $(E_{D}^{1})^{\prime}=(E_D^1)\cup\{\langle u,\bullet_a\rangle\}$,
\item $(E_{D}^{2})^{\prime}=(E_D^2)\cup\{\langle u,\bullet_b\rangle\}$,
\item $(E_D^{i})^{\prime}=E_D^{i}$, for each $i>2$,
\item $E^{\prime}_{A}=E_{A}\cup\{\langle\bullet_a,p_1\rangle,\langle\bullet_b,p_2\rangle, \langle\bullet_b,p_3\rangle\}$,
  \item $l^{\prime}=l/V^{\prime}$,  
  \item $L^{\prime}=L/(E_{D}^{i})^{\prime}_{i\in\mathcal{O}_{\Gamma}^{i}}[\langle u,\bullet_a\rangle\leftarrow \bar{c_1}-\bar{p_0};\langle u,\bullet_b\rangle\leftarrow\bar{b_1}\lor\bar{d_1}]$, and finally,
  \item $P^{\prime}=P[\langle\bullet_a,p_1\rangle\leftarrow 1;\langle\bullet_b,p_2\rangle\leftarrow 2;\langle\bullet_b,p_3\rangle\leftarrow 2]$.
\end{itemize}

In the left-hand side of the rule in figure~\ref{R0IE}, $p_i$, $i=1,3$, $u$ and $v$ are different nodes in the subgraph, such that $l(v)=l(u)$, the black arrows are deductive edges, which have as labels the bitstring representing the dependency set denoted by $L$. For example, $L(\langle p_1, u\rangle)=\bar{c}_1$ shows that the deductive edge $\langle p_1, u\rangle\in E_d^{0}$ is labeled by the dependency set $Sets(\bar{c}_1)$. The absence of a label on an edge indicates that the edge is unlabeled. A label's node is $\bullet$ whenever it is not relevant what is its label to read the rule. In this case the $\bullet$ is also used to denote the node. $\bullet$s label different nodes always. In figure~\ref{R0IE} the bullets label different nodes. In the set-theoretical semantics of the rules, explained in the previous paragraph we use $\bullet_a$ and $\bullet_b$ to reference the the two different nodes. Edges that belong to $E_{D}^{i}$ have the colour $i$; this is the red ordinal number $1,\ldots,n$ on a black deduction edge. The members of $E_A$, the ancestor edges, are coloured blue, and their labels under $P$ labelling function are red in the picture. For example, $\langle \bullet, p_1\rangle\in E_A$ and $P(\langle \bullet, p_1\rangle)=1$. Moreover, we have that $\langle u,\bullet\rangle\in E_D^{1}$ in the graph in the right-hand side of {\bf R0IE}.
In the sequel, we will find the label $\lambda$ assigned to some edges, i.e., those belonging to $E_D^{\lambda}$. As already mentioned in the paragraph after definition~\ref{DLDS}, the members of $E_D^{\lambda}$ are edges that have the dependency set calculated by the main Algorithm in appendix~\ref{Algorithm} that verifies if a {\bf DLDS} is valid. Moreover, a node in the left-hand side shows all the edges outgoing or incoming it. If there is no incoming edge drawn then there is no drawn at all.

There is a particular case that is present in some rules, namely the rules in figures~\ref{R0HE} to~\ref{R0HH},~\ref{Re2IH},~\ref{Re2HE},~\ref{Re2HH}, ~\ref{Rv2IH},~\ref{Rv2HE}, ~\ref{Re3XH} and~\ref{Rv3XH}. This particular feature has to done with the fact that at least one of collapsed nodes is labeled with a top-formula, that is, an hypothesis. In each of these rules, the right-hand side contains a marking $h$ indicating that at least of one of the collapsed nodes is an hypothesis. 

Regarding to understanding the remaining rules, we  do not need any additional explanation, and the reader should read them as explained in the precedent paragraphs. We advise the reader that all the rules in the graphical representation in the sequel assume that nodes and edges drawn in different positions are different always. For example, in figure~\ref{UM}, i.e {\bf R0IE}, $p_0$, $p_1$, $p_2$, $p_3$, $u$, $v$ and the two bullets ($\bullet$) below them are all pairwisely different nodes. Dashed lines represent possible paths in the graph.

\subsection{On the classification and type of rules}
\label{subsec:Rules-Classification}

Let $\beta$ be formula occurrence in an ND derivation in $M_{\imply}$, it is a hypothesis or the conclusion of a $\imply$-application or the conclusion of a $\imply$-Elim. Nine is the total number of possible pairs. Consider, for example,  the rules shown in figures ~\ref{UM},~\ref{DOIS},~\ref{TRES}, ~\ref{QUATRO} and~\ref{UM-EE}, i.e, rules {\bf R0IE}, {\bf R0HE}, {\bf }, {\bf } and {\bf } together with the remaining four cases. We consider that the full subgraph of the {\bf DLDS},  to which they apply,  determined by the set of nodes reachable from the nodes in their respective left-hand graph, does not have any collapse yet. For example in the context of the {\bf HC}-compression, when applying {\bf R0IE}, any node below the two bullets that are reachable from some of them are not collapsed node. These nine rules are what we call type-0 rules. In contrast, the rules in figure~\ref{CINCO} and~\ref{SEIS} are type-\Romannum{1} rules. They consider, i.e. have as a precondition, that their respective left-hand side graph represents a pair of nodes to collapse, such that exactly one of them is already the result of a previous collapse. Remember that the collapses follow the algorithm from the bottom up and left to right. There are three possible type-\Romannum{1} rules; depending on which rule, $\imply$-Intro or $\imply$-Elim, the right node to collapse is the conclusion of or if it is a hypothesis. The rules in figures~\ref{SETE} to~\ref{DEZ},~\ref{DEZESSEIS} and~\ref{DEZESSETE} are of type-\Romannum{2}. In contrast with type-0 and type-\Romannum{1}, their precondition is that (at least one of) the  nodes to be collapsed is target of an ancestor edge, i.e. a member of $E_A$. What we stated is equivalent to saying that their respective sons has already been collapsed before, according to the order of execution of algorithm~\ref{HorCom}. The type-\Romannum{2} rules are of two kind. Either they collapse only the vertexes or vertexes and edges of the {\bf DLDS}. The letter $v$ and $e$ indicates what is the kind in their respective rule names. There are nine possible combinations, as in type-0, for each kind. We do not find it necessary to show the remaining six cases to avoid very similar repeating rules. The type-\Romannum{2} rules are listed in definition below. They represent the collapse of two nodes that are targets of ancestor edges. In opposition to the type-\Romannum{1} rules, in the next level below the nodes that will be collapsed, there is no node resulting from an already collapsed pair of nodes. These already collapsed nodes occur below this next level. The total number of rules of the type-\Romannum{2} is 18; in the listing of rules i the sequel, some of then may be ommited, but we have to consider them too. We have also the type-\Romannum{3} rules that collapses one node that is already collapsed with other node not collapsed yet, in the context of the existence of A-edges arriving on the collapsed nodes.  These are the rules with names $\mathbf{R_e3XE}$, $\mathbf{R_e3XI}$, $\mathbf{R_e3XH}$, $\mathbf{R_v3XE}$, $\mathbf{R_v3XI}$, $\mathbf{R_v3XH}$.

We also consider the classification of the rules according to the effect that they have on the A-edges, i.e., if they move  ancestor edges downwards or upwards. The {\bf MDE}-rules, for {\bf M}oving {\bf D}own {\bf E}dges,  are used to collapse at least one top-node in a pair of nodes that have the same formula labeling them. We have already seen some of these rules in section~\ref{sec:motiv-example}, rules $\mathbf{R_v2HH}$, $\mathbf{R_e2HH}$ and $\mathbf{R_e2XH}$ are examples of {\bf MDE} rules. The {\bf MUE} rules,  for {\bf M}ove {\bf U}pwards {\bf E}dges, are those used to move the A-edge up, when collapsing nodes that are target of these edges.

\begin{definition}[Types of rules]
  The rules of compression are classified in types. The classification is the following:
  \begin{description}
  \item[Type-0] {\bf R0HH, R0HI, R0HE, R0IH, R0IE, R0II, R0EH, R0EI, R0EE};
  \item[Type-\Romannum{1}] $\mathbf{R1XH}, \mathbf{R1XE}, \mathbf{R1XI}$;
  \item[Type-\Romannum{2}] $\mathbf{R_v2EI}$, $\mathbf{R_e2EI}$, $\mathbf{R_v2EE}$, $\mathbf{R_e2EE}$, $\mathbf{R_v2EH}$, $\mathbf{R_e2EH}$, $\mathbf{R_v2II}$, $\mathbf{R_e2II}$, $\mathbf{R_v2IE}$, $\mathbf{R_e2IE}$, $\mathbf{R_v2IH}$, $\mathbf{R_e2IH}$, $\mathbf{R_v2HI}$, $\mathbf{R_e2HI}$, $\mathbf{R_v2HE}$, $\mathbf{R_e2HE}$, $\mathbf{R_v2HH}$, $\mathbf{R_e2HH}$;
    \item[Type-\Romannum{3}] $\mathbf{R_v3XH}$, $\mathbf{R_e3XH}$, $\mathbf{R_v3XE}$, $\mathbf{R_e3XE}$, $\mathbf{R_v3XI}$, $\mathbf{R_e3XI}$.     
    \end{description}
\end{definition}

Other dimension for classifying the compression rules is by the effect that they make in the A-edges.

\begin{definition}[Moving up and down A-edge rules]
  The rules of compression are also classified according their effects on the ancestor edges:
  \begin{description}
  \item[MUE- Moving Up Edges] $\mathbf{R_v2EI}$, $\mathbf{R_e2EI}$, $\mathbf{R_v2EE}$, $\mathbf{R_e2EE}$,$\mathbf{R_v2II}$, $\mathbf{R_e2II}$, $\mathbf{R_v2IE}$, $\mathbf{R_e2IE}$, $\mathbf{R_v2HI}$, $\mathbf{R_e2HI}$, $\mathbf{R_v2HE}$, $\mathbf{R_e2HE}$, $\mathbf{R_v2EH}$, $\mathbf{R_e2EH}$ $\mathbf{R_v2IH}$, $\mathbf{R_e2IH}$, $\mathbf{R_v3XE}$, $\mathbf{R_e3XE}$, $\mathbf{R_v3XI}$, $\mathbf{R_e3XI}$; $\mathbf{R_v3XH}$, $\mathbf{R_e3XH}$, $\mathbf{R_v3XE}$, $\mathbf{R_e3XE}$, $\mathbf{R_v3XI}$, $\mathbf{R_e3XI}$, $\mathbf{R_v3XH}$, $\mathbf{R_e3XH}$;    
  \item[MDE- Moving Down Edges] $\mathbf{R_v2HH}$, $\mathbf{R_e2HH}$, $\mathbf{R_e2XH}$, $\mathbf{R_v2XH}$;  
  \end{description}
  \end{definition}

%    {\bf Obs.} From the rules enumerated in the definition above, only the rules {\bf R0IE, R0HE, R0IH, R0HH} and {\bf R0EE} are depicted in this article. The type-\Romannum{1} rules when one of the nodes is a hypothesis could b

%==============================

%\newcounter{rule}
%\setcounter{rule}{1}

\begin{figure}[!h]
 \begin{minipage}{.47\textwidth}

\end{minipage}%
%\end{array}
%\]
 \caption{(a)$u$ and $v$ collapse   \hspace{2cm} (b) After collapse $HCom(u,v)$}
 \label{figI}\label{UM}\label{R0IE}
\end{figure}
\stepcounter{rule}

\begin{figure}
 \begin{minipage}{.47\textwidth}
 %
\end{minipage}%
%\end{array}
%\]
 \caption{(a)$u$ or $v$ assumptions coll.    \hspace{2cm} (b) After collapse $HCom(u,v)$ }
 \label{figIb}\label{DOIS}\label{R0HE}
\end{figure}
\stepcounter{rule}

% ======================= Rule =======================

\begin{figure}
 \begin{minipage}{.47\textwidth}
 %
\end{minipage}%
%\end{array}
%\]
 \caption{(a)$u$ and $v$ collapse   \hspace{2cm} (b) After collapse $HCom(u,v)$}
 \label{figIc}\label{TRES}\label{R0IH}
\end{figure}
\stepcounter{rule}

% ======================================== Rule =======================================

\begin{figure}
 \begin{minipage}{.47\textwidth}
 %
\end{minipage}%
%\end{array}
%\]
 \caption{(a)$u$ and $v$ collapse   \hspace{2cm} (b) After collapse $HCom(u,v)$}
 \label{figId}\label{QUATRO}\label{R0HH}
\end{figure}
\stepcounter{rule}

% ==================== Rule =================================================

\begin{figure}
 \begin{minipage}{.47\textwidth}
 %
\end{minipage}%
%\end{array}
%\]
 \caption{(a)$u$ and $v$ collapse   \hspace{2cm} (b) After collapse $HCom(u,v)$}
 \label{figIV}\label{CINCO}\label{R1XI}
\end{figure}
\stepcounter{rule}

% ===================================== Rule =================================

\begin{figure}
 \begin{minipage}{.47\textwidth}
 %
\end{minipage}%
%\end{array}
%\]
 \caption{(a)$u$ and $v$ collapse   \hspace{2cm} (b) After collapse $HCom(u,v)$}
 \label{figIV}\label{CINCO}\label{R1XE}
\end{figure}
\stepcounter{rule}

% ======================================Rule ==================================

\begin{figure}
 \begin{minipage}{.47\textwidth}
 %
\end{minipage}%
%\end{array}
%\]
 \caption{(a)$u$ and $v$ collapse   \hspace{2cm} (b) After collapse $HCom(u,v)$}
 \label{figIVb}\label{SEIS}\label{R1XH}
\end{figure}
\stepcounter{rule}

% ==============================Rule ======================================

\begin{figure}
%  %
\end{minipage}%
%\end{array}
%\]
 \caption{(a)$u$ and $v$ collapse   \hspace{2cm} (b) After collapse $HCom(u,v)$}
 \label{figII}\label{SETE}\label{Re2IE}
\end{figure}
\stepcounter{rule}

% ================== Rule  ====================================================

\begin{figure}
%  %
\end{minipage}%
%\end{array}
%\]
 \caption{(a)$u$ and $v$ collapse   \hspace{2cm} (b) After collapse $HCom(u,v)$}
 \label{figIIa}\label{OITO}\label{Re2IH}
\end{figure}
\stepcounter{rule}

%================================Rule Re2EH ======================================

\begin{figure}
%  %
\end{minipage}%
%\end{array}
%\]
 \caption{(a)$u$ and $v$ collapse   \hspace{2cm} (b) After collapse $HCom(u,v)$}
 \label{figIIa}\label{OITO}\label{Re2EH}
\end{figure}
\stepcounter{rule}

%================================= Rule  ==========================================

\begin{figure}
%  %
\end{minipage}%
%\end{array}
%\]
 \caption{(a)$u$ and $v$ collapse   \hspace{2cm} (b) After collapse $HCom(u,v)$}
 \label{figIIb}\label{NOVE}\label{Re2HE}
\end{figure}
\stepcounter{rule}

% ============================ Rule ================================================

\begin{figure}
%  %
\end{minipage}%
%\end{array}
%\]
 \caption{(a)$u$ and $v$ collapse   \hspace{2cm} (b) After collapse $HCom(u,v)$}
 \label{figIIy}\label{ONZE}\label{Rv2IE}
\end{figure}
\stepcounter{rule}

\begin{figure}
%  %
\end{minipage}%
%\end{array}
%\]
 \caption{(a)$u$ and $v$ collapse   \hspace{2cm} (b) After collapse $HCom(u,v)$}
 \label{figIIya}\label{DOZE}\label{Rv2IH}
\end{figure}
\stepcounter{rule}

% ============== Rule Rv2XH=====================

\begin{figure}
%  %
\end{minipage}%
%\end{array}
%\]
 \caption{(a)$u$ and $v$ collapse   \hspace{2cm} (b) After collapse $HCom(u,v)$}
 \label{figIIc}\label{DEZa}\label{Re2XH}
\end{figure}
\stepcounter{rule}

% ================================= Rule ==================================================

\begin{figure}
%  %
\end{minipage}%
%\end{array}
%\]
 \caption{(a)$u$ and $v$ collapse   \hspace{2cm} (b) After collapse $HCom(u,v)$}
 \label{figIIyb}\label{TREZE}\label{Rv2HE}
\end{figure}
\stepcounter{rule}

\begin{figure} [H]
%  %
\end{minipage}%
%\end{array}
%\]
 \caption{(a)$u$ and $v$ collapse   \hspace{2cm} (b) After collapse $HCom(u,v)$}
 \label{figIIIa}\label{DEZESSEIS}\label{Re3XH}
\end{figure}
\stepcounter{rule}

% ============================== Rule ====================================================

\begin{figure} [H]
%  %
\end{minipage}%
%\end{array}
%\]
 \caption{(a)$u$ and $v$ collapse   \hspace{2cm} (b) After collapse $HCom(u,v)$}
 \label{figIIIb}\label{DEZESSETE}\label{Rv3XI}
\end{figure}
\stepcounter{rule}

% ============================== Rule ====================================================

\begin{figure} [H]
%  %
\end{minipage}%
%\end{array}
%\]
 \caption{(a)$u$ and $v$ collapse   \hspace{2cm} (b) After collapse $HCom(u,v)$}
 \label{figIIIb}\label{DEZESSETE}\label{Rv3XH}
\end{figure}
\stepcounter{rule}

{\bf Obs}: In rules $\mathbf{R_e3XH}$, $\mathbf{R_v3XH}$ and $\mathbf{R_v2XH}$ the node $u$ in the left hand-side may be marked with $\hbar$, as in the right hand-side.

\begin{figure}[!h]
 \begin{minipage}{.47\textwidth}
 %
\end{minipage}%
%\end{array}
%\]
 \caption{(a)$u$ and $v$ collapse   \hspace{2cm} (b) After collapse $HCom(u,v)$}
 \label{figI-EE}\label{UM-EE}\label{R0EE}
\end{figure}
\stepcounter{rule}

{\bf Remark.} We have to remark that in some of the rules above, we use a shortened form of the labels assigned by the function $P$, the paths. This is due to have a lighter visualization of the rules. The paths are shortened by omitting sequences of zeros (0s) in the path. When showing concrete examples, as in the appendix, we do not use the shortened version. 

\begin{figure}
%  %
\end{minipage}%
%\end{array}
%\]
 \caption{(a)$u$ and $v$ collapse   \hspace{2cm} (b) After collapse $HCom(u,v)$}
 \label{figIIb}\label{NOVE}\label{Re2EE}
\end{figure}
\stepcounter{rule}

% ======= End Rule 9 =========================

\section{More on {\bf DGTD}, {\bf DLDS} and compression rules soundness preservation}
\label{sec:HC-Soundness}
We need some definitions concerning the relation to {\bf DLDS}s and the rules listed in the above subsection~\ref{subsec:Rules}. They are in the sequel.

\begin{definition}[$INS$-Incoming Deductive Edges of a node]
  Given a DLDS $\mathcal{D}$ of $\alpha$ from $\Gamma$ and a node $k\in\mathcal{D}$, the deductive in-degree of $k$ is defined as $INS(k)=\{f : f\in E_D^{i}, i\in \mathcal{O}(\Gamma\cup\{\alpha\})^{0}\land target(f)=k\}$.
\end{definition}

\begin{definition}[$OUTS$-Outcoming Deductive Edges from a node]
  Given a DLDS $\mathcal{D}$ of $\alpha$ from $\Gamma$ and a node $k\in\mathcal{D}$, the deductive out-degree of $k$ is defined as $OUTS(k)=\{f : f\in E_D^{i}, i\in \mathcal{O}(\Gamma\cup\{\alpha\})^{0}\land source(f)=k\}$.
  \end{definition}

Note that for any node $k$, both sets, $INS(k)$ and $OUTS(k)$, do not take into account the ancestor edges in their definition. We remember that the members of $E_A$ are not deductive edges. However,  they play an important, altough auxiliary role in the logical reading of any DLDS.  A trivial obeservation is that there is a natural map from {\bf DGTD}s to {\bf DLDS}s.

\begin{definition}
  Let $\mathcal{T}=\langle V,E_{D},E_{d},r,l,L\rangle$ be a {\bf DGTD}. Let  $\mathcal{O}_{S}$ be the order on  the range of $l$, provided by $\mathcal{T}$ itself and $\Gamma$ the set of leaves in $\mathcal{T}$. Let  $Dag(\mathcal{T})$ be $\langle V , (E_{D}^{i})_{i\in\mathcal{O}_{\Gamma}^{i}} , E_{A}, r, l, L, P\rangle$, where $E_D^{0}=E_D$,  $E_D^{i}=\emptyset$, for all $i\neq 0$ and $E_A=\emptyset$, $P=\emptyset$.
\end{definition}

It is easy to verify that $Dag(\mathcal{T})$ is well-defined, and hence it is a {\bf DLDS}, for every {\bf DGTD} $\mathcal{T}$. Thus, we have the mapping $Dag$ from {\bf DGTD}s to {\bf DLDS}s. When reading a {\bf DGTD} from top to bottom in a tree-like Natural Deduction derivation, there is at most one path from any top-formula occurrence to any other formula occurrence in the derivation. The following fact is an easy consequence of $Dag$'s definition above.

\begin{proposition}\label{def:Dag-of-a-DGTD} Let $\mathcal{T}$ be a {\bf DGTD}. For every pair of nodes $v$ and $u$ in $\mathcal{T}$, there is a bijection between the paths from $v$ to $u$, in $\mathcal{T}$, and 0-paths, i.e., using only members of $E_D^{0}$,  from $v$ to $u$ in $Dag(\mathcal{T})$. Moreover, the dependency sets, assigned by $L$, in both structures, are equal for every edge $\langle v,u\rangle\in E_D^0$. 
\end{proposition}

 From the definition of the mapping $Dag$, we can see that there is no path in the {\bf DLDS} $Dag(\mathcal{T})$ with colors different from $0$, due to $E_A^{i}=\emptyset$, for all $i\neq 0$. Moreover, there are no paths in $E_A$, for $E_A=\emptyset$.

The following definition shows how the information stored in the component $P$, the seventh one,  the last, of any {\bf DLDS} is used as a relative address for nodes in it. It uses:

\begin{definition}
  $el(\{e\})=e$ and $el(S)=\bot$, if $S$ is not $\{a\}$ for some $a$.
  \end{definition}

\begin{definition}[relative address of a node]\label{address-of-a-node}
  Let $\mathcal{D}$ be a DLDS of $\alpha$ from $\Gamma$ and let $\gamma\in\mathcal{O}(\Gamma\cup\{alpha\})^{0})^{\star}$ we say that $\gamma$ is the address of a node $v\in \mathcal{D}$ relative to a node $u\in \mathcal{D}$ iff the following algorithm~\ref{relative-address-of-a-node} returns $v$ on input $\gamma$,  $\mathcal{D}$ and $u$. The underlying idea is that $\gamma$ provides information on every branching  in the path from $u$ downwards $v$. Each ordinal from left to right in $\gamma$ indicates which branch to take in.
  \end{definition}

\begin{fullwidth}[width=\linewidth+2cm,leftmargin=-1cm,rightmargin=-1cm] 
  \begin{algorithm}[H]
%    \addtocounter{algorithm}{-1}
    \caption{finding a node from its relative address and origin of the path}
    \begin{algorithmic}[1]
      %      \algrestore{checkPaths}
     \Require{$u$, the origin, $\mathcal{D}$, the DLDS, and the relative address $\gamma$}
     \Let{b}{$u$}
     \Let{glues}{$\gamma$}
     \While{$glues\neq\epsilon$}
     \If{$size(OUTS(b))==1$}
     \Let{g}{$el(OUTS(b))$}
     \Let{b}{$target(g)$}
     \ElsIf{$size(OUTS(b))>1\land size(\{e/(e\in OUTS(b))\land (color(e)=head(\gamma))\})=1$}
     \Let{g}{$el(\{e/(e\in OUTS(b))\land (color(e)=head(\gamma))\})$}
     \Let{b}{$target(g)$}
     \Let{glues}{$rest(\gamma)$}
     \Else
     \State Return{\;false}
     \EndIf
     \EndWhile
     \State Return{\;b}
     \end{algorithmic}\label{relative-address-of-a-node}
    \end{algorithm}
\end{fullwidth}

For defining when a DLDS corresponds to a valid derivation we need the definition of Deductive path below.

\begin{definition}[Deductive path]
Given two nodes $v_1$ and $v_2$ in a {\bf VLDS} $\mathcal{D}=\langle V , (E_{D}^{i})_{i\in\{\bar{\lambda}\}\cup\mathcal{O}_{\Gamma}^{i}} , E_{A}, r, l, L, P\rangle$, we call a path $e_1,e_2,\ldots,e_n$ from $v_1$ to $v_2$ a deductive path, iff, for each $p=1,\ldots, n$, $e_p\in \bigcup_{i\in\{\bar{\lambda}\}\cup\mathcal{O}_{\Gamma}^{i}} E_D^{i}$. In particular, if  $e_1,e_2,\ldots,e_n$ is a deductive path from $v_1$ to $v_2$  and there is $i\neq 0$, such that $e_j\in E_D^{i}$ or $e_j\in E_D^{\bar{\lambda}}$, for some $0\leq j\leq n$, then the path is a mixed deductive path from $v_1$ to $v_2$.    
\end{definition}

Given a {\bf DLDS} $\mathcal{D}=\langle V , (E_{D}^{i})_{i\in\mathcal{O}_{\Gamma}^{i}} , E_{A}, r, l, L, P\rangle$ and a node $w\in V$, we define \[
Pre(w)=\{v:\mbox{There is a deductive path from $v$ to $w$}\}
\]
, the set of nodes that are linked to $w$ by some deductive path. Moreover we have the set of top nodes of a DLDS.
\[
Top(w)=\{v:\mbox{$v\in Pre(w)$ and, there is no $v^{\prime}\in V$, $\langle v^{\prime},v\rangle\in (E_{D}^{i})_{i\in\mathcal{O}_{\Gamma}^{i}}$ or $v$ is marked as hypothesis}\}
\]
Finaly, the set of full deductive paths reaching to $w\in V$ is:
\[
DedPaths(w)=\{\langle e_1,\ldots,e_n\rangle:\mbox{$e_1\ldots e_n$ is a deductive path, $source(e_1)\in TopNode(w)$ and $target(e_n)=w$}\}
\]

We introduce the relation $\sim$ between dependency sets.

\begin{definition}
  For any pair of dependency sets $\bar{b}$ and $\bar{c}$, $\bar{b}\sim \bar{c}$ holds, if and only if, $\bar{c}=\bar{b}$ or $\bar{c}=\lambda$ or $\bar{b}=\lambda$.
\end{definition}

\begin{definition}\label{def:Flow}
  Given a {\bf DLDS} $\mathcal{D}=\langle V , (E_{D}^{i})_{i\in\mathcal{O}_{\Gamma}^{i}} , E_{A}, r, l, L, P\rangle$ and a node $w\in V$, we define $Flow(\mathcal{D},w)$ as a function from $Pre(w)$ into $\wp((\mathcal{O}_{\Gamma}^{0})^{*}\times \mathcal{B}(\mathcal{O}_{S}))$, such that:
  \[
  Flow(\mathcal{D},w)(v)=
  \]
  \\
  \[
  \left\{
  \begin{array}{ll}
    \{(\vec{b(l(v))}, P(\langle v^{\prime},v\rangle)):\mbox{$\langle v^{\prime},v\rangle\in E_A$, $v^{\prime}\in V$}\} & \mbox{if $v\in Top(w)$, there is $v^{\prime}\in V$, $\langle v^{\prime},v\rangle\in E_A$, } \\
    \{(\vec{b(l(v))}, 0)\} & \mbox{if $v\in Top(w)$ and $\not\exists v^{\prime}\in V$, $\langle v^{\prime},v\rangle\in E_A$ }  \\
%    \{defining\}  & \mbox{if $v\not\in Top(w)$ and $\exists v^{\prime}\in V$, $\langle v^{\prime},v\rangle\in E_A$ }  \\
    \begin{array}{c}
      \left\{(\vec{b_1}\lor \vec{b_2},p):\fbox{\parbox{6cm}{ $(v_1,v_2,v)\in\imply_E$ and \\ $(b_i,[o_i|p])\in Flow(\mathcal{D},w)(v_i)$,  \\ and  $\langle v_i,v\rangle\in E_D^{o_i}$, and, \\ $b_i\sim L(\langle v_i,v\rangle)$, $i=1,2$, \\ OR $(v_1,v_2,v)\in\imply_E$ and \\
          $(b_i,[0|p])\in Flow(\mathcal{D},w)(v_i)$, $i=1$ or $i=2$, and \\
          $(b_j,\emptyset)\in Flow(\mathcal{D},w)(v_j)$, $j\neq i$, and \\
          $\langle v_i,v\rangle\in E_D^0$, $\langle v_j,v\rangle\in E_D^0$, and \\
          $b_i\sim L(\langle v_k,v\rangle)$, $k=1,2$,}}\right\} \\
      \cup \\
      \left\{(\vec{b^{\prime}}-\vec{\alpha},p):\fbox{\parbox{6cm}{ $(v^{\prime},v)\in\imply_I$ and \\ $(b^\prime,[o^\prime|p])\in Flow(\mathcal{D},w)(v^\prime)$  and \\  $\vec{b^{\prime}}\sim L(\langle v^\prime,v\rangle)$ and, \\ $\langle v^\prime,v\rangle\in E_D^{o^\prime}$, and $l(v)$=``$\alpha\imply l(v^\prime)$''}}\right\}  \\
      \cup \\
      \left\{(\vec{b(l(v))}, 0 ):\mbox{$v$ is marked with $\hbar$ and $\not\exists v^\prime$, $\langle v^\prime,v\rangle\in E_A$}\right\} \\
      \cup\\
      \left\{(\vec{b(l(v))}, P(\langle v^{\prime},v\rangle)):\mbox{$v$ is marked with $\hbar$ and $\langle v^{\prime},v\rangle\in E_A$}\right\}\\
      \cup\\
      \left\{(\vec{b^{\prime}}-\vec{\alpha},P(\langle v_a,v\rangle)):\fbox{\parbox{6cm}{ $(v^{\prime},v)\in\imply_I$ and \\ $(b^\prime,0)\in Flow(\mathcal{D},w)(v^\prime)$  and \\  $v_a\in V$, $\langle v_a,v\rangle\in E_A$  and, \\ $P(\langle v_a,v\rangle)=[j\mid p]$ and $\langle v,v^{\prime\prime}\rangle\in E_{D}^j$ \\ $v^{\prime\prime}\in Pre(w)$ or $v^{\prime\prime}=w$ \\ and $v_a\in Pre(w)$ \\ $\vec{b^{\prime}}\sim L(\langle v^\prime,v\rangle)$, and \\  $l(v)$=``$\alpha\imply l(v^\prime)$'', and \\
       $\langle v^{\prime},v\rangle\in E_D^{0}$}}\right\}  \\
      \cup\\
      \left\{(\vec{b_1}\lor \vec{b_2},j):\fbox{\parbox{6cm}{ $(v_1,v_2,v)\in\imply_E$ and \\ $(b_i,0)\in Flow(\mathcal{D},w)(v_i), i=1,2$  \\ and  $v_a\in V$, $\langle v_a,v\rangle\in E_A$, and \\ $P(\langle v_a,v\rangle)=[j\mid p]$ and $\langle v,v^{\prime}\rangle\in E_{D}^j$ \\
          $v^{\prime}\in Pre(w)$ or $v^{\prime}=w$ \\ and $v_a\in Pre(w)$ \\
          $\langle v_i,v\rangle\in E_D^0$ $i=1,2$, and \\
      $b_k\sim L(\langle v_k,v\rangle)$, $k=1,2$}}\right\} \\
    \end{array}
      & \mbox{otherwise}
  \end{array}\right.
  \]
\end{definition}

In the end of section~\ref{sec:motiv-example} we use the function {\bf Flow} in a compressed {\bf DLDS} to illustrate how it works by calculating the dependency sets inside it.  In appendix~\ref{appendix:ExemploFlow} we also illustrate how the function {\bf Flow}  provides the Dependency sets of a compressed DLDS. {\bf Flow} is  used to in a condition that is  part of the validity criteria for DLDS below.   

\begin{definition}\label{def:ValidDLDS}
  Given a structure $\mathcal{D}=\langle V , (E_{D}^{i})_{i\in\mathcal{O}_{\Gamma}^{i}} , E_{A}, r, l, L, P\rangle$, we say that it is a
  valid {\bf DLDS}, iff, the following conditions hold on it:
  \begin{description}
  \item[Color-Acyclicity] For each $i\in\mathcal{O}_{\Gamma}^{i}$, $E_D^{i}$ does not have cycles;
  \item[Leveled-Colored] The rooted sub-dag $\langle V , (E_{D}^{i})_{i\in\mathcal{O}_{\Gamma}^{i}} , r\rangle$ is leveled;
  \item[Ancestor-Edges] For each $\langle v_1, v_2\rangle\in E_A$, the level of $v_1$ is smaller than the level of $v_2$;
  \item[Ancestor-Backway-Information] For each $\langle v_1, v_2\rangle\in E_A$, $P(\langle v_1, v_2\rangle)$ is the relative address of $v_1$ from $v_2$;
  \item[Simplicity] The rooted sub-dag $\langle V , (E_{D}^{i})_{i\in\mathcal{O}_{\Gamma}^{i}} , r\rangle$ is a simple graph, i.e, for each pair of nodes $v_1$ and $v_2$, there is at most an $i\in\mathcal{O}_{\Gamma}^{i}$, such that $\langle v_1, v_2\rangle\in E_D^{i}$;
  \item[Ancestor-Simplicity] The sub-dag $\langle V, E_A\rangle$ is a simple graph;
  \item[Non-Nested-Ancestor-Edges] For each $\langle v_1,v_2\rangle\in E_A$, there is no $w$ in the path from $v_2$ to $v_1$, determined by $P(\langle u,v\rangle\in E_A)$, such that $\langle w,z\rangle\in E_A$, for some $z\in E_A$.
  \item[CorrectRuleApp] For each $w\in V$, $Flow(\mathcal{D},w)(v)$ is well-defined for each $v\in Pre(w)$. Moreover, for each $w$ and $v$,
    $Flow(\mathcal{D},w)(v)$, with $v\in Pre(w)$, we have:
    \begin{itemize}
    \item If $Flow(\mathcal{D},w)(v)=\{(\vec{b},p)\}$ then $OUT(v)=\{\langle v,v^\prime\rangle\}$ and the color of $\langle v,v^\prime\rangle$
      is $head(p)$, i.e., $\langle v,v^\prime\rangle\in E_D^{head(p)}$, and $\vec{b}=L(\langle v,v^\prime\rangle)$, and;
    \item If $Flow(\mathcal{D},w)(v)\neq\emptyset$ and it is not a singleton either then for each \linebreak
      $\Phi_{i}=\{(\vec{b},p)\in Flow(\mathcal{D},w)(v):\mbox{$head(p)=i$}\}$:
      \begin{enumerate}
      \item If $\Phi_{i}\neq\emptyset$  then there is only one $v^{\prime}$ $\langle v,v^{\prime}\rangle\in E_D^i$ and if $\Phi_i=\{(\vec{b},p)\}$ then $L(\langle v, v^{\prime}\rangle)=\vec{b}$ else $L(\langle v, v^{\prime}\rangle)=\lambda$, and;
        \item If  $\Phi_{i}=\emptyset$ then there is no $v^{\prime}\in V$, such that, $\langle v,v^{\prime}\rangle\in E_D^{i}$
      \end{enumerate}
      \end{itemize}

%%     $S=\{(v,\gamma):\mbox{$\gamma=P(\langle w,v\rangle)$ and $\langle w,v\rangle\in E_A$}\}$ and $S\neq\emptyset$, there is a partition $S=\bigcup_{p=1,k}S_p$, such that:
%%     \begin{enumerate}
%%     \item For all $w\in V$:
%%       \begin{enumerate}
%%       \item If there are
%%       \item If there is no
%%         \end{enumerate}

%%       \item For all pairs $(v,\gamma),(v^{\prime},\gamma^{\prime})\in S$, $last(\gamma)=last(\gamma^{\prime})$, and: 
%%     \item For all $p=1,k$, $\card{S_p}=2$ or $\card{S_p}=1$, and;
%%     \item If $\card{S_p}=\{(v_1,\gamma_1),(v_2,\gamma_2)\}$ then $len(\gamma_1)=len(\gamma_2)$
%%       \item $max(\{len(\gamma):\mbox{$(v,\gamma)\in S_p$ and $p=1,k$}\})\leq h$, where $h$ is the height of $\mathcal{D}$
%%       \item $unfold(V_{w},\bigcup_{p=1,k}\{S_p\})$ is a greedy Natural Deduction derivation of $l(w)$ from $l(S)$. The definition of  $unfold$ can be found in appendix~\ref{app:unfold};
%%         \end{enumerate}
%%   \item[correctRuleApp-I-Elim] For each $w\in V$, if $INS(w)=2$, $OUT(w)=1$, $\pair{v_1}{w},\pair{v_2}{w}\in E_D^0$ then $v_1,v_2$ and $w$ form a valid application of an $\imply$-Elim;
%%     \item[correctRuleApp-I-Intro]\label{verifyruleappI} For each $w\in V$, if $INS(w)=1$, $OUT(w)=1$, $v\in E_D^0$ then $v$ and $w$ form a valid application of an $\imply$-Intro;
        
  \end{description}

\end{definition}

 It is worth noting that in item {\it correctRuleApp}, the verification that a rule application is correct involves, among other things, finding out that the premises agree with the conclusion and checking that the dependency sets are correctly assigned, this is the main role of function $Flow$.

 Each of the items in definition~\ref{def:ValidDLDS} is an invariance property that should be  preserved by all compression rules applications. This is what theorem~\ref{theo:Soundness} says. The lemmas in sub-section~\ref{subsec:lemmata} prove that the HC rules preserve each condition in the definition~\ref{def:ValidDLDS}. The proof of theorem~\ref{theo:Soundness} uses all of them.

\subsection{HC rules and {\bf DLDS} validity preservation}
\label{subsec:lemmata}

In sub-section~\ref{subsec:Rules-Classification} we discuss the types of the rules, namely, type-0, type-\Romannum{1}, type-\Romannum{2} and type-\Romannum{3} listings point out which rule belongs to each of the types. Below we provide applicability conditions for each of the types.

\begin{definition}[Rule applicability conditions for type-0 rules]
A type-0 rule is applicable to a pair $u,v\in V$ in a {\bf DLDS} $\mathcal{D}=\langle V , (E_{D}^{i})_{i\in\{\bar{\lambda}\}\cup\mathcal{O}_{\Gamma}^{i}} , E_{A}, r, l, L, P\rangle$, iff, $l(u)=l(v)$, there is no $e\in E_A$, such that $targek(e)$ is one of the nodes in the left-hand side of the rule, and, for all $e\in (E_{D}^{i})_{i\in\{\bar{\lambda}\}\cup\mathcal{O}_{\Gamma}^{i}}$, if $target(e)$ is some node in the left-hand of the rule then $e\in E_D^{0}$\end{definition}

\begin{definition}[Rule applicability conditions for type-\Romannum{1} rules]
  A type-\Romannum{1} rule is applicable to a pair $u,v\in V$ in a {\bf DLDS} $\mathcal{D}=\langle V , (E_{D}^{i})_{i\in\{\bar{\lambda}\}\cup\mathcal{O}_{\Gamma}^{i}} , E_{A}, r, l, L, P\rangle$, iff, $l(u)=l(v)$, (1) there is no $e\in E_A$, such that $targek(e)$ or $source(e)$ is one of the nodes linked to $v$ in the left-hand side of the rule, and; (2)  $u$ is a node that it is already the result of a collaping, i.e., the nodes linked to $u$ are sources or targets of an ancestor rule ($E_A$), and, for each $e\in OUTS(u)$, there is $i\neq 0$, such that $e\in E_{D}^{i}$.\end{definition}

The rules of type-\Romannum{1} are rules in figures~\ref{CINCO}, ~\ref{SEIS} and~\ref{R1XI} are all the rules of type-\Romannum{1}. Only these rules satisfy the condition above.

\begin{definition}[Rule applicability conditions for type-\Romannum{2} rules]
  A type-\Romannum{2} rule is applicable to a pair $u,v\in V$ in a {\bf DLDS} $\mathcal{D}=\langle V , (E_{D}^{i})_{i\in\{\bar{\lambda}\}\cup\mathcal{O}_{\Gamma}^{i}} , E_{A}, r, l, L, P\rangle$, iff, $l(u)=l(v)$, both $u$ and $v$ are target of an edge $e\in E_A$, i.e.,  $targek(e)=v$ or $target(e)=u$.
\end{definition}

For example, rules in figures~\ref{Re2IE} to~\ref{Rv2IH} are of this kind.

\begin{definition}[Rule applicability conditions for type-\Romannum{3} rules]
  A type-\Romannum{2} rule is applicable to a pair $u,v\in V$ in a {\bf DLDS} $\mathcal{D}=\langle V , (E_{D}^{i})_{i\in\{\bar{\lambda}\}\cup\mathcal{O}_{\Gamma}^{i}} , E_{A}, r, l, L, P\rangle$, iff, $l(u)=l(v)$, exactly one of  $u$ or $v$ are target of an edge $e\in E_A$, i.e.,  $targek(e)=v$ or $target(e)=u$. If $v$ is the target of an $A$-edge, then $u$ is a node that is already collapsed by a rule of type-\Romannum{3}.
\end{definition}

All type-\Romannum{3} rules satisfy above condition. 

In this section, we show the preservation of each condition in the definition of validity of {\bf DLDS}, definition~\ref{def:ValidDLDS}.

We should observe that the {\bf HC} algorithm applies the {\bf HC} rules from the lowest to the highest levels and in each level from left to right.

\begin{definition}[Application of {\bf HC} rule in algorithmic position]  Given a {\bf DLDS} $\mathcal{D}$, we say that an application of a  {\bf HC} rule in $\mathcal{D}$ is in {\em algorithmic position} iff this application collapses two nodes, $v$ and $u$, in level $j$, such that, above $j$ there is no sub-graph originated from the collapse of two nodes by any application of any {\bf HC} rule and, there is no application that collapses two nodes by any {\bf HC} rule in level $j$ righter than both $v$ and $u$ either.
\end{definition}

{\bf Remark.} Due to the fact that the {\bf HC} algorithm~\ref{HorCom} applies in any level from left to right,  if a {\bf HC} application in a {\bf DLDS} $\mathcal{D}$ is in {\em algorithmic position} with respect to the nodes $v$ and $u$ in level $j$, then only the leftmost of $u$ and $v$ can be the result of a previous {\bf HC} rule application.    
The lemma below illustrates how the mechanism of the ancestor edges updating works.
Before this lemma statement, we need some definitions.

\begin{definition}[Partially Compressed DGTD]
  We say that a {\bf DLDS} $\mathcal{D}=\langle V , (E_{D}^{i})_{i\in\{\bar{\lambda}\}\cup\mathcal{O}_{\Gamma}^{i}} , E_{A}, r, l, L, P\rangle$ is a partially compressed DGTD, iff, there is a {\bf DGTD} $\mathcal{T}$, such that, $\mathcal{D}$ is the result of the application of some steps in line 3 of the algorithm~\ref{HorCom} to $\mathcal{T}$.
\end{definition}

Any {\bf DGTD} is a partially compressed DGTD, and the final result of the algorithm, the so-called totally compressed {\bf DGTD} is also a partially compressed {\bf DGTD}. We observe that the horizontal compression algorithm~\ref{HorCom} halts when submitted to any {\bf DGTD}, see theorem~\ref{theo:Termination}. Any partially compressed {\bf DGTD} that is not the final result of the application of the algorithm~\ref{HorCom} to it has an algorithmic position to collapse two nodes $u$ and $v$. For technical reasons, the following lemma considers only rules that are not {\bf MDE}. The format of a {\bf DLDS} that cannot serve as input to a non-{\bf MDE} rule has every level, but the last one, with no two nodes labelled with the same formula. This is what we obtain as an almost compressed {\bf DLDS}. This kind of {\bf DLDS} is called $\mathbf{MUE^{+}}$-compressed {\bf DLDS}. Only the top-formulas of the derivation may have repetitions in a same level. 

\begin{lemma}\label{lemma:DedPath}
  Let $\mathcal{D}=\langle V , (E_{D}^{i})_{i\in\{\bar{\lambda}\}\cup\mathcal{O}_{\Gamma}^{i}} , E_{A}, r, l, L, P\rangle$ be a valid partially compressed {\bf DGTD} and $u_{ap}$ and $v_{ap}$ nodes in level $j>0$, such that, there is at least one  compression rule that is not  {\bf MDE} that can be applied to $\mathcal{D}$ in algorithmic position to collapse $u_{ap}$ and $v_{ap}$. Consider a node $w$ in level smaller than $j$, such that, there is a mixed deductive path from the node  $u_{ap}$ to  $w$. Hence, if $w=source(e)$ and $e\in E_A$ then $target(e)=u_{ap}$ or there is $e^{\prime}\in E_D^{0}$, such that $target(e)=source(e^{\prime})$ and $target(e^{\prime})=u_{ap}$. Moreover, if $u_{ap}$ is marked as hypothesis then there is $e^{\prime}\in E_D^{0}\cup E_D^{\bar{\lambda}}$, such that, $target(e)=target(e^\prime)$ and $source(e^\prime)=u_{ap}$. %Figure~\ref{fig:SituationOnAncestorEdges} depicts these situations.
\end{lemma}

We should observe that the situations that the statement of the lemma involves 
hold for every lefthand side of the compression rules. %figure~\ref{fig:SituationOnAncestorEdges} depicts

{\bf Proof of lemma~\ref{lemma:DedPath}}
$\mathcal{D}$ is a partially compressed DGTD, so there is a {\bf DGTD} $\mathcal{T}$, such that, $\mathcal{D}$ is the result of the application of some steps in line 3 of the horizontal compression algorithm to $\mathcal{T}$. 
We prove the lemma by induction on the lexicographic pair $(I, II)$, where $I$ is the number of ancestor edges in $\mathcal{D}$, and $II$ is the number of steps executed by the horizontal compression algorithm~\ref{HorCom} on $\mathcal{T}$ to obtain $\mathcal{D}$. The base is $card(E_A)=0$ when there is no ancestor edge and no horizontal compression rule was applied. In this case, the statement trivially holds. For the inductive case, $card(E_A)=n>0$, there is a a rule that can be applied to $\mathcal{D}$ in algorithmic position to collapse $u_{ap}$ and $v_{ap}$, such that $u_{ap}$ and $v_{ap}$ are nodes in level $j>0$. Let $(card(E_A), II_{\mathcal{D}})$ be the lexicographic pair of $\mathcal{D}$. Thus, $II_{\mathcal{D}}$ is the number of horizontal compression rule applications that results in $\mathcal{D}$. So, some rule $R$ was applied to a previous {\bf DLDS} $\mathcal{D}^{\prime}$ resulting into $\mathcal{D}$. $R$ may or may not have created an ancestor edge in $\mathcal{D}$. Thus, we have two cases:
\begin{enumerate}
\item\label{SemAdicao} $card(E_{A})=card(E_{A}^{\prime})$, where $E_{A}^{\prime}$ is the set of ancestor edges in $\mathcal{D}^{\prime}$. In this case, the inductive hypothesis holds for $\mathcal{D}^{\prime}$, with pair $(card(E_A), II_{\mathcal{D}}-1)$. Since $\mathcal{D}^{\prime}$ is valid, due to the validity of $\mathcal{D}$, by inductive hypothesis $\mathcal{D}^{\prime}$ satisfies the property in the lemma statement.
\item\label{ComAdicao} $card(E_{A}^{\prime})<card(E_A)$. In this case, $R$ adds at least one ancestor edge to $\mathcal{D}^{\prime}$ to have $\mathcal{D}$, so the inductive complexity of the former is smaller than the latter. We observe that $\mathcal{D}^{\prime}$ is valid as a consequence of the validity of $\mathcal{D}$. Thus, by the inductive hypothesis, the property stated by the lemma holds for $\mathcal{D}^{\prime}$
\end{enumerate}

It remains to analyse the changes that the application of the rule $R$ makes in the ancestor edges of $\mathcal{D}^\prime$, case~\ref{SemAdicao} and the insertion of new ancestor edges by $R$, i.e. case~\ref{ComAdicao}. Separating the analysis according to the rules, we have to consider the type-0 and type-\Romannum{1} rules, since they are rules that only add ancestor edges; hence, we are in the case~\ref{ComAdicao} analysis. All the ancestor edges inserted in $\mathcal{D}^\prime$, resulting in $\mathcal{D}$ satisfy the property stated in the lemma.
The type-\Romannum{2} rules that are not {\bf MDE} change existing ancestor rules, they concern case~\ref{SemAdicao} analysis. We can inspect the changes determined by each of these rules and conclude that the property stated in the lemma also hold in this case. For the sake of showing the need for the hypothesis that $\mathcal{D}$ is valid, we describe in more detail rules {\bf R0IE} (case~\ref{ComAdicao} above) and  $\mathbf{R_e2IH}$ (case~\ref{SemAdicao} above) argumentations, in the following two paragraphs.

By inductive hypothesis, the property stated by the lemma holds for $\mathcal{D}^{\prime}$ and suppose that rule $R$ is {\bf R0IE} below in figure~\ref{fig:R0IE-Analysis}. We can observe that the dag in the lefthand side of the rule is a subgraph in $\mathcal{D}^{\prime}$, that $\mathcal{D}^{\prime}$ satisfies all validity conditions in definition~\ref{def:ValidDLDS} and the righhand side is subgraph of $\mathcal{D}$. It is easy to see that the lemma property holds for $\mathcal{D}$. The addition of the ancestor edges, in blue, after the application of {\bf R0IE} satisfies this condition. By the validity of $\mathcal{D}^{\prime}$ and the lemma property holding on $\mathcal{D}^{\prime}$, we can affirm that there is no $w$ in a level below $u$ (and $v$) and a  mixed deductive path from either one of $p_1$, $p_2$ or $p_3$, such that there is $e\in E_A$ and $source(e)=w$. If this is the case then $target(e)=u$, or $target(e)=\bullet$ or $target(e)=p_i$, for some $i$, since $\mathcal{D}^{\prime}$ satisfies the lemma property. However, by the conditions of applicability of {\bf R0IE} all the conditions on the existence of $w$, $e$ and their consequences are contradictory to the fact that {\bf R0IE} is applied to $\mathcal{D}^{\prime}$ resulting in $\mathcal{D}$. Thus, there is no $w$ and $e\in E_A$, such that $source(e)=w$. From this we conclude that the application of {\bf R0IE} to $\mathcal{D}^{\prime}$ results in $\mathcal{D}$ satisfiyng the lemma. 

\begin{figure}[!h]
 \begin{minipage}{.47\textwidth}
 \begin{tikzpicture}[->,>=stealth',shorten >=1pt,auto,node distance=1.5cm,semithick]
 \tikzstyle{every state}=[draw=none]
 \node[state] (q1) {$p_1$};
 \node[state] (q0) [above right of=q1] {$p_0$};
 \node[state] (q2) [below of=q1] {$u$};
 \node[state] (q3) [below of=q2] {$\bullet$};
 \node[state] (q6) [right of=q1] {$p_2$};
 \node[state] (q7) [below of=q6] {$v$};
 \node[state] (q8) [below of=q7] {$\bullet$};
 \node[state] (q11) [right of=q6] {$p_3$};

 \path (q0)  [dashed, bend right] edge [-] node {} (q1);
 \path (q1) edge node {$\bar{c_1}$} (q2)
 (q2) edge node [left] {$\bar{c}=\bar{c_1}-\bar{p_0}$} (q3)
 (q6) edge node {$\bar{b_1}$} (q7)
 (q7) edge node {$\bar{b_2}=\bar{b_1}\lor\bar{d_1}$} (q8)
 (q11) edge node {$\bar{d_1}$} (q7);
 \path (q2) edge [draw=none] node {$=$} (q7);
 \end{tikzpicture}
 \end{minipage}%
 \begin{minipage}{.15\textwidth}
    $\stackrel{\stackrel{HCom(u,v)}{\Longrightarrow}}{R0IE}$  
\end{minipage}% 
 \begin{minipage}{.2\textwidth}
\begin{tikzpicture}[->,>=stealth',shorten >=1pt,auto,node distance=1.5cm,semithick]
 \tikzstyle{every state}=[draw=none]
 \node[state] (q1) {$p_1$};
 \node[state] (q0) [above right of=q1] {$p_0$};
 \node[state] (q2) [below of=q1] {$u$};
 \node[state] (q3) [below of=q2] {$\bullet$};
 \node[state] (q6) [right of=q1] {$p_2$};
 \node[state] (q8) [below right of=q2] {$\bullet$};
 \node[state] (q11) [right of=q6] {$p_3$};

 \path (q0)  [dashed, bend right] edge [-] node {} (q1);
 \path (q1) edge node [left] {$\bar{c_1}$} (q2)
 (q2) edge node [left] {$\bar{c}=\bar{c_1}-\bar{p_0}$} node[pos=.9,right, red] {1} (q3)
 (q6) edge node [left] {$\bar{b_1}$} (q2)
 (q2) edge node {$\bar{b_2}=\bar{b_1}\lor\bar{d_1}$} node[pos=.9,right, red] {2} (q8)
 (q11) edge node [left, pos=0.5] {$\bar{d_1}$} (q2);
\path  (q8) [color=blue,bend right] edge node[right,red,pos=.7] {2} (q11)
 (q8) [color=blue,bend right] edge node[right,red,pos=.6] {2} (q6)
 (q3) [color=blue,bend left=120] edge node[left,red] {1} (q1);
\end{tikzpicture}
\end{minipage}%
%\end{array}
%\]
 \caption{(a)$u$ and $v$ collapse   \hspace{2cm} (b) After collapse $HCom(u,v)$}
 \label{fig:R0IE-Analysis}
\end{figure}

By the inductive hypothesis, the property stated by the lemma holds for $\mathcal{D}^{\prime}$ and suppose that rule $R$ is $\mathbf{R_e2IH}$ below in figure~\ref{fig:Re2IH-Analysis}. We can observe that the dag in the lefthand side of the rule is a subgraph in $\mathcal{D}^{\prime}$, that $\mathcal{D}^{\prime}$ satisfies all validity conditions in definition~\ref{def:ValidDLDS} and the righthand side is a subgraph of $\mathcal{D}$. By an argument analogous to the case of $\mathbf{R_e2IH}$ application above, we can conclude that there is no $w$ below and $e\in E_A$ below the $u$ (and $v$) level with $source(e)=w$, but the  ancestor edges (blue in the picture) in the lefthand side of figure~\ref{fig:Re2IH-Analysis}. This is also a consequence of the applicability conditions of rule $\mathbf{R_e2IH}$. Due to this, we can conclude that the application of $\mathbf{R_e2IH}$ to $\mathcal{D}^{\prime}$ results in $\mathcal{D}$ satisfying the lemma. The ancestor edges new in the righthand side of $\mathbf{R_e2IH}$, both of them satisfy the property of the lemma. In particular, the part of the property related to deductive edges labelled with the $\bar{\lambda}$ holds.

\begin{figure}[!h]
%  \begin{tabular}{@{}ccc@{}}
  %\[
  %\begin{array}{cc}
  \begin{minipage}{.47\textwidth}
 \begin{tikzpicture}[->,>=stealth',shorten >=1pt,auto,node distance=1.5cm,semithick]
 \tikzstyle{every state}=[draw=none]
 \node[state] (q1) {$p_1$};
 \node[state] (q0) [above right of=q1] {$p_0$};
 \node[state] (q2) [below of=q1] {$u$};
 \node[state] (q3) [below of=q2] {$\bullet$};
% \node[state] (q6) [right of=q1] {$p_2$};
 \node[state] (q7) [below of=q6] {$v$};
% \node[state] (q8) [below of=q7] {$\bullet$};
% \node[state] (q11) [right of=q6] {$p_3$};
 \node[state] (q31) [below of=q3] {$\bullet$};
 \node[state] (q81) [below right of=q3] {$\bullet$};

 \path (q0)  [dashed, bend right] edge [-] node {} (q1);
 \path (q1) edge node {$\bar{c_1}$} (q2)
 (q2) edge node [left] {$\bar{c}=\bar{c_1}-\bar{p_1}$} (q3)
% (q6) edge node {$\bar{b_1}$} (q7)
 (q7) edge node {$\bar{l(v)}$} (q3);
% (q11) edge node {$\bar{d_1}$} (q7);
 \path (q31) [dashed, bend left] edge [-]  node[pos=.95,right, red] {i} (q3);
 \path (q81) [dashed, bend right] edge [-] node[pos=.95,right, red] {j} (q3);
 \path (q2) edge [draw=none] node {$=$} (q7)
  (q31) [color=blue, bend left=120] edge node[left,red] {$s_1$} (q2)
  (q81) [color=blue, bend right=120] edge node[right,red] {$s_2$} (q7);
 \end{tikzpicture}
 \end{minipage}%
  \begin{minipage}{.15\textwidth}
  $\stackrel{\stackrel{HCom(u,v)}{\Longrightarrow}}{\mathbf{R_e2IH}}$  
  \end{minipage}%
 \begin{minipage}{.2\textwidth}
\begin{tikzpicture}[->,>=stealth',shorten >=1pt,auto,node distance=1.5cm,semithick]
 \tikzstyle{every state}=[draw=none]
 \node[state] (q1) {$p_1$};
 \node[state] (q0) [above right of=q1] {$p_0$};
 \node[state] (q2) [below of=q1] {$\stackrel{{\color{red} \hbar}}{u}$};
 \node[state] (q3) [below of=q2] {$\bullet$};
% \node[state] (q6) [right of=q1] {$\stackrel{{\color{red}j;s_2}}{p_2}$};
 % \node[state] (q8) [below right of=q2] {$\bullet$};
%  \node[state] (q11) [right of=q6] {$\stackrel{{\color{red}j;s_2}}{p_3}$};
 \node[state] (q31) [below of=q3] {$\bullet$};
 \node[state] (q81) [below right of=q3] {$\bullet$};

 \path (q0)  [dashed, bend right] edge [-] node {} (q1);
 \path (q1) edge node [left] {$\bar{c_1}$} (q2)
 (q2) edge node [left] {$\bar{\lambda}$} (q3);
% (q6) edge node [left] {$\bar{b_1}$} (q2)
% (q2) edge node {$\bar{b_2}=\bar{b_1}\lor\bar{d_1}$} (q3))
% (q11) edge node [left, pos=.1] {$\bar{d_1}$} (q2);
 %\path  (q8) [color=blue,bend right] edge node {} (q11)
 \path (q31) [dashed, bend left] edge [-]  node[pos=.95,right, red] {i} (q3);
 \path (q81) [dashed, bend right] edge [-] node[pos=.95,right, red] {j} (q3);
 \path % (q81) [color=blue,bend right] edge node {} (q6)
       (q31) [color=blue,bend left=135] edge node[left,red] {$[0;s_1]$} (q1)
       (q81) [color=blue, bend right] edge node[right,red] {$s_2$} (q2);  
      % (q81) [color=blue,bend right=120] edge node {} (q11);
\end{tikzpicture}
%\end{tabular}
\end{minipage}%
%\end{array}
%\]
 \caption{(a)$u$ and $v$ collapse   \hspace{2cm} (b) After collapse $HCom(u,v)$}
 \label{fig:Re2IH-Analysis}
\end{figure}

Finally, the argumentation for proving that all type-\Romannum{1}, type-\Romannum{2} and type-0 rules that are not {\bf MDE} also obtain $\mathcal{D}$ holding the lemma's property are a combination of the cases~\ref{ComAdicao} and~\ref{SemAdicao} analysis. Thus, we have that the property stated by the lemma holds in the inductive case too. This concludes the proof of the lemma. Moreover the type-\Romannum{3} rules $\mathbf{R_v3XE}$, $\mathbf{R_e3XE}$, $\mathbf{R_v3XI}$ and $\mathbf{R_e3XI}$ that are {\bf MUE} rules have arguments that are also combination of the cases~\ref{ComAdicao} and~\ref{SemAdicao} analysis.

\begin{center}
  Q.E.D.
\end{center}

The following lemmas correspond to the preservation of each condition in definition~\ref{def:ValidDLDS}. %We only remember that the Ancertor edge is not considered as a coloured edge in the terminology of {\bf DLDS}. Thus the following lemmas are only about cyclicity of deduction and coloured edges.

\begin{lemma}[Color-Acyclicity]\label{lemma:CA} 
Given a valid {\bf DLDS} $\mathcal{D}=\langle V , (E_{D}^{i})_{i\in\{\bar{\lambda}\}\cup\mathcal{O}_{\Gamma}^{i}} , E_{A}, r, l, L, P\rangle$, and $u$ and $v$ in level $j$, such that $l(u)=l(v)$.
  Let $R$ be any {\bf HC} rule, such that, $R$ applies in algorithmic position to $u$ and $v$ resulting in a {\bf DLDS} $\mathcal{D}^{\prime}$. Thus, condition {\em Color-Acyclicity} holds on $\mathcal{D}^{\prime}$
\end{lemma}

{\bf Proof.}
Let $\mathcal{D}=\langle V , (E_{D}^{i})_{i\in\{\bar{\lambda}\}\cup\mathcal{O}_{\Gamma}^{i}} , E_{A}, r, l, L, P\rangle$ be a valid {\bf DLDS}, $j$ be a level in $\mathcal{D}$and $R$ a {\bf HC} rule in algorithmic position that collapses $u,v\in V$. We prove that the resulting $\mathcal{D}^{\prime}$ {\bf DLDS} satisfies {\em Color-Acyclicity} by observing that:
\begin{itemize}
\item If $R$ is a type-0 rule, then both nodes, $u$ and $v$ do not result from collapses by any previous {\bf HC} rule application. If $u$ or $v$ were a result of any {\bf HC} rule application, then there would be ancestor edges arriving at least at one of them or at some of the nodes linked to them from above. Other case to consider is when $v$ or $u$ are marked with $h$ due to be the result of application of rules {\bf R0HE}, {\bf R0IH}, {\bf R0HH}, {\bf R1XH}, $\mathbf{R_e2IH}$, $\mathbf{R_v2IH}$, $\mathbf{R_e2HE}$, $\mathbf{R_v2HE}$, $\mathbf{R_e3XH}$,
  $\mathbf{R_v3XH}$. Observing that since $\mathcal{D}$ is a valid {\bf DLDS} then it satisfies {\em Color-Acyclicity}, so by inspection on each rule's resulting $\mathcal{D}^{\prime}$ we observe that the new edges, inserted by the right-hand side of the rule,  cannot form a cycle. The main reason is that the direction of the coloured deduction edges is from top to bottom, and they do not link a node to itself either. Moreover, any edge in $E_A$ direction is from the bottom-up. They do not form a cycle either. The same argumentation is used to type-\Romannum{1} rules that also creates A-edges.
\item The rules of type-\Romannum{2} and type-\Romannum{3} are similarly treated. We need to observe only that the left-hand side has no cycle for each colour $i$, then the graph resulting from the insertions and changings described by the right-hand side is also acyclic for each $i$. This argumentation also serves to the {\bf MUE} and {\bf MDE} that moves A-edges. 
  
    \end{itemize}
\begin{center}
  Q.E.D.
\end{center}

\begin{lemma}[Leveled-Colored]\label{lemma:LC} 
Given a valid {\bf DLDS} $\mathcal{D}=\langle V , (E_{D}^{i})_{i\in\{\bar{\lambda}\}\cup\mathcal{O}_{\Gamma}^{i}} , E_{A}, r, l, L, P\rangle$, and $u$ and $v$ in level $j$, such that $l(u)=l(v)$.
  Let $R$ be any {\bf HC} rule, such that, $R$ applies in algorithmic position to $u$ and $v$ resulting in a {\bf DLDS} $\mathcal{D}^{\prime}$. Thus, condition {\em Leveled-Colored} holds on $\mathcal{D}^{\prime}$
\end{lemma}

{\bf Proof.} This is a straightforward consequence: all rule applications only create coloured edges between subsequent levels. Moreover, the root is not affected by any rule application.
\begin{center}
  Q.E.D.
\end{center}

\begin{lemma}[Ancestor-Edges]\label{lemma:AE}
Given a valid {\bf DLDS} $\mathcal{D}=\langle V , (E_{D}^{i})_{i\in\{\bar{\lambda}\}\cup\mathcal{O}_{\Gamma}^{i}} , E_{A}, r, l, L, P\rangle$, and $u$ and $v$ in level $j$, such that $l(u)=l(v)$.
  Let $R$ be any {\bf HC} rule, such that, $R$ applies in algorithmic position to $u$ and $v$ resulting in a {\bf DLDS} $\mathcal{D}^{\prime}$. Thus, condition {\em Ancestor-Edges} holds on $\mathcal{D}^{\prime}$
\end{lemma}

{\bf Proof.} This follows immediatley from the fact that all rule application only creates ancestor edges from a node in level $l_1$ to a node in level $l_2$ and $l_1>l_2$.
\begin{center}
  Q.E.D.
\end{center}

\begin{lemma}[Ancestor-Backway-Information] \label{lemma:ABI}
Given a valid {\bf DLDS} $\mathcal{D}=\langle V , (E_{D}^{i})_{i\in\{\bar{\lambda}\}\cup\mathcal{O}_{\Gamma}^{i}} , E_{A}, r, l, L, P\rangle$, and $u$ and $v$ in level $j$, such that $l(u)=l(v)$.
  Let $R$ be any {\bf HC} rule, such that, $R$ applies in algorithmic position to $u$ and $v$ resulting in a {\bf DLDS} $\mathcal{D}^{\prime}$. Thus, condition {\em Ancestor-Backway-Information} holds on $\mathcal{D}^{\prime}$
\end{lemma}

{\bf Proof.} This is a straightforward consequence of all rule applications: (1) When it creates an ancestor-edge $e$, it labels it with the relative address of $source(e)$ from $target(e)$ as the label provided by the component $P$ of $\mathcal{D}$, and; (2) When it modifies the target or source of an ancestor edge, it updates the component accordingly $P$ of that labels the modified edge.

\begin{center}
  Q.E.D.
\end{center}

We provide the following definition based on  the analysis that each one of the listed rules does not preserve simplicity of A-edges.

\begin{definition}[A-edge simplicity destroyer rule] The following rules in some application cases can destroyer the simplicity of A-edges in the graph.
  $\mathbf{R_v2HH}$, $\mathbf{R_v2XH}$,$\mathbf{R_v2IH}$, $\mathbf{R_v2EH}$, $\mathbf{R_v3XH}$,   $\mathbf{R_e2HH}$, $\mathbf{R_e2XH}$,$\mathbf{R_eIH}$, $\mathbf{R_e2EH}$, $\mathbf{R_e3XH}$.
\end{definition}

We abbreviate {\em A-edge simplicity destroyer rule} as {\bf AESD-rule}.

\begin{lemma}[Simplicity and Ancestor-Simplicity] \label{lemma:SaAS}
Given a valid {\bf DLDS} $\mathcal{D}=\langle V , (E_{D}^{i})_{i\in\{\bar{\lambda}\}\cup\mathcal{O}_{\Gamma}^{i}} , E_{A}, r, l, L, P\rangle$, and $u$ and $v$ in level $j$, such that $l(u)=l(v)$.
  Let $R$ be any {\bf HC} rule, such that, $R$ {\bf is not} {\bf AESD} and it applies in algorithmic position to $u$ and $v$ resulting in a {\bf DLDS} $\mathcal{D}^{\prime}$. Conditions {\em Simplicity and Ancestor-Simplicity}  holds on $\mathcal{D}^{\prime}$
\end{lemma}

{\bf Proof.} Simplicity is a consequence of the applicability conditions for each type of rule. Note that the lefthand side of each compression rule figures out every deductive edge present in $\mathcal{D}$ in the algorithmic position determined by $j$, $u$ and $v$. On the right-hand side, do not place more than one deductive edge of the same colour between the same pair of nodes. This holds for every compression rule and ensures Simplicity in each application of the rules. A similar argument holds for Ancestor-Simplicity. We have to use lemma~\ref{lemma:DedPath}, however, to ensure that there is no ancestor edge with source below $u$ or $v$ and target in some node of the lefthand side. With this property provided by the lemma~\ref{lemma:DedPath}, we can ensure that the modifications and insertions of the ancestor edges in the righthand side of a compression rule application produce no more than one ancestor edge between the same pair of nodes.   

\begin{center}
  Q.E.D.
\end{center}

We note that lemma~\ref{lemma:SaAS} above does not hold, in general, if we relax the restriction that the applied HC-rule is not a {\bf AESD} rule. {\bf MDE} and {\bf AESD} rules apply on top-nodes that are target of A-edges, moving the A-edges down to have other node as target or keeping it as is.In the case of a {\bf MDE}-rule, the new target may be target of other A-edge. Thus, the inductive hypothesis when applying {\bf MDE} rules cannot hold. Observe the example in section~\ref{sec:motiv-example}, figure~\ref{fig:Final-Compression-Leftmost-A1-fifth-A1} where many applications of {\bf MDE} rules produce the same number of A-edges between the pair of nodes. Concerning the {\bf AESD} rules that are not {\bf MDE}, when two or more applications of {\bf AESD} are used in sequence, the corresponding A-edges that are kept may have equal sources and, with the collapse, they will have the same target. This is enough to destroy A-simplicity. However, we can observe that in this sequencial application, the configuration in the second application of the {\bf AESD} rule, the path that labels the A-edge, by the $P$ labelling function, is equal to the path that labels the A-edge that has the collapse node as target, just the previously collapsed. Withe the purposes of using the $A$-edge as source of information for reading corretly the resulting {\bf DLDS}, we do not need more than one $A$-edge. Thus, we can remove all, but only one of the $A$-edges. The final {\bf DLDS} is $A$-simple. Moreover, we have the following lemma. In figure~\ref{fig:Sequencing-AESD} we have this situation, but we show both $A$-edges, instead of removing one of them. The final {\bf DLDS} in the example in section~\ref{sec:motiv-example}shows also this phenomenon. 

\begin{lemma}[Unicity of A-edges labeling]\label{lemma:2AESD-Rules} Consider any sequential application of a pair of {\bf AESD} rules to an $A$-simple {\bf DLDS}. After the applications of these rules the {\bf DLDS} can be taken as a $A$-simple graph.
\end{lemma}

{\bf Proof} The proof is based on a discussion of one specific case. Consider that the sequence of {\bf AESD} rules $\mathbf{Re2IH}$ and $\mathbf{Re3XH}$ when applied to the configuration depicted by the leftmost graph in figure~\ref{fig:Sequencing-AESD}, inside a {\bf DLDS}. After the application of the mentioned rules, it  produces the rightmost graph. The two arrows convey the same information, that is, the path $s_2$ that is the address of $w$ from $u$. Thus, we only need one of the $A$-edges after application of the rules. This is automatically implemented by the fact that the set $A$-edges is a set, has no repetitions. 
There are ten {\bf AESD} rules and one {\bf hundred} of pairs. They are analogous to the case we illustrated here. For the purposes of this article the lemma is proved.  
\begin{center}
  Q.E.D.
\end{center}

Look at the example used in the lemma above, i.e., figure~\ref{fig:Sequencing-AESD}. Let $w$\footnote{the right and lower bullet in~\ref{fig:Sequencing-AESD}} be the source of the possibly many $A$-edges, say $v_1,\ldots,v_n$ that have top-nodes in the same level of $v$ and are premisses of the same conclusion    , the sequence of {\bf AESD} rule applications is as long as the number of top-nodes that are target of $A$-edges. Moreover, since the source node $w$ is shared by every $A$-edge, then the paths that go from each $v_i$, $i=1,\ldots,n$ and labels the many $A$-edges from each $v_i$ to $w$, carry the same logical information. Thus, as in lemma above, we can remove all, but only one of them, ending up with an $A$-simple graph, and hence a {\bf DLDS}. 
The full formalization of lemma~\ref{lemma:2AESD-Rules} is in progress and \cite{RobinsonLean} reports what has already been done.
The content of this paragraph can be seen as a proof for the following corollary of lemma~\ref{lemma:2AESD-Rules} above.

\begin{corollary}[{\bf AESD} rules preserve Flow]\label{coro:2AESD-Rules}
  Consider any sequential application of a pair of {\bf AESD} rules to an $A$-simple {\bf DLDS}. After the applications of these rules the function {\bf Flow} on the resulted {\bf Flow} is the restriction of the original {\bf Flow} function before the application of the sequence of {\bf AESD} rules.
  \end{corollary}

%{\color{red} Colocar aqui o lado esquerdo da regra Re2IH com duas setas do tipo s2 saindo od bullet mais embaixo a direita}

\begin{figure}
%  \begin{tabular}{@{}ccc@{}}
  %\[
  %\begin{array}{cc}
  \begin{minipage}{.3\textwidth}
 \begin{tikzpicture}[->,>=stealth',shorten >=1pt,auto,node distance=1.5cm,semithick]
 \tikzstyle{every state}=[draw=none]
 \node[state] (q1) {$p_1$};
 \node[state] (q0) [above right of=q1] {$p_0$};
 \node[state] (q2) [below of=q1] {$u$};
 \node[state] (q3) [below of=q2] {$u^{\prime}$};
% \node[state] (q6) [right of=q1] {$p_2$};
 \node[state] (q7) [below of=q6] {$v_1$};
 \node[state] (q8) [right of=q7] {$v_2$};
% \node[state] (q8) [below of=q7] {$\bullet$};
% \node[state] (q11) [right of=q6] {$p_3$};
 \node[state] (q31) [below of=q3] {$\bullet$};
 \node[state] (q81) [below right of=q3] {$w$};

 \path (q0)  [dashed, bend right] edge [-] node {} (q1);
 \path (q1) edge node {$\bar{c_1}$} (q2)
 (q2) edge node [left] {$\bar{c}=\bar{c_1}-\bar{p_1}$} (q3)
% (q6) edge node {$\bar{b_1}$} (q7)
 (q7) edge node [left,pos=.2] {{\scriptsize $\bar{l(v_1)}$}} (q3);
 \path (q8) edge node [left,pos=.2] {{\scriptsize $\bar{l(v_2)}$}} (q3);
% (q11) edge node {$\bar{d_1}$} (q7);
 \path (q31) [dashed, bend left] edge [-]  node[pos=.95,right, red] {i} (q3);
 \path (q81) [dashed, bend right] edge [-] node[pos=.95,right, red] {j} (q3);
 \path (q2) edge [draw=none] node {$=$} (q7);
 \path (q7) edge [draw=none] node {$=$} (q8) 
  (q31) [color=blue, bend left=120] edge node[left,red] {$s_1$} (q2)
 (q81) [color=blue, bend right=100] edge node[right,red] {$s_2$} (q7)
 (q81) [color=blue, bend right=120] edge node[right,red] {$s_2$} (q8);
 \end{tikzpicture}
 \end{minipage}%
  \begin{minipage}{.1\textwidth}
   . \\\\\\\\\\\\\\\\\\\\\\\\
    $\stackrel{{\scriptsize \mathbf{R_e2IH}}}{\Rightarrow}$
  \end{minipage}%
 \begin{minipage}{.3\textwidth}
\begin{tikzpicture}[->,>=stealth',shorten >=1pt,auto,node distance=1.5cm,semithick]
 \tikzstyle{every state}=[draw=none]
 \node[state] (q1) {$p_1$};
 \node[state] (q0) [above right of=q1] {$p_0$};
 \node[state] (q2) [below of=q1] {$\stackrel{{\color{red} \hbar}}{u}$};
 \node[state] (q3) [below of=q2] {$\bullet$};
% \node[state] (q6) [right of=q1] {$\stackrel{{\color{red}j;s_2}}{p_2}$};
 % \node[state] (q8) [below right of=q2] {$\bullet$};
 %  \node[state] (q11) [right of=q6] {$\stackrel{{\color{red}j;s_2}}{p_3}$};
  \node[state] (q8) [right of=q2] {$v_2$};
 \node[state] (q31) [below of=q3] {$\bullet$};
 \node[state] (q81) [below right of=q3] {$\bullet$};

 \path (q0)  [dashed, bend right] edge [-] node {} (q1);
 \path (q1) edge node [left] {$\bar{c_1}$} (q2)
 (q2) edge node [left] {$\bar{\lambda}$} (q3);
% (q6) edge node [left] {$\bar{b_1}$} (q2)
% (q2) edge node {$\bar{b_2}=\bar{b_1}\lor\bar{d_1}$} (q3))
% (q11) edge node [left, pos=.1] {$\bar{d_1}$} (q2);
 %\path  (q8) [color=blue,bend right] edge node {} (q11)
  \path (q8) edge node [left,pos=.2] {{\scriptsize $\bar{l(v_2)}$}} (q3);
 \path (q31) [dashed, bend left] edge [-]  node[pos=.95,right, red] {i} (q3);
 \path (q81) [dashed, bend right] edge [-] node[pos=.95,right, red] {j} (q3);
  \path (q2) edge [draw=none] node {$=$} (q8);
 \path % (q81) [color=blue,bend right] edge node {} (q6)
       (q31) [color=blue,bend left=135] edge node[left,red] {$[0;s_1]$} (q1)
       (q81) [color=blue, bend right] edge node[right,red] {$s_2$} (q2)
       (q81) [color=blue, bend right] edge node[right,red] {$s_2$} (q8);  
      % (q81) [color=blue,bend right=120] edge node {} (q11);
\end{tikzpicture}
%\end{tabular}
 \end{minipage}%
 \begin{minipage}{.1\textwidth}
   . \\\\\\\\\\\\\\\\\\\\\\\\   
 $\stackrel{\mathbf{R_e3XH}}{\Rightarrow}$  
  \end{minipage}%
 \begin{minipage}{.25\textwidth}
\begin{tikzpicture}[->,>=stealth',shorten >=1pt,auto,node distance=1.5cm,semithick]
 \tikzstyle{every state}=[draw=none]
 \node[state] (q1) {$p_1$};
 \node[state] (q0) [above right of=q1] {$p_0$};
 \node[state] (q2) [below of=q1] {$\stackrel{{\color{red} \hbar}}{u}$};
 \node[state] (q3) [below of=q2] {$\bullet$};
% \node[state] (q6) [right of=q1] {$\stackrel{{\color{red}j;s_2}}{p_2}$};
 % \node[state] (q8) [below right of=q2] {$\bullet$};
%  \node[state] (q11) [right of=q6] {$\stackrel{{\color{red}j;s_2}}{p_3}$};
 \node[state] (q31) [below of=q3] {$\bullet$};
 \node[state] (q81) [below right of=q3] {$\bullet$};

 \path (q0)  [dashed, bend right] edge [-] node {} (q1);
 \path (q1) edge node [left] {$\bar{c_1}$} (q2)
 (q2) edge node [left] {$\bar{\lambda}$} (q3);
% (q6) edge node [left] {$\bar{b_1}$} (q2)
% (q2) edge node {$\bar{b_2}=\bar{b_1}\lor\bar{d_1}$} (q3))
% (q11) edge node [left, pos=.1] {$\bar{d_1}$} (q2);
 %\path  (q8) [color=blue,bend right] edge node {} (q11)
 \path (q31) [dashed, bend left] edge [-]  node[pos=.95,right, red] {i} (q3);
 \path (q81) [dashed, bend right] edge [-] node[pos=.95,right, red] {j} (q3);
 \path % (q81) [color=blue,bend right] edge node {} (q6)
       (q31) [color=blue,bend left=135] edge node[left,red] {$[0;s_1]$} (q1)
       (q81) [color=blue, bend right=90] edge node[right,red] {$s_2$} (q2);
 \path (q81) [color=blue, bend right=45] edge node[right,red] {$s_2$} (q2);  
      % (q81) [color=blue,bend right=120] edge node {} (q11);
\end{tikzpicture}
%\end{tabular}
\end{minipage}%
%\end{array}
 %\]
 \caption{Applying rules $\mathbf{Re2IH}$ and $\mathbf{Re3XH}$ in sequence, lemma~\ref{lemma:2AESD-Rules}}
 \label{figIIa}\label{OITO}\label{fig:Sequencing-AESD}
\end{figure}

\begin{lemma}[Non-Nested-Ancestor-Edges preservation]\label{lemma:NNAEP}
  Given a valid {\bf DLDS} $\mathcal{D}=\langle V , (E_{D}^{i})_{i\in\{\bar{\lambda}\}\cup\mathcal{O}_{\Gamma}^{i}} , E_{A}, r, l, L, P\rangle$, and $u$ and $v$ in level $j$, such that $l(u)=l(v)$.
  Let $R$ be any {\bf HC} rule, such that, $R$ applies in algorithmic position to $u$ and $v$ resulting in a {\bf DLDS} $\mathcal{D}^{\prime}$. Condition {\em Non-Nested-Ancestor-Edges} holds on $\mathcal{D}^{\prime}$
  \end{lemma}

{\bf Proof.} Let $\mathcal{D}=\langle V , (E_{D}^{i})_{i\in\{\bar{\lambda}\}\cup\mathcal{O}_{\Gamma}^{i}} , E_{A}, r, l, L, P\rangle$ be a valid {\bf DLDS}, $j$ be a level in $\mathcal{D}$ and $R$ a {\bf HC} rule in algorithmic position that collapses $u,v\in V$. We prove that the resulting $\mathcal{D}^{\prime}$ {\bf DLDS} satisfies {\em Non-Nested-Ancestor-Edges} by observing that lemma~\ref{lemma:DedPath} ensures the property that any node $w$ in a mixed deductive path from $u$ ($v$), below the level $j$ that is source of an ancestor edge $e\in E_A$, $source(e)=w$, $target(e)$ is $u$ ($v$), or linked to $u$ ($v$) by a deductive edge of color $0$ or $\bar{\lambda}$. The application of any {\bf HC} rule either moves the target above or below one level or, adds a new edge in a mixed deductive path without sources of $w$. This was already argumented in the proof of lemma~\ref{lemma:DedPath}. Hence, the modifications caused by the application of any {\bf Rule} do not introduce nesting. The same observation holds for the {\bf AESD} rules too. Specifically, for the {\bf MDE} rules,  since when they move downwards an A-edge, the new target takes part in the old path. The indictive hypothesis ensures that this path has no nesting.

\begin{center}
  Q.E.D.
\end{center}

\begin{lemma}[CorrectRuleApp] \label{lemma:CRAppII}
Given a valid {\bf DLDS} $\mathcal{D}=\langle V , (E_{D}^{i})_{i\in\{\bar{\lambda}\}\cup\mathcal{O}_{\Gamma}^{i}} , E_{A}, r, l, L, P\rangle$, and $u$ and $v$ in level $j$, such that $l(u)=l(v)$.
  Let $R$ be any {\bf HC} rule, such that, $R$ applies in algorithmic position to $u$ and $v$ resulting in a {\bf DLDS} $\mathcal{D}^{\prime}$. Thus, condition {\em CorrectRuleApp} holds on $\mathcal{D}^{\prime}$
\end{lemma}

{\bf Proof.} By hypothesis, $\mathcal{D}$ is valid. Thus, $Flow(\mathcal{D},w)(v)$ is well-defined, for every $w\in D$ and $v\in Pre(w)$. Moreover, item [{\bf CorrectRuleApp}] holds, that is, for every $w$ and $v\in Pre(w)$ both items in {\bf CorrectRuleApp} holds. We have to prove that $\mathcal{D}^{\prime}$ also satisfies these two items of {\bf CorrectRuleAPP}, for each possible application of $R$ in $u$ and $v$ in $\mathcal{D}$, yielding $\mathcal{D}^{\prime}$. Due to readability and article's size concerns, we will not show all \total{rule} cases detailed case-analysis proof to catch and agree with a validity of this mathematical result. At a higher-level of presentation, we show here a typical and complex enough case in detail. Let us consider that $R$ is $\mathbf{R_v3XE}$ as it is shown in figure~\ref{fig:Rv3XE-Analysis} below, and that $w_u$ is the conclusion node which has $u$ as a premiss, and analogously, for $w_v$ regarded to $v$. Thus, in this particular case, we have that $Flow(\mathcal{D}^{\prime},w_v)(v)=\{(\bar{b}_2,[0;k;s_3])\}$, and we can verify that $\bar{b}_2=L(v,w_v)$,
$Flow(\mathcal{D},w_u)(u)=\{(\bar{f}_1\lor\bar{f}_2, [0;\ldots;i;s_1]),(\bar{c}_1-\vec{l(p_0)},[0;\ldots;j,s_2])\}$, considering that $\mathcal{D}$ is valid and hence, $l(u)=l(p_0)\imply l(p_1)$ and $(p_1,u)\in \imply_I$, as well $(p_{1a},p_{1b},u)\in\imply_E$. The item {\bf CorrectRuleApp} holds in $\mathcal{D}$, since $\langle u,w_u\rangle\in E_D^{0}$ and $L(\langle u,w_u\rangle)=\lambda$, for $card(Flow(\mathcal{D},w_u)(u))=2$, and then, due to subitem 1 of the second item of [{\bf CorrectRuleApp}] holds. After the application of $\mathbf{R_v3XE}$, see the right-hand side of figure~\ref{fig:Rv3XE-Analysis}, we have the following analysis concerning the validity of $\mathcal{D}^{\prime}$, remembering that $l(u)=l(v)$.
\[
Flow(\mathcal{D}^{\prime},w_u)(u)=\{(\bar{f}_1\lor\bar{f}_2, [0;\ldots;i;s_1]),(\bar{c}_1-\vec{l(p_0)},[0;\ldots;j,s_2]),
(\bar{b}_1\lor \bar{d}_1,[1;k;s_3])
\}
\]
What is justified by:
\[
\begin{array}{ll}
  Flow(\mathcal{D}^{\prime},w_u)(p_{1a})=\{(\vec{l(p_{1a})},[0;0;\ldots;i;s_1])\} \;\; and & \;\;
  Flow(\mathcal{D}^{\prime},w_u)(p_{1b})=\{(\vec{l(p_{1b})},[0;0;\ldots;i;s_1])\} \\
  Flow(\mathcal{D}^{\prime},w_u)(p_{1})=\{(\bar{c}_1,[0;0;\ldots;j;s_2])\}  & \\
  Flow(\mathcal{D}^{\prime},w_u)(p_{2})=\{(\vec{l(p_{2})},[0;1;k;s_3])\} \;\; and & \;\;
  Flow(\mathcal{D}^{\prime},w_u)(p_{3})=\{(\vec{l(p_{3})},[0;1;k;s_3])\} \;\;
  \end{array}
\]
With, $\bar{f}_1=\vec{l(p_{1a})}$, $\bar{f}_2=\vec{l(p_{1b})}$, $\bar{b}_1=\vec{l(p_{2})}$ and $\bar{d}_1=\vec{l(p_{3})}$. Thus, the  conditions on [{\bf CorrectRuleApp}] hold for $\mathcal{D}^{\prime}$, observing the the nodes other then $w_v$, $u$, $w_u$ and their node premisses are not changed. So, the conditions stated by  {\bf CorrectRuleApp} hold on them, since $\mathcal{D}$ satisfies it. 

%% we have three ancestor edges that do not change and preserve the CorrectRuleApp property trivially. Moreover, the ancestor edge with target $v$, by $R$ application, in fact {\bf R18}, gives rise to a new deductive path formed by the ancestor edges arriving to $p_2$ and $p_3$ respectively, instead of the component in the left-hand side formed only by the ancestor edge arriving into $v$. The 

%% By induction on the number of HC compression rules already applied in $\mathcal{D}$. Thus,  as $\mathcal{D}$ is valid by inductive hypothesis then {\bf CorrectRuleApp} holds for $\mathcal{D}$. As a trivial consequence, $Flow(\mathcal{D},w)(v)$ is well-defined for any $w\in D$ and $v\in Pre(w)$. With the sake of having a clearer explanation, w.l.o.g,  Then $Flow(\mathcal{D},w)(w_u)$ and $Flow(\mathcal{D},w)(w_v)$ are well-defined for any $w\in D$, such that, $w_v,w_u\in Pre(w)$. By the definition of $Flow$, the fact that $R$ applies in algorithmic position to $u$ and $v$ and the inductive hypothesis implies that $Flow(\mathcal{D}^{\prime},w_u)(u)=Flow(\mathcal{D},w_u)(u)=$       and $Flow(\mathcal{D}^{\prime},w_v)(v)=Flow(\mathcal{D},w_v)(v)$. 
%%  For example, 

\begin{figure} [H]
%  \begin{tabular}{@{}ccc@{}}
  %\[
  %\begin{array}{cc}
  \begin{minipage}{.45\textwidth}
 \begin{tikzpicture}[->,>=stealth',shorten >=1pt,auto,node distance=1.5cm,semithick]
 \tikzstyle{every state}=[draw=none]
 \node[state] (q1) {$p_1$};
 \node[state] (q0) [above right of=q1] {$p_0$};
 \node[state] (q1a)[left of=q1] {$p_{1a}$};
 \node[state] (q1b)[above of=q1a] {$p_{1b}$};
 \node[state] (q2) [below of=q1] {${u}$};
 \node[state] (q3) [below of=q2] {$\bullet$};
 \node[state] (q3L) [below of=q3] {$\bullet$};
 \node[state] (q6) [right of=q1] {$p_2$};
 \node[state] (q7) [below of=q6] {$v$};
 % \node[state] (q8) [below of=q7] {$\bullet$};
 \node[state] (q8) [below of=q7] {$\bullet$};
 \node[state] (q11) [right of=q6] {$p_3$};
 \node[state] (q31) [below left of=q3L] {$\bullet$};
 \node[state] (q81) [below of=q3L] {$\bullet$};
 \node[state] (q91) [below right of=q3L] {$\bullet$};

 \path (q0)  [dashed, bend right] edge [-] node {} (q1);
 \path (q1) edge node {$\bar{c_1}$} (q2);
 \path (q1a) edge node [left] {$\bar{f_1}$} (q2)
 (q1b) edge node {$\bar{f_2}$} (q2)
 (q2) edge node [left] {$\lambda$} node[pos=.95,right, red] {} (q3)
 (q6) edge node {$\bar{b_1}$} (q7)
 (q7) edge node {$\bar{b_2}=\bar{b_1}\lor\bar{d_1}$}  (q8)
 (q11) edge node {$\bar{d_1}$} (q7);
  \path (q3) [dashed] edge [-] node {} (q3L);
 \path (q31) [dashed, bend left] edge [-]  node[pos=.95,right, red] {i} (q3L);
 \path (q81) [dashed] edge [-] node[pos=.95,right, red] {j} (q3L);
 \path (q91) [dashed, bend left] edge [-] node[pos=.95,right, red] {k} (q8);
 % \path (q81) [dashed, bend left] edge [-] node[pos=.95,right, red] {j} (q8);
 \path (q2) edge [draw=none] node {$=$} (q7)
 (q31) [color=blue, bend left] edge node[left,red,pos=.95] {{\scriptsize $[0,\ldots;i;s_1]$}} (q1b)
 (q31) [color=blue] edge node[right,pos=.7,red] {{\scriptsize $[0,\ldots;i;s_1]$}} (q1a)
 (q81) [color=blue, bend right=120] edge node[right,pos=.9,red] {{\scriptsize $[0,\ldots;j;s_2]$}} (q1)
 (q91) [color=blue, bend right=120] edge node[right,pos=.05,red] {{\scriptsize $[0,k;s_3]$}} (q7);
 \end{tikzpicture}
 \end{minipage}%
  \begin{minipage}{.05\textwidth}
  $\stackrel{\stackrel{HCom(u,v)}{\Longrightarrow}}{\mathbf{R_v3XE}}$  
  \end{minipage}%
 \begin{minipage}{.05\textwidth}% \scale{.8}{
\begin{tikzpicture}[->,>=stealth',shorten >=1pt,auto,node distance=1.5cm,semithick]
 \tikzstyle{every state}=[draw=none]
 \node[state] (q1) {$p_1$};
 \node[state] (q0) [above right of=q1] {$p_0$};
 \node[state] (q1a)[left of=q1] {$p_{1a}$};
 \node[state] (q1b)[above of=q1a] {$p_{1b}$};
 \node[state] (q0) [above right of=q1] {$p_0$};
 \node[state] (q2) [below of=q1] {$u$};
 \node[state] (q3) [below of=q2] {$\bullet$};
 \node[state] (q3L) [below of=q3] {$\bullet$};
 \node[state] (q6) [right of=q1] {$p_2$};
  \node[state] (q8) [below right of=q2] {$\bullet$};
  \node[state] (q11) [right of=q6] {$p_3$};
 \node[state] (q31) [below of=q3] {$\bullet$};
 %\node[state] (q81) [below right of=q3L] {$\bullet$};
 \node[state] (q31) [below left of=q3L] {$\bullet$};
 \node[state] (q81) [below of=q3L] {$\bullet$};
 \node[state] (q91) [below right of=q3] {$\bullet$};

 \path (q0)  [dashed, bend right] edge [-] node {} (q1);
 \path (q1) edge node {$\bar{c_1}$} (q2)
 (q1a) edge node [left] {$\bar{f_1}$} (q2)
 (q1b) edge node {$\bar{f_2}$} (q2)
 (q2) edge node [left] {$\lambda$} node[pos=.95,right, red] {} (q3)
 (q6) edge node [right, pos=.3] {$\bar{b_1}$} (q2)
 (q2) edge node {$\bar{b_2}=\bar{b_1}\lor\bar{d_1}$} node[pos=.95,right, red] {1} (q8)
 (q11) edge node [right, pos=.2] {$\bar{d_1}$} (q2);
 %\path  (q8) [color=blue,bend right] edge node {} (q11)
   \path (q3) [dashed] edge [-] node {} (q3L);
 \path (q31) [dashed, bend left] edge [-]  node[pos=.95,right, red] {i} (q3L);
 \path (q81) [dashed] edge [-] node[pos=.95,right, red] {j} (q3L);
 \path (q91) [dashed, bend right] edge [-] node[pos=.95,right, red] {k} (q8);
 \path (q91) [color=blue,bend right=120] edge node[right,pos=.7,red] {{\scriptsize $[0;1;k;s_3]$}} (q6)
 (q31) [color=blue, bend left] edge node[left,red,pos=.95] {{\scriptsize $[0;\ldots;i;s_1]$}} (q1b)
 (q31) [color=blue] edge node[right,pos=.7,red] {{\scriptsize $[0;\ldots;i;s_1]$}} (q1a)
 (q81) [color=blue, bend right=120] edge node[right,pos=.95,red] {{\scriptsize $[0;\ldots;j;s_2]$}} (q1)
% (q91) [color=blue, bend right=120] edge node[right,pos=.15,red] {$k;s_3$} (q7);
 (q91) [color=blue, bend right=135] edge node[right,pos=.3,red] {{\scriptsize $[0;1;k;s_3]$}} (q11);
\end{tikzpicture}
%\end{tabular}
\end{minipage}%
%\end{array}
%\]
 \caption{(a)$u$ and $v$ collapse   \hspace{2cm} (b) After collapse $HCom(u,v)$}
 \label{fig:Rv3XE-Analysis}
\end{figure}

The other cases, concerning the other~\total{rule}-1 rules, take part in the forthcoming formalization of this compressing algorithm and formal proof of the theorem in this article, see~\cite{RobinsonLean}.
We are working in  a more formal presentation, with  an integral and more detailed argument with a formalization in the LEAN ITP. \cite{RobinsonLean} shows initial steps in the formalization of what we have in this article/report. 

\begin{center}
  Q.E.D.
\end{center}

Lemmas~\ref{lemma:SaAS} is the only lemma that does not work for every HC-rule. Ancestor simplicity is not preserved, in general, by {\bf MDE} rules. It is preserved by all other HC-rules. Thus, if we do not consider {\bf MDE} rules, we have the following theorem~\ref{theo:Soundness}.

\begin{theorem}\label{theo:Soundness}
  The set of compression rules that are not {\bf ASED} preserves validity of any valid {\bf DLDS};
\end{theorem}

{\bf Proof.} We show for each one of the rules presented in the previous section that if a {\bf DLDS} $\mathcal{D}$ is valid, then so is the $\mathcal{D}$  after the application of the rule, by lemmas~\ref{lemma:CA},~\ref{lemma:LC}, ~\ref{lemma:AE}, ~\ref{lemma:ABI}, ~\ref{lemma:SaAS}, ~\ref{lemma:NNAEP}, ~\ref{lemma:CRAppII}, ~\ref{lemma:CRAppIElimIIntro}. Observe that lemma~\ref{lemma:SaAS} is the only in the list of lemmas that is restricted to rules that are not {\bf MDE}.

\begin{center}
  Q.E.D.
\end{center}

A formal proof of the above lemma would be a very long case analysis, such that, each case has to check each of the items in the definition of validity of {\bf DLDS}  in definition~\ref{def:ValidDLDS}. Due to the tedious repetition of cases, we postponed this formalization to a further article, conveyed with the help of an Interactive Theorem Prover. In~\cite{RobinsonLean} it is report the initial step forward this formalization.

Look at the example at the beginning of this article, figures~\ref{fig:summary-MUE} and~\ref{fig:summary-MUE}. Due to the order of HC-rules application, the compression algorithm  applied all the (appliable) {\bf MUE} rules before the first {\bf MDE} rule application. This is a consequence of the fact that no {\bf MDE} rule can be applied before all the other rules cannot be applied anymore. Remember that the order of application of the rules is bottom-up and right to left. The pattern for the application of any of the {\bf MDE} rules appears in the {\bf DLDS} only after some rule has been applied. 

%{\color{red}   ESTOU AQUI ACIMA PARA DEFINIR OS ESCOPOS das regras MDE}

A {\bf MDE} rule is only applied when the two nodes that will be collapsed by the application are both A-egde targets and top-nodes. The {\bf MUE} rules repeatly move the ancestor edges up until they do not apply anymore  when only some {\bf MDE} rule can be applied,  i.e., we reach firstly a $\mathbf{MUE}^{+}$-compressed {\bf DLDS}.  
We can state the following theorem that uses the definition below.

\begin{definition}[Partially $\mathbf{MUE}^{+}$-compressed {\bf DGTD}] A partially $\mathbf{MUE}^{+}$-compressed {\bf DGTD} is a a partially compressed {\bf DGTD} obtained by the sole application of {\bf MUE} HC-compression rules to a {\bf DGTD}.
\end{definition}

\begin{theorem}\label{theo:MUE-rules}
  Let $\mathcal{D}=\langle V , (E_{D}^{i})_{i\in\{\bar{\lambda}\}\cup\mathcal{O}_{\Gamma}^{i}} , E_{A}, r, l, L, P\rangle$ be a valid partially $\mathbf{MUE}^{+}$-compressed {\bf DGTD}. Consider the HC compression algorithm, restricted only to {\bf MUE} rules, applied to $\mathcal{D}$. The algorithm stops after a finite number of steps and obtains a valid {\bf DLDS} $\mathcal{D}^{\prime}$ that has no level with a pair of top-nodes labelled with the same formula, except possibly for top-nodes occurring in the same level and labelled with the same formula. 
\end{theorem}

{\bf Proof.} Since {\bf DGTD}s are {\bf DLDS} too, by theorem~\ref{theo:Soundness} $\mathcal{D}^{\prime}$ is a valid {\bf DLDS}. Each time, a {\bf MUE} rule applies to a valid {\bf DLDS}, it yields a {\bf DLDS} with one pair of nodes labelled with the same formula nodes less. Type-0 rules create A-edges, Type-\Romannum{1} rules creates A-edges too. Type-\Romannum{2} rules, but the {\bf MDE} rules, move the A-edges up along the derivation. The combination of these {\bf MUE} rules, plus the Type-\Romannum{3} rules, for example  $\mathbf{R_v3XE}$, $\mathbf{R_e3XE}$, $\mathbf{R_v3XI}$, $\mathbf{R_e3XI}$; collapse all nodes labelled with the same formula in levels lower than the top-nodes connected to them. At this stage no {\bf MUE} rule can be applied, for there is no configuration that matches the left handside of these rules. The only nodes labelled with the same formula in the same level are the top-nodes. Thus,  $\mathcal{D}^{\prime}$ is a  valid $\mathbf{MUE}^{+}$-compressed {\bf DLDS}.

\begin{center}
  Q.E.D.
\end{center}

\begin{corollary}\label{co:MUE-compressed-format}
If $\mathcal{D}$ is a $\mathbf{MUE}^{+}$-compressed {\bf DLDS} the only targets of A-egdes in $\mathcal{D}$ are the top-formulas.
  \end{corollary}

Finally, using lemma~\ref{lemma:2AESD-Rules} and corollary \ref{coro:2AESD-Rules} we note that we can obtain a fully-compressed {\bf DLDS} by the application of the {\bf MDE} rules from bottom-up and left-to-right. 
{\bf MDE} rules collpase the top-nodes labelled with the same formula that occur in the same level. Thus, we have the followin theorem. 

\begin{theorem}\label{theo:Termination}
  If algorithm~\ref{HorCom} extended with a second round of {\bf MDE} rules applications is applied to a valid {\bf DLDS} then it eventually halts providing a {\bf DLDS} that has no level with
  two nodes labeled with a same formula.
\end{theorem}

%{\bf Proof.} Let $\mathcal{D}$ be a valid {\bf DLDS}. If $\mathcal{D}$ has no level with two nodes labelled with the same formula, then the algorithm halts. Suppose there are levels with two nodes labelled with the same formula, then there are the lowest level $l_0$ and a leftmost pair of nodes $u$ and $v$ labelled with this same formula. Since  aLgorithm~\ref{HorCom} spans $\mathcal{D}$ from the lowest level up and in each level from left to right, it will collapse the pair $u$ and $v$ using the appropriate rule among the rules of compression. Due to theorem~\ref{theo:Soundness} the result of the compression is a {\bf DLDS} $\mathcal{D}^{\prime}$ that has one less pair of collapsible nodes. This conclusion is because each rule does not add new pairs of collapsible nodes after its application. The algorithm then will eventually halt.

The root is labelled with the conclusion of the valid {\bf DLDS}. To have the information about the dependency set from which this conclusion depends, we have to consider the effect of the last rule on dependency sets represented by labels on the deductive edges incoming in $r$. Remember that the last rule can be a $\imply$-Intro or a $\imply$-Elim. We add a new node linked to the root $r$ with a (new) distinguished edge with the sole purpose of having it labelled with the final dependency set. We call this a grounded {\bf DLDS}.

\begin{definition}
  Let $\mathcal{D}=\langle V , (E_{D}^{i})_{i\in\mathcal{O}_{\Gamma}^{i}} , E_{A}, r, l, L, P\rangle$ be a valid {\bf DLDS}. Let  $\mathcal{D}^{\prime}$ extend $\mathcal{D}$ by adding a new node $g$, such that, $l(g)=\Ground$, a new deductive edge of color 0 $\langle r,g\rangle$, such that $L(\langle r,g\rangle)=\vec{b}(S)$, where:
  \begin{enumerate}
  \item If $l(r)=\alpha_1\imply\alpha_2$ and $\langle v,r\rangle$ is its only incoming edge, a deductive edge of color 0, then
    $r$ is the conclusion of a $\imply$-intro and $S=Set(L(\langle v,r\rangle))-\{\alpha_1\}$; and
  \item If $r$ is the conclusion of a $\imply$-Elim and $\langle v_1,r\rangle$ and $\langle v_2,r\rangle$ are its only incoming edges then $S=Set(L(\langle v_1,r\rangle))\cup Set(L(\langle v_2,r\rangle))$
  \end{enumerate}
  We call $\mathcal{D}^{\prime}$ a {\bf grounded} {\bf DLDS}.
  \end{definition}

Note that only valid {\bf DLDS} can be grounded, by the force/will of the definition. The root of a grounded {\bf DLDS}, is, again, by the 
the force of definition, $r$, where $\langle r,g\rangle\in E_D^0$ and $l(g)=\Ground$. The node $g$ is the support for the ground only. We have the following corollary. We say that a {\bf DLDS} is compressed iff it has no collapsible pair of node in any level. 

\begin{corollary}\label{coro:DLDSofTautologies}
  Let $\alpha$ be any tautology in $M_{\imply}$. There is a compressed grounded {\bf DLDS} $\mathcal{D}=\langle V , (E_{D}^{i})_{i\in\mathcal{O}_{\Gamma}^{i}} , E_{A}, r, l, L, P\rangle$, such that $l(r)=\alpha$, $\Gamma$ is the set of sub-formulas of $\alpha$ and $L(\langle r,\Ground\rangle)=\vec{b}(\emptyset)$.
\end{corollary}

{\bf Proof.} If $\alpha$ is a $M_{\imply}$ tautology then there is a Natural Deduction proof of it. By the normalization of $M_{\imply}$ there is a normal ND proof. This proof has only occurrence of subformulas of $\alpha$. By lemma~\ref{Greedy} there is a greedy ND derivation of $\alpha$ having also only sub-formulas of $\alpha$ occurrences. Proposition~\ref{prop1} and corollary\ref{dischargecorolario} ensure the existence of a {\bf DGTD} $\mathcal{T}$ that is a proof of $\alpha$. Finally, from  proposition~\ref{def:Dag-of-a-DGTD} we can easily see that $Dag(\mathcal{T})$ is a valid {\bf DLDS} and its extension to a grounded {\bf DLDS} has $L(\langle r,\Ground\rangle)=\vec{b}(\emptyset)$. Thus, by theorem~\ref{theo:Termination} and theorem~\ref{theo:Soundness}, we have a compressed and grounded {\bf DLDS} with the conclusion depending of no assumption, i.e., it is a proof of $l(r)$.
\begin{center}
  Q.E.D.
\end{center}

Finally, we adjust the compression algorithm to obtain fully-compressed {\bf DLDS} from valid {\bf DGTD}.

\begin{algorithm}[H]
  \caption{Horizontal Compression}
  \label{HorComMDE}
  \begin{algorithmic}[1]
    \Require{A tree-like greedy valid derivation $\mathcal{D}$}
    \Ensure{The valid DLDS that is $\mathcal{D}$ fully-compressed} 
    \For{$lev$ from 1 to $h(\mathcal{D})$}
    \For{$u$ and $v$ at $lev$}
    \State $HCom_{MUE^{+}}(u,v)$
    \EndFor
    \EndFor
    \For{$lev$ from 1 to $h(\mathcal{D})$}
    \For{$u$ and $v$ at $lev$}    
    \State $HCom_{MDE}(u,v)$
    \EndFor
    \EndFor
  \end{algorithmic}
\end{algorithm}

\section{On the size of compressed Natural Deduction proofs}
\label{sec:Upper-HC}
%% \begin{definition}[Simple Colored Graph]
%%   Standard Definition
%%   \end{definition}

We start by a simple consequence of the validity of any {\bf DLDS}. 

\begin{lemma}\label{ordinals}
 Let $\mathcal{D}=\langle V , (E_{D}^{i})_{i\in\mathcal{O}_{\Gamma}} , E_{A}, r, l, L, P\rangle$ be a compressed grounded {\bf DLDS}, obtained by the application of the Horizontal Compression algorithm~\ref{HorComMDE}  to a valid {\bf DGTD}. The underlying graph of $\mathcal{D}$ with colors in $\mathcal{O}_{\Gamma}$ and $E_A$ is a simple graph.
\end{lemma}

{\bf Proof.} This is a trivial consequence of validity definition, since the fifth and sixth conditions on the validity is the simplicity of the full colored subgraph of $\mathcal{D}$, with colors in $\mathcal{O}_{\Gamma}$ and $E_A$,respectively.
\begin{center}
  Q.E.D.
  \end{center}

In lemma~\ref{ordinals} above, we consider only the full subgraph coloured with ordinals and the ancestor edges ($E_A$). However, some rules, such as rule~{\bf R\ref{SETE}}, create coloured edges labelled with the colour $\bar{\lambda}$, although no more than one by pair of vertexes. Thus, we should note that rules~{\bf R\ref{SETE}},~{\bf R\ref{OITO}},~{\bf R\ref{NOVE}},~{\bf R\ref{QUINZE}},~{\bf R\ref{DEZESSEIS}} do not create more than one edge labeled with $\bar{\lambda}$ between any pair of nodes. As already said, We consider $\bar{\lambda}$ as a colour. Besides that, the rules mentioned above are rules that collapse edges in a way that anytime a pair of edges is collapsed, the algorithm generates at most one $\bar{\lambda}$ coloured edge. The collapse of edges is a consequence of the collapse of their respective source and targets in a two-step action. Thus, the previous lemma extends to the following proposition~\ref{DLDSsimple}, where the only addition to the previous lemma~\ref{ordinals} is to consider $\bar{\lambda}$ as a colour.

\begin{proposition}\label{DLDSsimple}
Let $\mathcal{D}=\langle V , (E_{D}^{i})_{i\in\mathcal{O}_{\Gamma}} , E_{A}, r, l, L, P\rangle$ be a compressed grounded {\bf DLDS}, obtained by the application of the Horizontal Compression algorithm~\ref{HorComMDE} to a valid {\bf DGTD}. The underlying graph of $\mathcal{D}$ with colors in $\mathcal{O}_{\Gamma}^{\bar{\lambda}}$ and $E_A$ is a simple graph.
\end{proposition}

{\bf Proof.} This is a trivial consequence of the validity of the {\bf DGTD}, formalized in the lemmas above, and the fact that we can only apply the rule in figure~\ref{DEZ} when all the node in its graphical representation are pairwisely different. The focus is on the two bullets being different.  
\begin{center}
  Q.E.D.
  \end{center}

%% \begin{lemma}\label{MainLemma}
%%   Let $\mathcal{T}=\langle V,E_{D},E_{d},r,l\rangle$ be a tree-like derivation of $\alpha$ from $\Gamma$. Let $\mathcal{D}=\langle V^{\prime} , (E_{D}^{i})_{i\in\mathcal{O}_{\Gamma}} , E_{A}, r, l, L, P\rangle$ be the DLDS obtained from $\mathcal{T}$ by the application of the Horizontal Compression algorithm~\ref{HorCom} to it. If $w\in V$ is an $A$-source of at least two different $A$-Edges then the respective $A$-targets $u_1$ and $u_2$ are such that, if $u_1$ and $u_2$ are at the same level of the rooted dag $\mathcal{D}$ then $l(u_1)\neq l(u_2)$.
%% \end{lemma}

%% \begin{corollary}\label{coroum}
%%   If $u$ and $v$ are $A$-targets of $A$-edges $f_1$ and $f_2$, such that $l(u)=l(v)$ at the same level, then $source(f_1)\neq source(f_2)$.
%% \end{corollary}

%% \begin{corollary}\label{corodois}
%%   Let $w$ be the result of a collapse and target of a $\bar{\lambda}$-Edge that has $v\in V^{\prime}$, an hypothesis,  as source of this $\bar{\lambda}$-Edge, then if $w$ is $A$-target of two edges $f_1$ and $f_2$ then $source(f_1)\neq source(f_2)$ (see figure~\ref{DEZ}b).
%% \end{corollary}

\begin{lemma}\label{lemma:SizeofTheRelativeAddress}
Let $\mathcal{D}=\langle V , (E_{D}^{i})_{i\in\mathcal{O}_{\Gamma}} , E_{A}, r, l, L, P\rangle$ be a compressed grounded {\bf DLDS}, obtained by the application of the Horizontal Compression algorithm~\ref{HorComMDE} to a valid {\bf DGTD}, with height $h$. For each $\langle u,w\rangle\in E_A$, $len(P(\langle u,w\rangle))$ if defined is smaller than $h$.

  \end{lemma}

{\bf Proof.} In a valid {\bf DLDS}, each ancestor edge $\langle u,w\rangle\in E_A$ has its source $u$ in a level stricly smaller than $w$'s level, i.e., the target. Since $P(\langle u,w\rangle)$, according the validity of $\mathcal{D}$ is the relative address of $u$ from $w$, which goes downwards always, then the size of $P(\langle u,w\rangle)$ cannot be bigger than the difference between the levels of $w$ and $u$. Thus $len(P(\langle u,w\rangle))leq h$. Observe that no compression rule increases the height of the {\bf DLDS}.
\begin{center}
  Q.E.D.
  \end{center}

We state the following lemma without proof, since it is an easy consequence of the definition of $l$.

\begin{lemma}\label{lemma:SizeofTheLabels}
  Let $\mathcal{D}=\langle V , (E_{D}^{i})_{i\in\mathcal{O}_{\Gamma}} , E_{A}, r, l, L, P\rangle$ be a compressed grounded {\bf DLDS}, obtained by the application of the Horizontal Compression algorithm~\ref{HorComMDE} to a valid {\bf DGTD}. For each $\langle u,w\rangle\in E_D^{0}$, the size of $l(\langle u,w\rangle)$ the lengh of the bitstring that is size of $\Gamma$ plus one.
  \end{lemma}

\begin{definition}{Size of a {\bf DLDS}}
  Let $\mathcal{D}$ be a grounded {\bf DLDS} of $\alpha$ from $\Gamma$. The size of $\mathcal{D}$ is the lenght of the string obtained by the juxtaposition of the the strings (words) in the alphabet of ordinals derived from $\Gamma$ plus $\lambda$, and the punctuation marks: $\{,\},\rangle,\langle$ representing each component of $\mathcal{D}$.
\end{definition}

We observe that in fact, the punctuation marks do not need to be taken into account in the complexity analysis. Thus, the followin results do not take them into account.

\begin{proposition}[Upper-bound on simple $DLDS$]\label{prop:SizeSimple}
  Let $\mathcal{D}$ be a {\bf DLDS} with height $h$, with $m$ as the size of the set of node labels, such that, has a simple graph as the underlying colored graph. The size of $\mathcal{D}$ has upper-bound $\mathcal{O}(h\times m^4)$.
    \end{proposition}

\begin{proof}
 Recall that $\mathcal{D}=\langle V , (E_{D}^{i})_{i\in\mathcal{O}_{\Gamma}^{i}} , E_{A}, r, l, L, P\rangle$, $size(\Gamma)=m$;
 $size(V)\leq h\times m$; considering the labelling of vertexes $size(labeled(V))\leq h\times m^2$, i.e., lemma~\ref{lemma:SizeofTheLabels}.  
 Considering the colors of the the deduction edges,  for each color $i=0,m$, $size(E_{D}^{i})\leq size(V)^2$; hence $size(labeled(E_D^{i}))\leq size(V)^{2}\times m = (h^2\times m^2)\times m= h^2\times m^3$. Considering all colors we have $\size(\bigcup_{i=0,m}E_D^{i})\leq (m+1)\times h^2\times m^3=h^2\times m^4+h^2\times m^3$. By lemma~\ref{lemma:SizeofTheRelativeAddress} we have that $size(P(\langle u,v\rangle))\leq h$, for each $\langle u,v\rangle\in E_A$. Thus, we have that $size(E_A)\leq size(V)\times size(V)\times h\leq (h\times m)^2\times h= (h\times m)^2$. Finaly we have that $size(\mathcal{D})\leq \mathcal{O}(h\times m^4)$
  \end{proof}
\begin{center}
  Q.E.D.
  \end{center}

\begin{corollary}[Upper-bound on compressed grounded {\bf DLDS}]\label{coro:SizeDLDS}
  Let $\Pi$ be a proof of $\alpha$ in $M_{\imply}$ with height $h$. Let $m$ be the number of formulas of occurrying in $\Pi$. Then there is a compressed grounded {\bf DLDS} $\mathcal{D}=\langle V , (E_{D}^{i})_{i\in\mathcal{O}_{\Gamma}^{i}} , E_{A}, r, l, L, P\rangle$, such that $l(r)=\alpha$, $L(\langle r,\Ground\rangle)=\vec{b}(\emptyset)$ and $size(\mathcal{D})\leq \mathcal{O}(h\times m^4)$.
\end{corollary}

{\bf Proof.} From corollary~\ref{coro:DLDSofTautologies} we have the existence of $\mathcal{D}$ and from proposition~\ref{prop:SizeSimple} we have the upperbound. 
\begin{center}
  Q.E.D.
  \end{center}

\section{On the experimental compression rate of the algorithm}
\label{sec:Experimental}
\begin{figure}[H]

    \includegraphics[width=18cm]{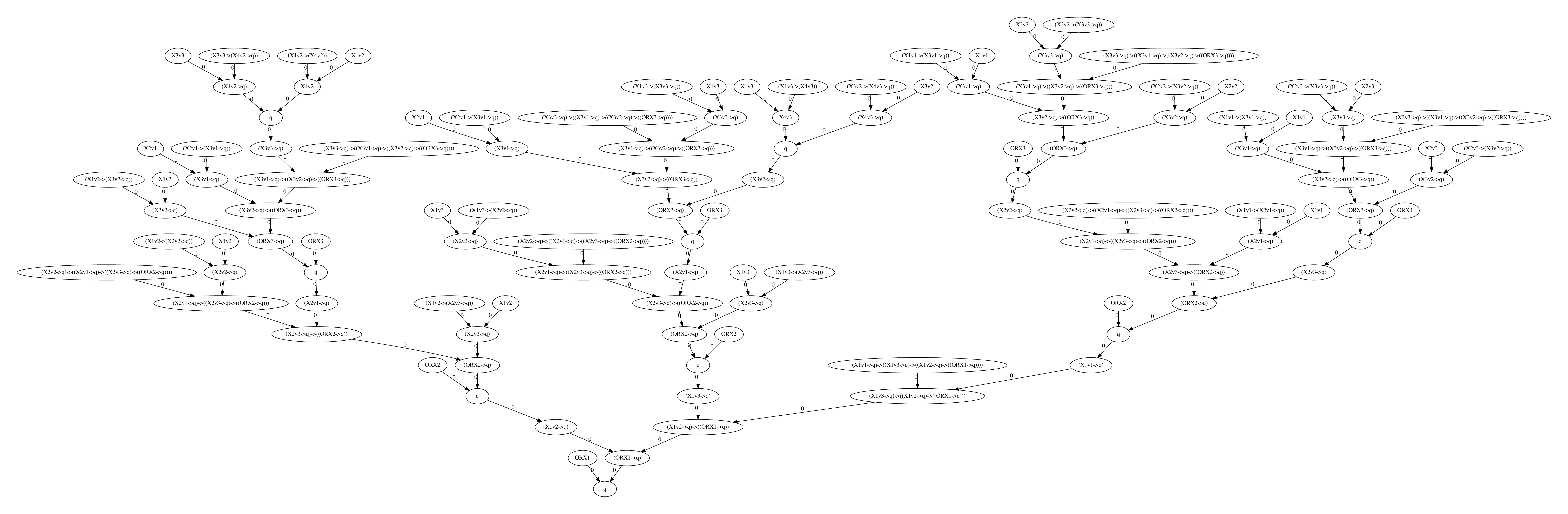}

  \caption{Natural Deduction brute force proof of non-hamiltonicity of the graph in figure~\ref{G3}}
    \label{G3full}
\end{figure}

In 2019, one of the authors implemented horizontal compression (HC) using Python2.7.15 and the GRAPHVIZ library to store and manipulate graphs. In his master dissertation, he implemented Natural Deduction proofs and {\bf DLDS}  in GRAPHVIZ. Some classes of formulas with huge proofs were submitted to the HC algorithm. We developed a program to generate purely implicational Natural Deduction proofs of non-hamiltonicity of graphs to have specimens submitted to the HC implementation. Fibonacci based Natural Deduction purely implicational formulas with exponential size were compressed by the HC algorithm and compared to Huffman compression algorithm, one of the benchmarks
in compression of strings. In 2019 we could not find a robust compression algorithm for proofs. Thus, we report here the comparison with a string-based algorithm in the case of Huffman compression. We show a qualitative comparison of a non-hamiltonicity proof of size almost
$3^3$ formulas with its HC compressed {\bf DLDS} graph. We can  see in figures~\ref{G3full} and~\ref{G3compressed}. We know that a graph with three nodes is quite small. Its proof of non-hamiltonicity is a Natural Deduction proof with almost $3^3$ formulas. Anything bigger cannot be shown for the sake of qualitative analysis. The set of HC-rules in 2019 was not the same that is shown here. The current set is an optimzed and more adequate set of rules, mainly to have a simpler proof of soundness preservation.

\begin{figure}[H]
{\small
\begin{tikzpicture}[]
\node(a) at (2,0) {$v1$};
\node(b) at (4,0) {$v3$};
\node(c) at (3,2) {$v2$};
\path[->] (a) edge (b);
\path[->] (a) edge (c);
\end{tikzpicture}
}
\caption{Graph G3}
\label{G3}
\end{figure}
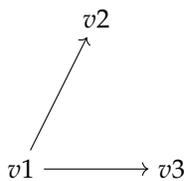

\begin{figure}[H]

    \includegraphics[width=18cm]{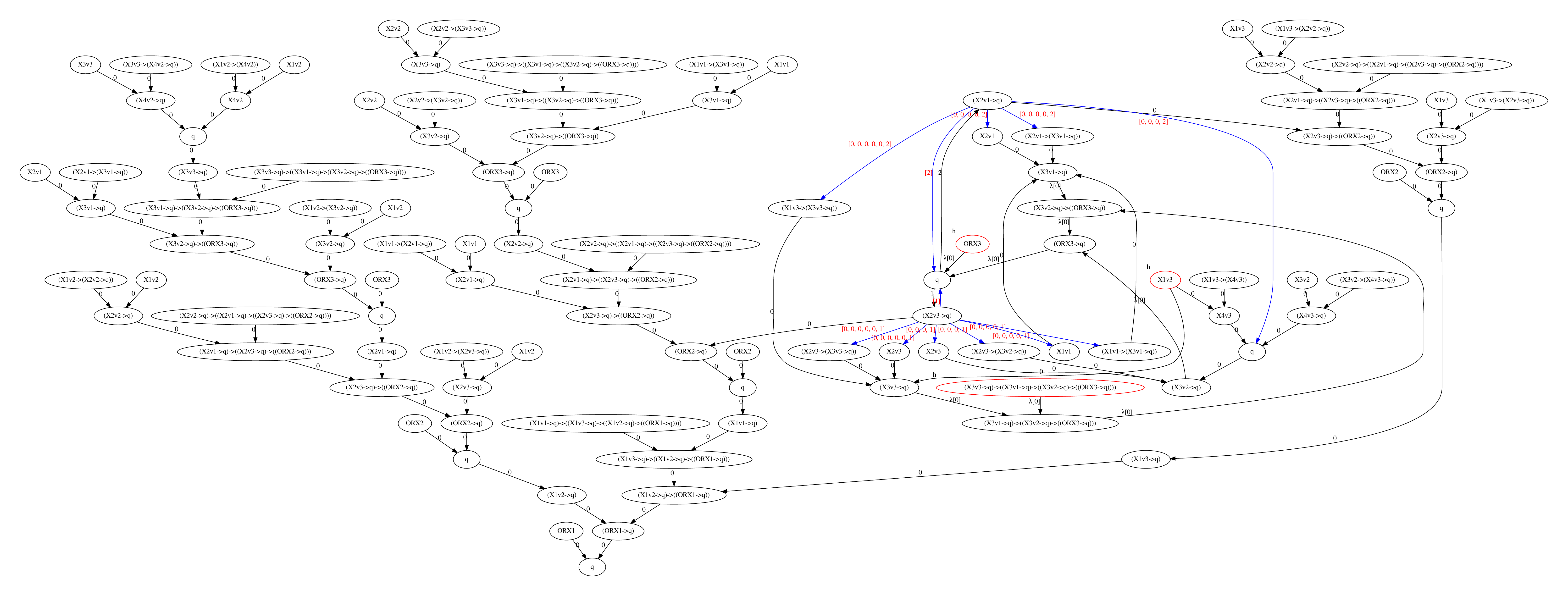}

  \caption{HC compressed non-hamiltonicity proof for the graph in figure~\ref{G3}}
    \label{G3compressed}
\end{figure}

\begin{figure}[H]

    \includegraphics[width=20cm]{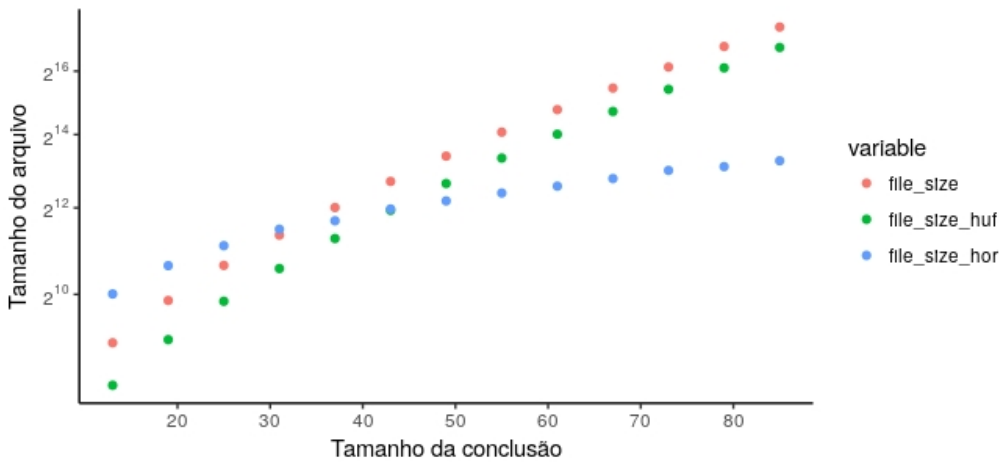}

  \caption{Compression comparison between Huffman and HC compression for big tautologies}
    \label{fig:CompressionRate}
\end{figure}

In the compressed proof shown in figure\~ref{G3compressed}, we can see only three collapse nodes, the red ones, which generated some ancestor edges, and the blue ones. In general,  as big the proof is, as redundant it is, see\cite{IfCologHaeusler} and the HC algorithm works better. Figure~\ref{fig:CompressionRate} shows the comparison of the compression rate of the Huffman with the horizontal compression (HC) rate.  

\begin{figure}[H]

    \includegraphics[width=15cm]{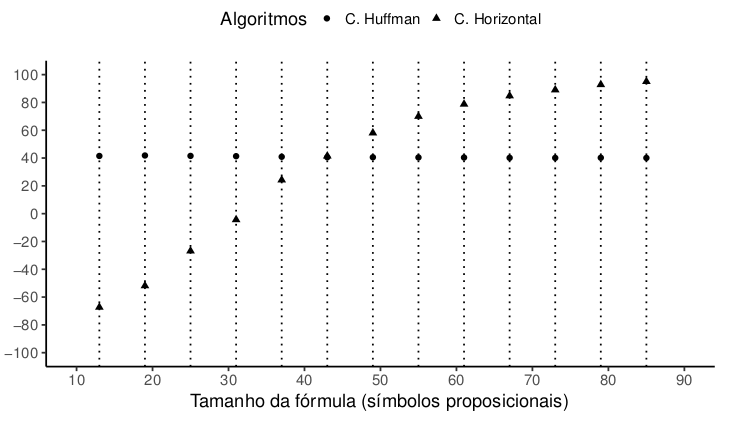}

  \caption{Compression rate comparison between Huffman and HC compression for big tautologies}
    \label{fig:RateCompression}
\end{figure}

We can see in figure~\ref{fig:CompressionRate} how HC is better than Huffman when the proofs are bigger, and more redundant henceforth. In figure~\ref{fig:RateCompression} we can effectively compare the comnpression rate between both algorithms.

\begin{figure}[H]

    \includegraphics[width=15cm]{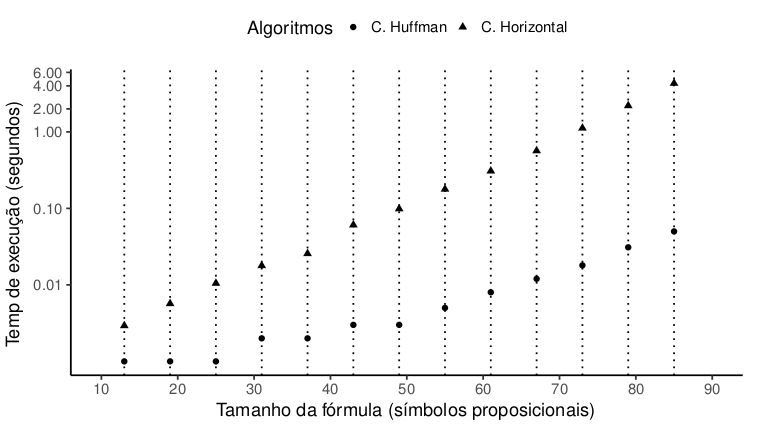}

  \caption{Comparison between Huffman and HC compression time for big tautologies of the class Fibonacci(n)}
    \label{fig:Timeelapse}
\end{figure}

Finally, in figure~\ref{fig:Timeelapse} we compare the time used by both algorithms for compressing the proofs of the class Fibonacci(n) that was used in this basic quantitative  study.

\section{Conclusion and further research}

We think we succeeded to a high degree in almost all of the contributions listed in the second paragraph of the introduction. One of the primary purposes of this report is to provide a comprehensive technical presentation of the Horizontal Compression method to compact propositional proofs in $M_{\imply}$. Using this report, other researchers or we can use a prover assistant to formalize the proof of soundness preservation, termination and even the polynomial upper bound. We think that maybe the last one because having traditional estimative techniques on data structures might not need formal proof. The experimental part lacks a comparison with Ziv-Lempel method. It will follow soon. We advocate that the compression of proofs and their corresponding efficient verification is a research goal {\it per se}. It can be taken independently from our theoretical proofs already published. Concerning them, we currently work to improve our arguments to have shorter and more intuitive ones. The fact that nowadays we are concerned with assisted proofs on the compression method shown here is already a research goal {\it per se}. Currently we are working on a LEAN formalization of the soundness preservation of {\bf HC} concerning the $MUE$ rules, i.e., the Moving Up Edges. This allows us to conclude that the compressed {\bf DLDS} obtained is a simple graph, except for the leaves, i.e., the nodes having Ancestor rules arriving at them. In this way we cannot ensure that this kind of almost simple {\bf DLDS} cannot have a polynomial size bound on the number of nodes of the its underlying graph. It may have a leaf $u$ with the cardinality of the cardinality of the set $\{\langle v,u\rangle : \langle v,u\rangle\in E_A\}$ bigger than any polynomial on the number of nodes of the underlying graph of the {\bf DLDS}. It is important to note, on the other hand, that the size of a the set above, when $u$ is not a leaf, is empty, so the underlying graph, without the leaves is trivially simple, including for edges in $E_A$. This is a direct consequence of the definition of validity of {\bf DLDS}s.

\noindent\textbf{Acknowledgments}
 {\small  This work arose in the context of term- and proof-compression research
supported by the ANR/DFG projects \negthinspace \emph{HYPOTHESES} and \emph{%
BEYOND LOGIC} [DFG grants 275/16-1, 16-2, 17-1] and the CNPq project\emph{\
Proofs: Structure, Transformations and Semantics} [grant 402429/2012-5].
The first author, Haeusler,  wants to thank very much prof. Lew Gordeev, his co-author and
co-responsible for many of the ideas in a first version of {\bf HC} compression method on Natural Deduction proofs.
Nowadyas we intend to use the {\bf HC} algorithm in a alternative proof, more concrete, way to prove what we proved in \cite{GH2020,GH2022,GH2022arxiv} and that we presented in Logic Coloquium 2021, see abstract in \cite{GHLogicColoquium2021}. 
We
would like to thank L. C. Pereira and all colleagues in PUC-Rio for their
contribution as well as P. Schroeder-Heister (EKUT) and M. R. F. Benevides
(UFRJ) For their support of these projects. Special thanks go to S. Buss and R.
Dyckhoff for their valuable comments during the preparation of \cite{GH2019}}. We thank Robinson Callou
 for accepting the chalenge to formalize the results presented in this article in $L\exists\forall N$. One of the
 authors is grateful to Leonardo de Moura for the constant motivation to have the formalization of the {\bf HC} compression in some ITP. There is a huge group of people that should be listed in the ackowlegdements. There is no list of names in this part of the article to avoid any mistaken omission.

 \bibliography{NDDerivationsCompression}

\begin{thebibliography}{RCMF22}

\bibitem[GH19]{GH2019}
L.~Gordeev and E.~H. Haeusler.
\newblock Proof compression and np versus pspace.
\newblock {\em Studia Logica}, 107(1):55--83, 2019.

\bibitem[GH20]{GH2020}
L.~Gordeev and E.~H. Haeusler.
\newblock Proof compression and np versus pspace ii.
\newblock {\em Bulletin of the Section of Logic}, 49(3):213--230, 2020.

\bibitem[GH22a]{GHLogicColoquium2021}
L.~Gordeev and E.~H. Haeusler.
\newblock On proof theory in computational complexity.
\newblock In {\em Report on the Logic Coloquim 2021}, 2022.
\newblock The Logic Coloquium Proceedings of Abstracts is published as an
  article in \cite{LogicColoquium2021}.

\bibitem[GH22b]{GH2022arxiv}
L.~Gordeev and E.~H. Haeusler.
\newblock On proof theory in computational complexity: overview, 2022.

\bibitem[GH22c]{GH2022}
Lew Gordeev and Edward~Hermann Haeusler.
\newblock {P}roof {C}ompression and {NP} {V}ersus {PSPACE} {II}: {A}ddendum.
\newblock {\em Bulletin of the Section of Logic}, page 9 pp., Jan. 2022.

\bibitem[Hae22]{IfCologHaeusler}
Edward~Hermann Haeusler.
\newblock Exponentially huge natural deduction proofs are redundant:
  Preliminary results on m\({}_{\mbox{{\(\supset\)}}}\).
\newblock {\em {FLAP}}, 9(1):287--326, 2022.

\bibitem[Hud93]{Hudelmaier}
J.~Hudelmaier.
\newblock An $o\left( n\log n\right) $-space decision procedure for
  intuitionistic propositional logic.
\newblock {\em Journal of Logic and Computation}, 3:1--13, 1993.

\bibitem[Pra65]{Prawitz}
D.~Prawitz.
\newblock {\em Natural deduction: A proof-theoretical study}.
\newblock Almqvist \& Wiksell, 1965.

\bibitem[RCMF22]{RobinsonLean}
Edward Hermann~Haeusler Robinson Callou M.C.~Filho, Jefferson de Barros~Santos.
\newblock Towards a proof in lean about the horizontal compression of dag-like
  derivations in minimal purely implicational logic.
\newblock In {\em Pre-Proceedings of the LFSA2022}, pages 8--23, 2022.

\end{thebibliography}
 \bibliographystyle{alpha}

 \appendix
\section{A polynomial time algorithm to verify validity of {\bf DLDS}}
\label{Algorithm}

In this section we present a (polynomial in time) algorithm to verifiy whether a {\bf DLDS} is valid or not. In fact, the algorithm verifies whether its input, a {\bf DLDS}, is a (partial sometimes) result of the compressing method shown in~\ref{sec:HC} when applied, even partially,  to any valid {\bf DLDS}. As a greedy tree-like derivation is a trivially valid {\bf DLDS} by the soundness of Natural Deduction for $M_{\imply}$, the correctness of the algorithm follows. However, any discussion on correctness is postponed to section~\ref{sec:HC-Soundness}. The algorithms described below runs on an input of type {\bf DLDS}, i.e., an element of $\langle V,(E_{D}^{i})_{i\in\mathcal{O}_{\Gamma}^{0}},E_{A},r,l,L,P\rangle$ as defined in definition~\ref{DLDS}, in the context of a set $\Gamma$ of formulas and a (fixed) linear ordering $\mathcal{O}_{\Gamma}^{0}$ on $\Gamma$ having two additional ordinals 0 and $\alpha$. Typically, $\Gamma$ is the set of subformulas of a purely implicational formula. The algorithm uses the type of {\bf Flows}, namely, $(\bigcup_{i\in\mathcal{O}_{\Gamma}^{0,\lambda}}E_{D}^{i})\longrightarrow (\mathcal{O}_{\Gamma}^{0,\lambda}\longrightarrow\mathcal{B}(\mathcal{O}_{S})\times (\mathcal{O}_{\Gamma}^{0,\lambda})^{\star})$ to take care of the propagation of dependency between levels of the underlying rooted graph of the {\bf DLDS}. To facilitate the readability, we call {\bf colors} to the type $\mathcal{O}_{\Gamma}^{0,\lambda}$, i.e., the type of {\bf Flow} is denoted by $(\bigcup_{i\in colors}E_{D}^{i})\longrightarrow (colors\longrightarrow\mathcal{B}(colors)\times colors^{\star})$.

The reader can observe that the algorithm below computes the function $Flow$ from top downwards. Returning an error, if the $Flow$ cannot be defined in some node of the {\bf DLDS} or returns the dependency set of the root, that should be empty in order to the {\bf DLDS} be valid.

\begin{fullwidth}[width=\linewidth+2cm,leftmargin=-1cm,rightmargin=-1cm] 
\begin{algorithm}[H]\label{checkDLDS}
  \caption{Verification that a DLDS $\mathcal{D}$ is a valid derivation from $\Delta\subseteq\Gamma$}
%  \algsetup{indent=1cm}
    \begin{algorithmic}[1]
      \Require{$\mathcal{D}=\langle V,(E_{D}^{i})_{i\in\mathcal{O}_{\Gamma}^{0}},E_{A},r,l,L,P\rangle$ and $\Delta\subseteq\Gamma$}
      \Require{$\langle V,\bigcup_{i\in \mathcal{O}_{\Gamma}^{0,\lambda}} E_{D}^{i}\rangle$ with root $r$}
      \Statex {\bf Uses} $S_1,S_2:Flow$
      \State $CheckRootedDagLevel(V,(E_{D}^{i})_{i\in\mathcal{O}_{\Gamma}^{0,\lambda}},E_{A},r)$
      \Let{$S_1$}{$\emptyset$}
      \Let{$S_2$}{$\emptyset$}
      \State $AddConclusionDummyEdge(\langle r,conc\rangle; \mathcal{D}; L(\langle r,conc\rangle)=\overline{\Delta}\rangle)$
      \Let{$Lev$}{$height(\mathcal{D})$}
      \For{$ln=Lev\;downto\;0$}
      \For{$p\;at\;level\;ln$}
      \For{$c\in colors$}
       \State {\bf try} $Premisses$ $\gets$ $Premiss(S_1,p,c)$ {\bf end try}
       \State {\bf catch} $(exception\;{\color{red}Invalid Premisses)}$ {\bf end catch}
       \State {\bf if} Hyp(p) {\bf then}
       \State \hspace{\indensp} {\bf if} (c==0) {\bf then}
       \State \hspace{2\indensp} {\bf if} $Premisses==\emptyset$ {\bf then}
       \State \hspace{3\indensp} $S_2(0)$ $\gets$ $(\overline{l(p)},\epsilon)$
       \State \hspace{2\indensp} {\bf else}
       \State \hspace{3\indensp} {\bf thrown} {\color{red} InvalidException}(``Vertex'' p `` as a regular (not related to $\lambda$-Edges) hypothesis cannot have incoming edges or premisses '')                    
       \State \hspace{\indensp} {\bf else}
       \State \hspace{2\indensp} {\bf if} $Premisses==\emptyset$ {\bf then}
       \State \hspace{3\indensp} {\bf if} $\exists g\in E_A(target(g)=p)$ {\bf then}
       \State \hspace{4\indensp} {\bf for} $g\in \{g: target(g)=p\}$ {\bf do}
       \State \hspace{5\indensp} {\bf try} $S(hd(P(g))$ $\gets$ $(\overline{l(p)},tail(P(g)))$ {\bf end try} {\bf catch} $(exception\;{\color{red}Invalid color/path})$ {\bf end catch}
       \State \hspace{4\indensp} {\bf end for}
       \State \hspace{3\indensp} {\bf else}
       \State \hspace{4\indensp} $S_2(c)$ $\gets$ $\emptyset$
       \State \hspace{2\indensp} {\bf else}
       \State \hspace{3\indensp} $S_2(c)$ $\gets$ $Update(Premiss,p,c,S_1)$
       \State {\bf else}
       \State \hspace{\indensp} {\bf if} (c==0) {\bf then}
       \State \hspace{2\indensp} {\bf if} $Premisses==\emptyset$ {\bf then}
       \State \hspace{3\indensp} {\bf thrown} {\color{red} InvalidException}(``Hypothesis should not have premisses'')
       \State \hspace{2\indensp} {\bf else}
       \State \hspace{3\indensp} $S_2(c)$ $\gets$ $Update(Premiss,p,c,S_1)$
       \State \hspace{\indensp} {\bf else}
       \State \hspace{2\indensp} {\bf if} $Premisses==\emptyset$ {\bf then} 
       \State \hspace{3\indensp} $S_2(c)$ $\gets$ $\emptyset$
       \State \hspace{2\indensp} {\bf else}
       \State \hspace{3\indensp} $S_2(c)$ $\gets$ $Update(Premiss,p,c,S_1)$
       \EndFor
    \algstore{HVerification}      

    \end{algorithmic}
\end{algorithm}
\end{fullwidth}

\begin{fullwidth}[width=\linewidth+2cm,leftmargin=-1cm,rightmargin=-1cm]
\begin{algorithm}[H]
   \addtocounter{algorithm}{-1}
    \caption{Verification that a DLDS $\mathcal{D}$ is a valid derivation}
    \begin{algorithmic}[1]
      \algrestore{HVerification}
             \For{$f\in out(p)$}
       \State {\bf if} $color(f)==0$ {\bf then}
       \State \hspace{\indensp} $S_1(f)(0)$ $\gets$ $S_2(0)$
       \State {\bf elseif} $color(f)>0$ {\bf then}
       \State \hspace{\indensp} {\bf if} $S_1(f)(color(f))==\emptyset$ {\bf then}
       \State \hspace{2\indensp} $S_1(f)(color(f))$ $\gets$ $S_2(color(f))$
       \State \hspace{\indensp} {\bf else}
       \State \hspace{2\indensp} {\bf thrown} {\color{red} InvalidException}(``$S_1$ already defined, should not pass by initial check of decorated rooted dag'')
       \State {\bf elseif} $color(f)==\lambda$ $\land$ $out(p)==1$ {\bf then}   HERE HERE
       \State \hspace{\indensp} {\bf for} $c^{\prime}\in colors$ {\bf do}
       \State \hspace{2\indensp} $S_1(f)(c^{\prime})$ $\gets$ $S_2(c^{\prime})$
       \State \hspace{\indensp} {\bf end for}
       \State {\bf else}
       \State \hspace{\indensp} {\bf thrown} {\color{red} InvalidException}(``undefined color of edge or more than one $\lambda$-edge going out a vertex'')
       \EndFor
       \State $S_2$ $\gets$ $\emptyset$
       \EndFor
       \EndFor
       \State Checking whether the assumption set for the conclusion of the derivation contains only the hypothesis (non-discharged assumptions)
       \State {\bf if} $S_1(\langle r,conc\rangle)(0)==(\Delta,\epsilon)\land \forall c(S_1(\langle r,conc\rangle)(c)==\emptyset)$ {\bf then}
       \State \hspace{\indensp} print(``$\mathcal{D}$ is a valid derivation of $l(r)$ from $\Delta$'')
       \State {\bf else}
       \State \hspace{\indensp} print(``$\mathcal{D}$ is NOT a valid derivation of $l(r)$ from $\Delta$'')
    \end{algorithmic}
\end{algorithm}
\end{fullwidth}

The following algorithm is used as a procedural unit in algorithm~\ref{checkDLDS}. As a matter of better readability we call $DedEdges$ the set
$(\bigcup_{i\in colors^{-\lambda}}E_{D}^{i})$, i.e., the set of deductive edges of any color, but $\lambda$. 

\begin{fullwidth}[width=\linewidth+2cm,leftmargin=-1cm,rightmargin=-1cm] 
\begin{algorithm}[H]\label{retrievePremiss}
  \caption{Retrieves premisses related to color $c$ of a rule that has node $p$ as conclusion}
%  \algsetup{indent=1cm}
  \begin{algorithmic}[1]
      \Statex $Premiss(S,p,c)$
      \Require{$\mathcal{D}=\langle V,(E_{D}^{i})_{i\in\mathcal{O}_{\Gamma}^{0}},E_{A},r,l,L,P\rangle$ {\bf Global}}
      \Require{$S:Flow$, $p\in V$ and $c\in colors$}
      \Ensure{A mapping $F:colors^{-\lambda}\rightarrow (DedEdges)^2\cup DedEdges$}
      \State {\bf if} $c>0$ {\bf then}
      \State \hspace{\indensp} {\bf if} $Suitable(\{f:(S(f)(c)\neq\bot)\land target(f)=p\},p)$ {\bf then}
      \State \hspace{2\indensp} $Return(\langle c, \{f:(S(f)(c)\neq\bot)\land target(f)=p\}\rangle)$
      \State \hspace{\indensp} {\bf else}
      \State \hspace{2\indensp} {\bf thrown} {\color{red} InvalidException}(``incoming edges are not suitable premisses'')
      \State {\bf elseif} $(c==0)$ {\bf then}
      \State \hspace{\indensp} {\bf if} $Hyp(p)$ {\bf then}
      \State \hspace{2\indensp} {\bf if} $\{f:(S(f)(c)\neq\bot)\land target(f)=p\}==\emptyset$ {\bf then}
      \State \hspace{3\indensp} $Return(\emptyset)$
      \State \hspace{2\indensp} {\bf else}
      \State \hspace{3\indensp} {\bf thrown} {\color{red} InvalidException}(``A hypothesis has no premisses of color 0'')      
      \State \hspace{\indensp} {\bf else}
      \State \hspace{2\indensp} {\bf if} $\{f:(S(f)(c)\neq\bot)\land target(f)=p\}==\emptyset$ {\bf then}
      \State \hspace{3\indensp} {\bf thrown} {\color{red} InvalidException}(``Only hypothesis have empty set of premisses'')            
      \State \hspace{2\indensp} {\bf else}
      \State \hspace{3\indensp} $Return(\langle c,\{f:(S(f)(c)\neq\bot)\land target(f)=p\}\rangle)$
      \State {\bf elseif} $c==\lambda$ {\bf then}
      \State \hspace{\indensp} $F$ $\gets$ $\emptyset$
      \State \hspace{\indensp} {\bf for} $co\in colors-\{\lambda\}$ {\bf do}
      \State \hspace{2\indensp} $F(co)$ $\gets$ $\{f:(S(f)(co)\neq\bot)\land target(f)=p    \}$
      \State \hspace{\indensp} {\bf end for}
      \State \hspace{\indensp} $Return(F)$
      \State {\bf else}
      \State \hspace{\indensp} {\bf thrown} {\color{red} InvalidException}()
    \end{algorithmic}
\end{algorithm}
\end{fullwidth}

The following algorithm is used as a functional unit in algorithm~\ref{checkDLDS} too.
It returns the partial {\bf Flow} associated to the color $c$, using the premisses $Ps$ to update the {\bf Flow} $S$ resulting in the partial {\bf Flow}. Verification of conformity (validity) of $Ps$,  $S$ and the
$DLDS$ in the context (global data) is also performed.

\begin{fullwidth}[width=\linewidth+2cm,leftmargin=-1cm,rightmargin=-1cm] 
\begin{algorithm}[H]\label{UpdatePartialFlow}
  \caption{Given presmisses $Ps$, a color $c$ and a {\bf Flow} $S$ returning the partial flow defined on $c$ obtained by updating the effect of the associated inference from the given data, including the case of $\lambda$-edge}
%  \algsetup{indent=1cm}
  \begin{algorithmic}[1]
      \Statex $UpdatePartialFlow(Ps:Premisses, p:node, c:color, S:Flow)$
      \Require{$\mathcal{D}=\langle V,(E_{D}^{i})_{i\in\mathcal{O}_{\Gamma}^{0}},E_{A},r,l,L,P\rangle$ {\bf Global}}
      \Require{$S:Flows$, $p\in V$ and $c\in colors, Ps\in Set(DedEdges)$}
      \Ensure{A pair $F:\mathcal{B}(\mathcal{O}_{S})\times (\mathcal{O}_{\Gamma}^{0,\lambda})^{\star}$}
      \State {\bf if} $\exists!x((source(x)=p)\land x\in E_D^{c}\land c\neq\lambda)$ {\bf then}
      \State \hspace{\indensp} g $\gets$ $\iota x((source(x)=p)\land x\in E_D^{c}\land c\neq\lambda)$
      \State \hspace{\indensp} {\bf if} $card(Ps)==2$ {\bf then}
      \State \hspace{2\indensp} {\bf if} $Is-\imply-E(Ps,g,S)$ {\bf then}
      \State \hspace{3\indensp} $F(c)$ $\gets$ $(\pi_1(S(minor(Ps))(c))\oplus \pi_1(S(major(Ps))), tail(\pi_2(S(minor(Ps))(c))))$
      \State \hspace{2\indensp} {\bf else}
      \State \hspace{3\indensp} {\bf thrown} {\color{red} InvalidException} (``Wrong Application of $\imply$-E rule. Cannot Update Flow '')
      \State \hspace{\indensp} {\bf elseif} $card(Ps)==1$ {\bf then}
      \State \hspace{2\indensp} {\bf if} $Is-\imply-I(Ps,g,S)$ {\bf then}
      \State \hspace{3\indensp} $F$ $\gets$ $(\pi_1(S(el(Ps))(c))-Antecedent(l(p)), tail(\pi_2(S(el(Ps))(c))))$      
      \State \hspace{2\indensp} {\bf else}
      \State \hspace{3\indensp} {\bf thrown} {\color{red} InvalidException} (``Wrong Application of $\imply$-I rule. Cannot Update Flow '')
      \State \hspace{\indensp} {\bf else}
      \State \hspace{\indensp} {\bf thrown} {\color{red} InvalidException} (``Wrong Application of rule. Cannot Update Flow '')
      \State {\bf elseif} $\exists!x((source(x)=p)\land x\in E_D^{\lambda})$ {\bf then}
      \State \hspace{\indensp} g $\gets$ $\iota x((source(x)=p)\land x\in E_D^{\lambda})$      
%      \State \hspace{\indensp} {\bf for} $co\in colors-\{\lambda\}$ {\bf do}
%      \State \hspace{2\indensp} $Prem$ $\gets$ $\{f:S(f)(c)\neq\bot\land target(f)=p\}$
      \State \hspace{2\indensp} {\bf if} $card(Prem)==2$ {\bf then}
      \State \hspace{3\indensp} {\bf if} $Is-\imply-E-\lambda(Prem,g,S)$ {\bf then}
      \State \hspace{4\indensp} $F$ $\gets$ $(\pi_1(S(minor(Prem))(c))\oplus \pi_1(S(major(Prem))), tail(\pi_2(S(minor(Prem))(c))))$
      \State \hspace{3\indensp} {\bf else}
      \State \hspace{4\indensp} {\bf thrown} {\color{red} InvalidException} (``Wrong Application of $\imply$-E rule. Cannot Update Flow '')
      \State \hspace{2\indensp} {\bf elseif} $card(Prem)==1$ {\bf then}
      \State \hspace{3\indensp} {\bf if} $Is-\imply-I-\lambda(Prem,g,S)$ {\bf then}
      \State \hspace{4\indensp} $F$ $\gets$ $(\pi_1(S(el(Prem))(c))-Antecedent(l(p)), tail(\pi_2(S(el(Prem))(c))))$      
      \State \hspace{3\indensp} {\bf else}
      \State \hspace{4\indensp} {\bf thrown} {\color{red} InvalidException} (``Wrong Application of $\imply$-I rule. Cannot Update Flow '')
      \State \hspace{2\indensp} {\bf else}
      \State \hspace{3\indensp} {\bf thrown} {\color{red} InvalidException} (``Wrong Application of rule. Cannot Update Flow '')
%      \State \hspace{\indensp} {\bf end for}
      \State {\bf else}
      \State \hspace{\indensp} {\bf thrown} {\color{red} InvalidException} (``Wrong Application of rule. Cannot Update Flow '')
      \end{algorithmic}
  \end{algorithm}
\end{fullwidth}

{\bf Observation.} The following definitions are used in algorithm~\ref{UpdatePartialFlow};
\begin{itemize}
\item $el(\{e\})=e$
\item $Is-\imply-E(Ps,g,S): EqStrings(l(source(minor(Ps)))\imply l(source(g)), l(source(major(Ps)))) \land (\pi_1(S(minor(Ps))(co))\oplus \pi_1(S(major(Ps))(co)) == L(g)) \land \pi_2(S(major(Ps))(c0))=\pi_2(S(minor(Ps)(co)))$
\item $Is-\imply-E-\lambda(Ps,g,S): EqStrings(l(source(minor(Ps)))\imply l(source(g)), l(source(major(Ps))))  \land (\pi_2(S(major(Ps))(c0))=\pi_2(S(minor(Ps)(co))))$
\item $Is-\imply-I(Ps,g,S): EqStrings(l(source(el(Ps))), Sucedent(l(source(g)))) \land (\pi_1(S(el(Ps))(co))-Antecedent(l(sourge(g))))==L(g))$
\item $Is-\imply-I(Ps,g,S): EqStrings(l(source(el(Ps))), Sucedent(l(source(g))))$
\end{itemize}

\section{An example using the function $Flow$}\label{appendix:ExemploFlow}

\begin{figure}[H]
  \includegraphics[width=12cm]{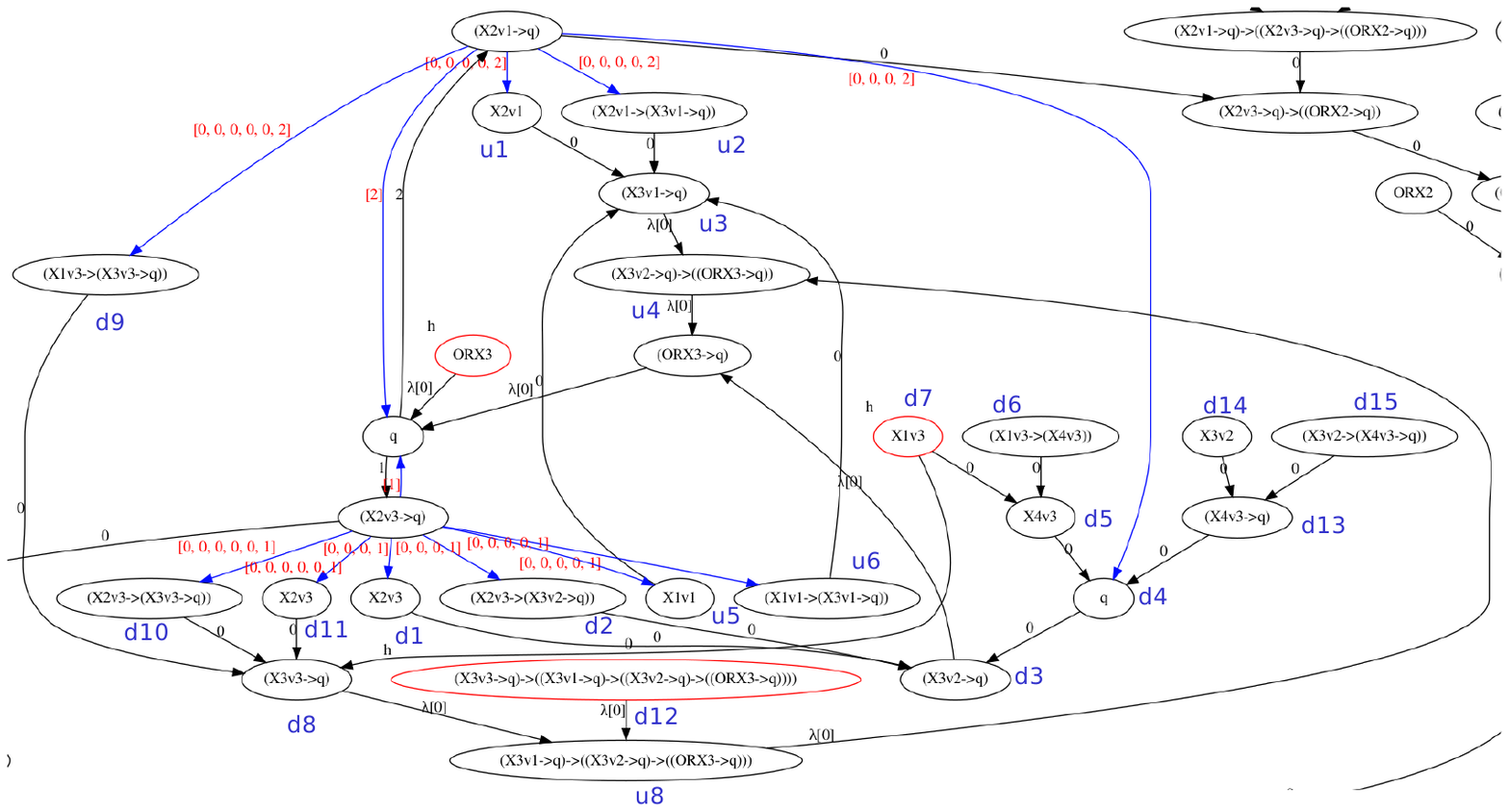}
  \caption{Part of the compressed DLDS from figure~\ref{G3} used in the explanation of $unfold$}
    \label{annotatedG3}
\end{figure}

%% \documentclass{article}
%% \usepackage[latin1]{inputenc}
%% \usepackage{float}
%% \usepackage{graphicx}
%% \usepackage{amsmath}
%% \usepackage[margin=0.25in]{geometry}
%% \newcommand{\imply}{\supset}
%% \begin{document}
Consider the portion of the (compressed) {\bf DLDS} $\mathcal{D}$ shown in figure~\ref{annotatedG3}. Note that the vertical positions of the nodes may have nothing to do with the usual downwards way of reading N.D. derivations. For example, the node labeled with ``$X2v1-> q$'', in the highest vertical position in the figure,  is the conclusion of an $\imply$-Intro rule that has the node labeled with $q$ as premiss. Observe that $q$ appears in a position below its conclusion.  
To illustrate the use  of the function $Flow$, definition~/ref{def:Flow}, we consider the node $w$, labeled with $ORX3->q$. We advice the reader that we only illustrate the definition of the function {\bf Flow} in this appendix. The verification depends on the labeling function $L$ that is not represented in the picture below due to a clearer visualization of the {\bf DLDS}. The computer program that generates this example produces a messy graph if we show the labeling function $L$. 

By definition~\ref{def:Flow}, we have the following values for $Flow(\mathcal{D},w)$, considering that the name of the nodes are displayed near each one of them in figura~\ref{annotatedG3}. With the sake of showing a cleaner explanation on this example, instead of using bitstrings for representing the dependency sets we use the informal and shorter set representation.

  \begin{align}
     Flow(\mathcal{D},w)(u_1)&=\{(\{X2v1\},00002)\}
    \label{eqn:1} \\
%  \end{align} 
%  \begin{align}
    Flow(\mathcal{D},w)(u_2)&=\{(\{X2v1\imply(X3v1\imply q)\},00002)\}
    \label{eqn:2} \\
%  \end{align} 
%  \begin{align}
    Flow(\mathcal{D},w)(u_6)& =\{(\{X1v1\imply(X3v1\imply q)\},00001)\}  
    \label{eqn:3} \\
%  \end{align} 
%  \begin{align}
    Flow(\mathcal{D},w)(u_5)& =\{(\{X1v1\},00001)\} 
    \label{eqn:4} \\
    Flow(\mathcal{D},w)(u_3)& =\{(\{X1v1,X1v1\imply(X3v1\imply q)\},0001),(\{X2v1,X2v1\imply(X3v1\imply q)\},0002)\}
  \end{align}

  For each node $u\in Pre(w)$, each $(\vec{b},p)\in Flow(\mathcal{D},w)(u)$ of type $(\mathcal{O}_{\Gamma}^{0})^{*}\times \mathcal{B}(\mathcal{O}_{S})$ is such that, $\vec{b}$ is or represents the dependency set that should label, using $L$, the unique deductive edge $\langle u,v\rangle$ of color $head(p)$ in $\mathcal{D}$. The function $Flow$ recursively calculates this dependency set along the set of nodes that are in all deductive paths from the top nodes reaching to $w$. For example, in $\mathcal{D}$, $u_1$ is a top node labeled with $X2v1$ and is target of an ancestor edge labeled with path $00002$. This label, in a valid {\em DLDS}, is the relative address of the source of the ancestor edge, the node just above left $u_1$ in figure~\ref{annotatedG3}, and labeled with $X2v1\imply q$. Thus, the pair $(\{X2v1\},00002)$ carries the information that the edge $\langle u_1,u_3\rangle$ of color $0=head(00002)$ is or should be labeled by $L$ with the bitstring that represents $\{X2v1\}$. An important observation is that when the {\em DLDS} is a tree-like deduction, $Flow(\mathcal{D},w)(v)=\{(\vec{b},\emptyset)\}$ for every $w$ and $v$ and $\vec{b}$ the usual dependency set for tree-like deductions. However, when we have already collapsed nodes, as it is the case of $u_3$, we have to create a consistent flow of information. Note that $Flow(\mathcal{D},w)(u_3)$, above, is not a singleton, it has two elements, they are respectively related to each derivation that share the node $u_3$ after the collapse. The pair $(\{X1v1,X1v1\imply(X3v1\imply q)\},00001)$ has to do with the subderivation that has $u_1$ and $u_2$ as premisses and $u_3$ as conclusion, and  relative address going to the formula $X2v1\imply q$, while the pair $(\{X2v1,X2v1\imply(X3v1\imply q)\},00002)$ goes to $X2v3\imply q$ and has premisses in $u_5$ and $u_6$ and shared  conclusion $u_3$. The set $Flow(\mathcal{D},w)(u_3)$ is, hence, a two-element set. The number of elements of $Flow(\mathcal{D},w)(v)$ reflects how many different ways of deducting $v\in Pre(w)$ there are in $\mathcal{D}$. 

  {\bf About the label $\lambda$}. We noted, in the above paragraph,  that $Flow(\mathcal{D},w)(u_3)$ has two elements and hence is related with two different dependency sets, one is $\{X1v1,X1v1\imply(X3v1\imply q)\}$ and the other is $\{X2v1,X2v1\imply(X3v1\imply q)\}$. In this case, the {\em DLDS} cannot have both as labels assigned by $L$ to the edge $\langle u_3,u_4\rangle$, we use the label $\vec{\lambda}$ to indicate that the dependency set should be  considered dynamically to be used downwards in a verification of validity. As a consequence,  the {\em DLDS} validity conditions that involve the function $Flow$, item [CorrectRuleApp] in definition~\ref{def:ValidDLDS}, has some validity verification cases related on the $\lambda$. We have to observe that the label $\lambda[0]$ that appears in figure~\ref{annotatedG3} labeling some edges is used as an aggregation of the labels $\lambda$ and the $0$. This is due only to a better visualization and has not to do with any technical or any other deeper meaning. The following values of $Flow$ show the results in $Pre(w)$. Observe that the local validity of the rule, i.e., whether the node represents a valid $\imply$ introduction or elimation rule is verified by the $Flow$ function. In the case that the rule is not correctly applied the function $Flow$ is not well-defined.

  \begin{align}
     Flow(\mathcal{D},w)(d_{10})&=\{(\{X2v3\imply(X3v3\imply q)\},000001)\}
    \label{eqn:11} \\
%  \end{align} 
%  \begin{align}
    Flow(\mathcal{D},w)(d_{11})&=\{(\{X2v3\},000001)\}
    \label{eqn:21} \\
%  \end{align} 
%  \begin{align}
    Flow(\mathcal{D},w)(d_9)& =\{(\{X1v3\imply(X3v3\imply q)\},000002)\}  
    \label{eqn:31} \\
%  \end{align} 
%  \begin{align}
    Flow(\mathcal{D},w)(d_7)& =\{(\{X1v3\},\emptyset)\} 
    \label{eqn:41} \\
    Flow(\mathcal{D},w)(d_8)& =\{(\{X1v3,X1v3\imply(X3v3\imply q)\},00002),(\{X2v3,X2v3\imply(X3v3\imply q)\},00001)\}
    \label{eqn:51} \\
    Flow(\mathcal{D},w)(d_{12})&=\{((\{X3v3\imply q)\imply((X3v1\imply q)\imply((X3v2\imply q)\imply ((ORX3\imply q))))\},\emptyset)\}
     \label{eqn:61} \\
     Flow(\mathcal{D},w)(u_8)& =\{(\{X1v3,X1v3\imply(X3v3\imply q), \nonumber \\
     &(X3v3\imply q)\imply((X3v1\imply q)\imply((X3v2\imply q)\imply ((ORX3\imply q))))\},0002), \nonumber \\
     & (\{X2v3,X2v3\imply(X3v3\imply q), \nonumber\\
     &(X3v3\imply q)\imply((X3v1\imply q)\imply((X3v2\imply q)\imply ((ORX3\imply q))))\},0001)\}
      \label{eqn:71} \\
     Flow(\mathcal{D},w)(u_4)& =\{(\{X2v1,X2v1\imply(X3v1\imply q),X1v3,X1v3\imply(X3v3\imply q), \nonumber\\
     & \;\;\;\;\;\;\;\;(X3v3\imply q)\imply((X3v1\imply q)\imply((X3v2\imply q)\imply ((ORX3\imply q))))\},002), \nonumber\\
     & \;\;\;\;\;\;(\{X1v1,X1v1\imply(X3v1\imply q), \nonumber\\
     & \;\;\;\;\;\;\;X2v3,X2v3\imply(X3v3\imply q), \nonumber\\
     & \;\;\;\;\;\;\;(X3v3\imply q)\imply((X3v1\imply q)\imply((X3v2\imply q)\imply ((ORX3\imply q))))\},001) \nonumber\\
     & \;\;\;\;\}
     \label{eqn:91} 
  \end{align}

The value returned by $Flow(\mathcal{D},w)(u_4)$, in line~\ref{eqn:91}, represents two different deductions of $ORX3\imply q$. In the sequel, each deduction is associated to its respective place in the original (uncompressed) Natural Deduction derivation, by following the paths , here $001$ and $002$. The first is a deduction of $X2v3\imply q$ and the second is a deduction of $X2v1\imply q$. The node labeled with $q$ is from where the mentioned paths diverge to each respective conclusion.  

\end{document}